\documentclass[12pt]{article}
\usepackage[utf8]{inputenc}
\usepackage{enumerate,amsmath,amsthm, amsfonts, float, amssymb,physics, graphicx, authblk, marvosym, hyperref, listings, setspace, placeins, caption, dsfont, setspace, authblk}
\usepackage[usenames, dvipsnames]{color}
\def\prodn{{\prod_{i = 1}^n}}

\usepackage[left=1in, right=1in, top=1in, bottom = 1in]{geometry}

\usepackage{natbib}

\usepackage[toc,page]{appendix}
\usepackage{times}
\usepackage{ragged2e}
\usepackage{multirow}

\date{}

\title{A Pseudo-likelihood Approach to Under-5 Mortality Estimation}
\author[1]{Taylor Okonek}
\author[2]{Katherine Wilson}
\author[2,3]{Jon Wakefield}
\affil[1]{Department of Mathematics, Statistics, and Computer Science, Macalester College, 1600 Grand Ave, Saint Paul, MN 55105, USA}
\affil[2]{Department of Biostatistics, University of Washington, 3980 15th Ave NE, Box 351617, Seattle, WA 98195, USA}
\affil[3]{Department of Statistics, University of Washington, Padelford Hall, NE Stevens Way, Seattle, WA 98195, USA}
\affil[ ]{\textit{tokonek@macalester.edu, wilsonkl@uw.edu, jonno@uw.edu}}

\begin{document}
	
	\maketitle 
	
	\begin{abstract}
		Accurate and precise estimates of the under-5 mortality rate (U5MR) are an important health summary for countries. However, full survival curves allow us to better understand the pattern of mortality in children under five. Modern demographic methods for estimating a full mortality schedule for children have been developed for countries with good vital registration and reliable census data, but perform poorly in many low- and middle-income countries (LMICs). In these countries, the need to utilize nationally representative surveys to estimate the U5MR requires additional care to mitigate potential biases in survey data, acknowledge the survey design, and handle the usual characteristics of survival data, for example, censoring and truncation. In this paper, we develop parametric and non-parametric pseudo-likelihood approaches to estimating child mortality across calendar time from complex survey data. We show that the parametric approach is particularly useful in scenarios where data are sparse and parsimonious models allow efficient estimation. We compare a variety of parametric models to two existing methods for obtaining a full survival curve for children under the age of 5, and argue that a parametric pseudo-likelihood approach is advantageous in LMICs. We apply our proposed approaches to survey data from four LMICs.
	\end{abstract} 

	\RaggedRight
	
\section{Introduction}

Estimates of child mortality rates for specific age groups at a national and subnational level provide important information on the health of a country and inform targeted public health interventions. Historically, estimates of interest have been the neonatal mortality rate (NMR: probability of dying before age 1 month), the infant mortality rate (IMR: probability of dying before age 1 year), and the under-5 mortality rate (U5MR: probability of dying before the age of 5). While these summaries give a rough picture of the pattern of mortality under the age of 5, they do not constitute a \textit{complete} pattern of mortality before the age of 5. As such, producing a full, continuous survival curve for children under the age of 5 is of interest for informing targeted interventions \citep{verhulst2022divergent, guillot2022modeling}, and quantifying the differences in mortality patterns between countries.

Modern demographic methods for estimating a full mortality schedule for children under the age of 5 have been developed in a high-income country setting where vital registration information is readily available \citep{guillot2022modeling, eilerts2021age, verhulst2022divergent}. One such method is the log-quad model \citep{guillot2022modeling}, which uses the Human Mortality Database (HMD) \citep{barbieri2015data} to obtain a continuous curve quantifying the relationship between age and the (log) probability of dying before a given age. This approach uses the HMD to obtain parameter values, which are plugged into the log-quad model's formula to obtain full, continuous curves. \cite{guillot2022modeling} note that the patterns of mortality that are estimated from the model are importantly different from the observed data in LMICs. \cite{eilerts2021age} and \cite{verhulst2022divergent} note that sub-Saharan African and south Asian countries typically observe higher levels of the child mortality rate (CMR: the probability of dying between ages 1 and 5 given survival to age 1) for a given IMR when compared to high-income countries. \cite{verhulst2022divergent} call this a ``very late" pattern of under-5 mortality.   

Another popular method that makes use of HMD life tables is the Singular Value Decomposition (SVD) approach described in \cite{clark2019general}. Here, the information from HMD life tables are compressed into three or four principal components that summarise observed full mortality schedules over an entire lifetime. Although this approach can be used more generally with other lifetables (see \citet{alexander2017flexible}, for example), the SVD approach used in conjunction with HMD life tables as in \citet{clark2019general} is intended to produce all-age mortality schedules at a yearly scale. As this requires the assumption of a constant mortality hazard within yearly age groups, this specific application of SVD is not well-suited for estimating child mortality, since a constant hazard between ages 0 and 1 year is an unrealistic assumption.

In addition to different patterns of under-5 mortality in LMICs compared to high-income countries, the data sources available in LMICs typically differ from high-income countries. In most high-income countries, vital registration and reliable census information are readily available, hence the mortality data is more granular and potentially subject to fewer biases than are present in data from LMICs. In countries without vital registration data or reliable census information, we instead rely on nationally representative surveys. In many LMICs, the Demographic and Health Surveys (DHS) are considered the most reliable source of information for such outcomes, and they are conducted with reasonably high frequency (the aim is every 5 years). To date, DHS has conducted more than 400 surveys in over 90 countries, and is one of the primary data sources used in the production of child mortality estimates by the UN Inter-agency Group for Child Mortality Estimation (IGME) \citep{alkema2014global}. Survey-weighted estimates of health outcomes with variance estimates that account for the survey design are preferred when there is enough data to obtain such estimates with high precision. 

As noted in \citet{hill1995age}, \cite{lawn2008four}, and \cite{guillot2022modeling}, surveys such as the DHS may be subject to biases in addition to other data limitations. One example of bias is age-heaping, where more children are recorded as having died at particular ages than is truly the case. In DHS surveys, this often occurs at age 12 months (see Section A.2 of the supplementary material for examples). Additionally, the ages at death of most children are not observed exactly (i.e., censored). This combined with the need to appropriately account for survey weights and potential biases from age-heaping form statistical modeling challenges that are unique to surveys in LMICs; all of these challenges have not yet been addressed simultaneously in the literature.

An additional challenge specific to U5MR estimation is distinguishing between cohort- and period-estimates of mortality. When estimating U5MR, we typically want to obtain period-specific estimates rather than cohort-specific estimates, as the most recent, cohort-specific estimates of U5MR we could obtain will always be five years in the past. Period estimates are for ``synthetic" children, where the usual approach envisages a cohort of children that live their first five years of life all in a single time period. This is opposed to a real cohort of children who are born in one time period and move through time (periods) as they age. The concept of synthetic people (children or otherwise) allows us to provide estimates of demographic indicators such as life expectancy or U5MR that are a reasonable summary of the current state of the mortality pattern. In practice, when estimating a demographic indicator for synthetic children, we consider what a real child would contribute to each period \textit{as though} they were a synthetic child. As detailed in Section \ref{sec:survival_framework}, in a survival analysis framework this corresponds to treating the time period as a time-varying covariate. While existing methods have made use of this approach in a discrete survival setting \citep{mercer2015space}, none have \textit{explicitly} formulated the problem as that of a time-varying covariate in continuous time.

In this paper we propose a pseudo-likelihood estimate of full mortality schedules for children under the age of 5 in LMICs, that takes full advantage of the granularity of the data available while accounting for both the survey design and potential biases in the surveys. Rather than assume a model based on data from high-income countries, we instead deal with DHS data directly to obtain both an estimate of the survival curve in LMICs at a national level. These methods can flexibly incorporate a variety of parametric distributions, and are readily extendable to subnational estimates. In Section \ref{sec:survival_framework} we reframe the production of period estimates for under-5 mortality rates in LMICs using continuous survival models for mortality with a time-varying covariate representation, accounting for potential censoring. In Section \ref{sec:proposedapproach} we outline our proposed methodology in addition to two existing methodologies currently used in child mortality estimation. Section \ref{sec:application} contains an application to four LMICs, a discussion of model validation, and results. We conclude with a discussion of benefits of our proposed approach, limitations, and future work in Section \ref{sec:discussion}.

\section{Survival framework}
\label{sec:survival_framework}

To begin, we define notation that is common in the demography and statistics literature, and is used throughout this paper. Mortality is typically estimated as a risk (probability) that a child dies before a certain age. Let $X$ denote the survival time. We denote the probability that a child died between the ages of $x$ and $x + n$, given that they survived until at least age $x$, as ${}_nq_x = \Pr(X < x + n \mid X > x)$. With age given in months, as we do throughout, we therefore denote U5MR as ${}_{60}q_0$, IMR as ${}_{12}q_0$, and NMR as $_1q_0$. 

We treat mortality as a time-to-event outcome in a survival framework. In this framework our estimand of interest is the survival curve $S(x)$, or the probability of surviving to at least age $x$. We can directly translate quantities ${}_xq_0$ to a survival curve via $S(x) = 1 - {}_xq_0$, and ${}_nq_x = 1 - S(x + n)/S(x)$.

\begin{figure}[!h]
	\centering
	\includegraphics[scale = 0.35]{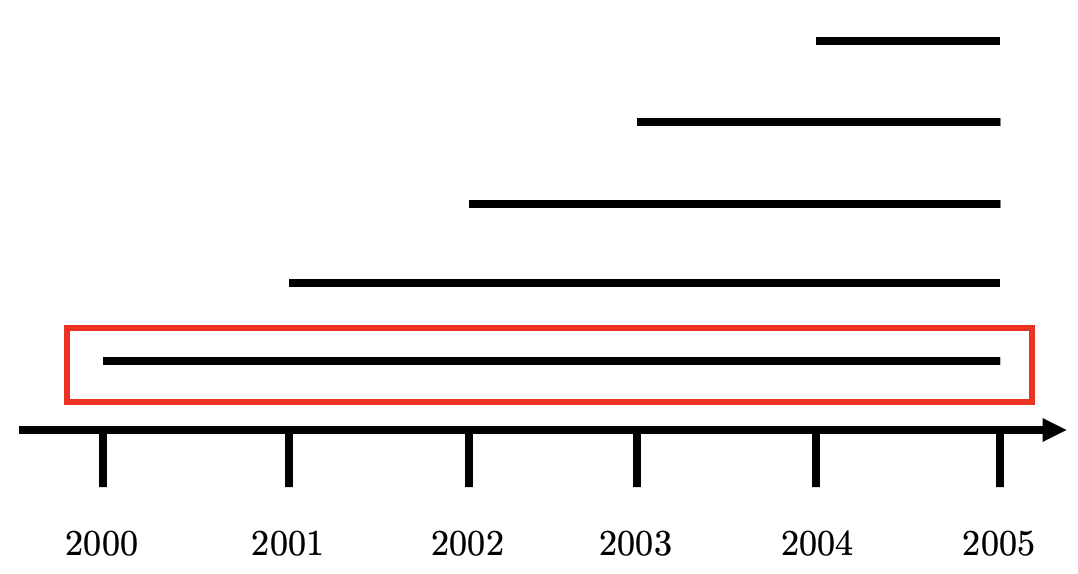}
	\includegraphics[scale = 0.35]{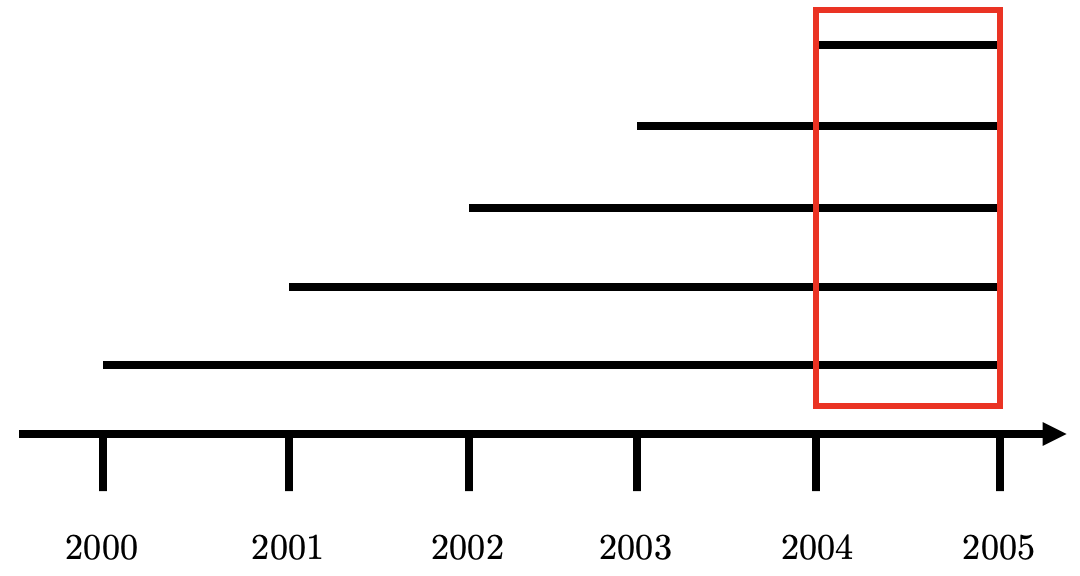}
	\caption{Left-hand side: Potential lifespans of observations used to obtain a cohort estimate of U5MR for the cohort born in 2000. Right-hand side: Potential lifespans of observations used to obtain a period estimate of U5MR for the year 2004. All children are assumed to be born on January first of a given year. Horizontal lines indicate the potential lifespans of children up to January 1st, 2005.}
	\label{fig:period_vs_cohort}
\end{figure}

An important distinction in demography, that again has a survival analysis flavor, is period versus cohort estimates. The age-period-cohort distinction is subtle but well-documented (see \citet{carstensen2007age}, for example). Importantly, the subset of data used to estimate cohort and period estimates differs. In Figure \ref{fig:period_vs_cohort}, we illustrate this difference. For simplicity in this explanation, we assume all children are born on January 1st of a given year. We see that the data used to obtain a \textit{cohort estimate} of U5MR for the cohort born in 2000 consists of only children born in the year 2000. Note that we will always be five years behind schedule in terms of estimate production because we need to observe the full, first five years of a cohort before calculating cohort U5MR.

The data used to obtain a period estimate of U5MR for the year 2004 contains data from five distinct cohorts: cohorts 2000, 2001, 2002, 2003, and 2004, as seen on the right-hand side of Figure \ref{fig:period_vs_cohort}. Of note, when obtaining cohort estimates, both age and time align, whereas when obtaining period estimates, age and time are distinct. This is because the age of a synthetic child is not directly tied to time as we observe it. Therefore, we let ${}_xq_{n,p}$ vary by period $p$ in addition to the varying by age. For example, we may write the probability a child dies between the ages of 1 and 2 in the year $2001$ as ${}_{12}q_{12,2001}$, where the deaths that inform this estimate must come from the cohort of children born in $2000$ who survive until at least age 1. 

It is important to note that when computing period estimates, some of the data will be subject to left-truncation. In general, left-truncation, also known as late entry, occurs if an individual is not at risk of experiencing the event prior to a certain left-truncation time. If not dealt with, left-truncation can induce selection bias, as individuals who experience the event prior to left-truncation time would inherently not be included in the data. In our applications, left-truncation occurs when an individual is not at risk of dying in a specific age band, within a specific time period because they are \textit{older} than that specific age band before the period begins. Truncation accounts for the potential bias that would be introduced into our estimate from the individuals who were born in the earlier cohorts, yet died before our time period of interest. As an example, in Figure \ref{fig:period_vs_cohort} (right panel), all individuals born at the beginning of the year 2000 \textit{who are still alive by 2004} would be subject to left-truncation at age 4 when computing their contribution to the period U5MR estimate for 2004. If we compute the period estimate of ${}_{60}q_{0, 2004}$ using five, discrete values, ${}_{12}q_{48,2004}$, ${}_{12}q_{36,2004}$, ${}_{12}q_{24,2004}$, ${}_{12}q_{12,2004}$, ${}_{12}q_{0, 2004}$, for each cohort born in 2000 through 2004, respectively, left-truncation is dealt with implicitly through the conditional probability structure of ${}_nq_x$, as we will see in the discrete hazards approach \citep{allison2014event, mercer2015space} described in Section \ref{sec:discretehaz}. This accounts for the artificially smaller risk set that each individual contributes to the period estimate, based on the age at which they enter a given period.

If we are interested in obtaining estimates of U5MR for multiple periods across time, this truncation structure can be incorporated into a model by treating period as a time-varying covariate. This is done implicitly in a discrete hazards approach, but can be done explicitly in the pseudo-likelihood approach we propose that allows us to use continuous survival models for age. The discretely-categorized variable, period, is treated as a covariate that changes \textit{through} time, and is simply an indicator for which synthetic cohort we are considering.

A final piece of the survival framework is how we deal with censored observations. Not all children die before 5 years of age. Since we are interested in our application in estimating a continuous survival curve for children \textit{only up to the age of 5}, all children who do not die before 5 years of age will be right-censored at that age, since they can no longer be at risk of dying \textit{under} the age of 5 at later ages. Children who die in a time period later than the period in which they were born also contribute right-censored observations to those earlier time periods. A survival framework also allows us to deal with interval censoring, where we know only that an event has occurred for an individual between two ages. DHS surveys contain daily observed death dates for children who died before the age of 1 month, monthly, interval-censored observations for children who died between 1 month and 24 months (e.g., we may observe that a child that dies between the ages of 2 and 3 months), and yearly, interval-censored observations for children who died after 2 years of age. There are some rare exceptions in the data where the DHS records more detailed information for particular children. Interval censoring can be appropriately addressed by discretely categorizing observations, as is done in some of the existing approaches described in Section \ref{sec:proposedapproach}, but can also be addressed in the continuous survival framework that we propose.

\section{Methods}
\label{sec:proposedapproach}

Historically, survival methods have been extended to the survey setting, in the context of the Cox proportional hazards model \citep{binder1992fitting, lin2000fitting, breslow2007weighted, breslow2008az}. Such methods address the survey aspect of the data via a pseudo-likelihood approach to estimation \citep{binder1983variances}, in which we weight each individual's likelihood contribution by their sampling weight, and maximize the pseudo-likelihood to give weighted (pseudo) MLEs. The variance of the estimates is computed via sandwich estimation. A brief description of the general approach to pseudo-likelihood estimation described in \citet{binder1983variances} is given in Section B of the supplementary material.

Bootstrap and jackknife procedures have been developed for variance estimation for various complex survey designs, including the two-stage, stratified cluster design common to DHS surveys. To obtain bootstrapped variance estimates, $n_h - 1$ clusters are sampled with replacement within strata $h$, where $n_h$ is the number of clusters in strata $h$ \citep{rao1988resampling}. Pointwise confidence intervals may then be constructed using percentiles of the bootstrap samples. A jackknife procedure for the same setting is described in \cite{pedersen2012child}.

For the remainder of this section, we describe two proposed approaches for estimating continuous survival curves for synthetic children across multiple time periods. Both methods are based on pseudo-likelihood---one nonparametric, and one parametric---and are novel in the context of child mortality estimation in LMICs. Existing approaches are described in Sections \ref{sec:logquad} and \ref{sec:discretehaz}.

\subsection{Nonparametric approach}
\label{sec:turnbull}

The classic and most popular nonparametric estimate of a survival curve is the Kaplan-Meier estimator \citep{kaplan1958nonparametric}. Let $t_i$ be a time when at least one event (death) occurred, $d_i$ be the number of events that occurred at time $t_i$, and $n_i$ be the number of children who have not had an event or been censored up to time $t_i$. Then the Kaplan-Meier estimator of the survival curve at time $t$ is
\begin{align*}
	\hat{S}(t) = \prod_{i:t_i \leq t} \left( 1 - \frac{d_i}{n_i} \right).
\end{align*}

Under noninformative right censoring and left truncation, the Kaplan-Meier estimator is the nonparametric maximum likelihood estimator (NPMLE) of the survival curve. However, the Kaplan-Meier estimator, in it's simplest form, is unsuitable for interval censored data. A generalization of the Kaplan-Meier estimator to arbitrarily truncated and censored observations is the Turnbull estimator \citep{turnbull1976empirical}. The identifiability of the Turnbull estimator for the interval censoring case we consider was proven by \cite{wang1994}. In Section C.1 of the supplementary material we introduce notation for the Turnbull estimator and describe the estimator alongside an example for our motivating application. Incorporating survey weights into the Turnbull estimator is straightforward. This extension to the Turnbull estimator is detailed in Section C.1 of the supplementary material, and produces a survey-weighted, NPMLE for arbitrarily truncated and censored data with survey weights. 

While this approach can produce point estimates for child mortality survival curves in an LMIC context, uncertainty quantification is not straightforward. First, it should be noted that due to the fixed structure of interval censoring present in DHS data (as noted at the end of Section \ref{sec:survival_framework}), the Turnbull estimator will never, in practice, allow us to obtain any information about the survival curve \textit{between} age groups defined by this structure. In theory, since deaths are recorded daily in the first month of life and children can die in any given month, given enough data the Turnbull estimator would produce what is essentially a complete survival curve with the only information missing being between 24 hour periods. However, since child deaths are rare, we end up with large gaps of information in the survival curves produced from this method when applied to DHS data.

Second, \cite{groeneboom1992information} note that, compared to the Kaplan-Meier estimator, the Turnbull estimator has less appealing asymptotics. In the interval censoring case we consider in this paper, the estimator converges pointwise (i.e., at a fixed value $t$) at a rate of $(n \log(n))^{1/3}$ to a non-Gaussian distribution. The question of obtaining valid confidence bands for the Turnbull estimator remains an open statistical question. Though some have recommended using a bootstrap procedure for variance estimation (see \cite{sun2001variance}, for example), the coverage of these procedures is not well justified (and therefore not necessarily correct) due to the rate of convergence and non-Gaussian asymptotics. 

Although the bootstrap is not well-justified for the Turnbull estimator, we do use the bootstrap procedure appropriate for a two-stage, stratified sampling design from \cite{rao1988resampling}, described at the beginning of this section, to assist with model comparison in our application. In our model comparison approach, we treat the Turnbull estimator as a baseline estimate of the survival curve, and aim to determine whether a given parametric model is ``reasonably" close to the Turnbull estimator. Obtaining some measure of uncertainty for the Turnbull estimator facilitates this comparison.

As a well-justified variance estimator is not available for the Turnbull estimator, we do not recommend using the Turnbull estimator for \textit{official} estimates of full mortality schedules for children under the age of 5 in LMICs. It is especially important in scenarios where the data does not come from a census or other vital registration source to accurately quantify the uncertainty of estimates. The Turnbull approach does, however, produce a point estimate of the survival function, and therefore is a useful reference when assessing how well a parametric distribution summarizes the pattern of U5MR in LMICs, as its point estimates do not rely on parametric assumptions.

\subsection{Parametric approach}
\label{subsec:par_approach}

Suppose we have children $i = 1, \dots, n$. Let,
\begin{enumerate}
	\item $p = 1, \dots, P$: consecutive time periods, which may be single years or combinations of years (e.g. 1 or 5 year periods)
	\item $l_p$: length of period $p$, measured in the same units as age of child
	\item $y_p$: date at the start of time period $p$ 
	\item $b_i$: child's date of birth
	\item $a_{pi} = y_p - b_i$: the age the child would be at $y_p$
	\item $I_i$: an indicator that child $i$ is interval-censored. If $I_i = 1$, child $i$ is interval-censored. If $I_i = 0$, child $i$ is right-censored or has an exact death time
	\item $E_i$: an indicator that child $i$'s death is exactly observed. If $E_i = 1$, then $I_i = 0$, and if $E_i = 0$, then $I_i$ could be $0$ or $1$ 
	\item $t_{i}$: child's age at right-censoring or age at death
	\item $t_{0i}$: child's age at beginning of interval censoring, if child is interval censored
	\item $t_{1i}$: child's age at end of interval censoring, if child is interval censored
	\item $\tilde{p}_i$: if $E_i = 1$, the period in which that child died
	\item $U_{x_i}(p) = \left\{ p: a_{pi} > -l_p, a_{pi} < x_i \right\}$. $U_{x_i}(p)$ is the set of periods for which child $i$ is alive and at risk of dying, where $x_i$ is one of $t_i$, $t_{0i}$, or $t_{1i}$ where appropriate
\end{enumerate}

\noindent Let $F_{\boldsymbol{\theta}}$ denote the CDF for the specified parametric distribution, and $H_{\boldsymbol{\theta}}$ the corresponding cumulative hazard function, dependent on a set of unknown parameters $\boldsymbol{\theta}$. In the case of simple random sampling, the likelihood for all individuals in our dataset across all time periods can be written as
{\small
	\begin{align*}
		L(\boldsymbol{\theta}) & = \prod_{i = 1}^n L_i(\boldsymbol{\theta}) \\
		& = \prodn \left[ 1 - F_{\boldsymbol{\theta}, i} (t_i) \right]^{1 - I_i} \left[  F_{\boldsymbol{\theta}, i} (t_{1i}) - F_{\boldsymbol{\theta}, i} (t_{0i}) \right]^{I_i} [f_{\boldsymbol{\theta}, i}(t_i)]^{E_i}, \\
		& = \prodn \underbrace{\left[ \exp(-H_{\boldsymbol{\theta},i} (t_i)) \right]^{1 - I_i}}_{right-censored} \underbrace{\left[  \exp(-H_{\boldsymbol{\theta},i} (t_{0i})) - \exp(-H_{\boldsymbol{\theta},i} (t_{1i})) \right]^{I_i}}_{interval-censored} \underbrace{[\exp(-H_{\boldsymbol{\theta},i}(t_i)) h_{\boldsymbol{\theta}, \tilde{p}_i}(t_i)]^{E_i}}_{exact},
	\end{align*}
}
where
\begin{align*}
	H_{\boldsymbol{\theta},i}(x_i) = \sum_{U_{x_i}(p)} \int_{\text{max}\{ a_{p_i}, 0\}}^{\text{min}\{ x_i, a_{p_i} + l_p \}} h_{\boldsymbol{\theta},p}(u) du,
\end{align*}
and $h_{\boldsymbol{\theta},p}(u)$ is a period-specific hazard function for a specified, parametric distribution.

To obtain survey-weighted estimates, we obtain pseudo-MLEs \citep{binder1983variances} of the distribution-specific parameters by maximizing the sum of log likelihood contributions for each individual observation multiplied by the survey weights. To obtain finite population variance estimates, we use a trick in which we treat our estimator as a weighted total, and use \texttt{R}'s \texttt{survey} package. The details of this calculation are given in Section C.2 of the supplementary material. 

As our proposed methodology focuses on providing a continuous, age-specific mortality curve for children under the age of 5, we focus on two existing methods that can provide this, modulo a few assumptions: the log-quad model \citep{guillot2022modeling}, and the discrete hazards model \citep{li2019changes, wu2021}. The latter requires the assumption that the discrete hazards ${}_nq_x$ estimated for each $x$ are constant within the interval $[x, x+ n)$ in order to obtain a full survival curve. 

Of note, \cite{scholey2019age} proposed a continuous, parametric approach model for infant mortality. There are similarities between it and our proposed approach, notably the use of a continuous hazard to assist in defining a survival curve for children. The method differs in its focus on the pattern of infant mortality as opposed to U5MR, the use of daily observed deaths from a high-income country which removes the need to account for interval-censored observations, and the use of data that does not come from a survey and therefore does not need to account for the survey design. The methods described in \cite{scholey2019age} serve as high income country analogues to our proposed methods, and we consider one family of hazards that produces the best fitting survival curve for U.S. data in our proposed methodology. 

The proposed approach can handle age heaping by lengthening the intervals in which children die, surrounding the time when age heaping is assumed to occur. In our application, we address age heaping at 12 months by interval-censoring observations recorded as having died between 6 and 18 months for that entire 12 month period, $[6, 18)$. We chose this window to capture a wide range of potential age-heaping surrounding 12 months, but other windows could instead be chosen, depending on assumptions about when age-heaping occurs. In aggregating data over these longer intervals, we will lose some precision in our estimate of the survival curve but should decrease bias due to age heaping, without needing to discard the information that age-heaped individuals provide for our estimates. We emphasize that the benefit of this straightforward approach to addressing age heaping is that the assumptions involved \textit{are made clear}, in this case, that the only age heaping in our data occurs between 6 and 18 months. Incorporating additional assumptions about where age heaping occurs is straightforward; we can include additional intervals surrounding the dates where age heaping is thought to occur (for example, ages 3-10 days for age heaping at 7 days). 




\subsection{Log-quad model}
\label{sec:logquad}

The log-quad model described in \cite{guillot2022modeling} builds on the approach in \cite{wilmoth2012flexible}, and can provide an estimate of a continuous survival curve from ages 0 to 5 using only an observed or previously estimated ${}_{60}\hat{q}_0$. Other optional inputs to the log-quad model include values ${}_xq_0$ for different ages $x$. Following \cite{clark2019general}, we call the model ``empirical" because the coefficients input to the model are not estimated during the modeling process, but instead are computed beforehand using data from the Under-5 Mortality Database (U5MD) \citep{guillot2022modeling}. The model specifies
\begin{align*}
	\log({}_xq_0) = a_x + b_x \log({}_{60}\hat{q}_0) + c_x \log({}_{60}\hat{q}_0)^2 + v_x k,
\end{align*}
where $x$ takes on one of the 22 values $\{7d,$ $14d,$ $ 21d,$ $28d,$ $2m,$ $3m,$ $4m,$ $5m,$ $6m,$ $7m,$ $8m,$ $9m,$ $10m,$ $11m,$ $12m,$ $15m,$ $18m,$ $21m,$ $2y,$ $3y,$ $4y,$ $5y\}$. The age-specific coefficients $\{a_x, b_x, c_x, v_x\}$ are provided in the U5MD, ${}_{60}\hat{q}_0$ is input to the model as a fixed covariate, and the parameter $k$ is an optional parameter describing whether the age pattern of mortality is ``early" or ``late." By early, we mean that NMR and IMR are higher than what is usually observed when compared to U5MR, and by late we mean that NMR and IMR are lower than what is usually observed when compared to U5MR, based on the patterns of mortality before the age of 5 in countries with highly reliable child mortality data, such as those included in the U5MD. When all 22 possible values for $x$ are supplied to the model, \citet{guillot2022modeling} propose an uncertainty band around the estimated survival curve, based on the deviation of the shape of the estimated curve from an overall average curve estimated using data from the HMD. The derivation of their uncertainty band relies on a few key assumptions---including that the deviations of the estimated curve from the overall average curve are independent across ages---that are detailed in Section C.3 of the supplementary material.

Multiple follow-up papers (e.g.  \cite{eilerts2021age, verhulst2022divergent}), as well as \cite{guillot2022modeling}, note that the log-quad model is generally unsuitable for use in LMICs, or in countries with (broadly) early or late patterns of child mortality. This is unsurprising given that the coefficients in the U5MD are estimated from high-income countries which likely have differing health care systems, and structural and programmatic support for decreasing child mortality. \cite{guillot2022modeling} also note that there are known biases in the data sources available in LMICs. One of these issues, age-heaping, can be addressed by excluding data. In DHS surveys especially, age heaping typically occurs at age 12 months. Rather than input all $22$ possible age groups into the model for estimating the $k$ parameter, the user may instead leave out a range of ages (\cite{guillot2022modeling} suggest 9 to 21 months) that they believe covers the ages where data is heaped. Note that this is distinct from treating deaths between the ages of 9 and 21 months as interval censored. The rationale for this approach is that in removing those deaths, the estimated curve will essentially smooth over any age heaping that occurs. A significant downside to this approach is that it involves throwing away useful information about the pattern of U5MR. 

While the log-quad model can address age-heaping, it has additional characteristics that may be unsuitable in LMICs. Due to its formulation, the log-quad model's prediction of U5MR is identical to the value of U5MR input as a covariate \textit{with zero uncertainty} (when $x = 5y$, the age-specific coefficients from the model are estimated as $\{a_x, b_x, c_x, v_x\} = \{0, 1, 0, 0\}$). In settings with reliable data, this may be a reasonable (even desirable) property. However, in LMICs where U5MR is often estimated with considerable uncertainty, we do not necessarily want our predicted value of U5MR to align perfectly with a point estimate, but rather to lie within a range of reasonable values defined by the confidence interval for U5MR. 

\subsection{Discrete hazards approach}
\label{sec:discretehaz}

The discrete hazards approach described in \cite{allison2014event} (as well as \cite{mercer2015space,li2019changes, wu2021}) formulates child mortality data in an explicit survival framework. This framework is currently used by the UN and DHS for estimating subnational U5MR in LMICs \citep{li2019changes, wu2021}. The discrete hazards model splits the time before the age of 60 months into $J$ discrete intervals $[x_1, x_2)$,$[x_2, x_3)$, $\dots$, $[x_J, x_{J+1})$ where $x_{j + 1} = x_{j} + n_j$, $x_1 = 0$. Then the U5MR can be computed as
\begin{align}
	\label{eq:discretehaz}
	{}_{60}q_0 = 1 - \prod_{j = 1}^J (1 - {}_{n_j}q_{x_j}).
\end{align}
\cite{mercer2015space} divide the first 60 months of life for individuals into six intervals, $J = 6$: $[0,1)$, $[1,12)$, $[12,24)$, $[24,36)$, $[36,48)$, $[48,60)$, where $(x_1, \dots, x_6) = (0, 1, 12, 24, 36, 48)$, $(n_1, \dots, n_6) = (1, 11, 12, 12, 12, 12)$. Data is tabulated into binomial counts indexed by age group $j$, and potentially indexed by time period $p$ as well, where the number of observations $y_{jp}$ corresponds to the number of deaths observed in that age group and time period, and the number at risk $N_{jp}$ corresponds to the number of children alive in that age group and time period. Note that by construction of the age intervals, we can also estimate NMR and IMR from this model. 

\cite{mercer2015space} fit a logistic regression model, 
\begin{align*}
	y_{jp} \mid N_{jp}, {}_{n_j} q_{x_j,p} & \sim \text{Binomial}(N_{jp}, {}_{n_j} q_{x_j,p}), \\
	\text{logit}({}_{n_j} q_{x_j,p}) & = \beta_{jp},
\end{align*}
where $\beta_{jp}$ is an age-period specific intercept. Pseudo-MLEs of $\beta_{jp}$ are obtained by fitting this model in \texttt{R}'s \texttt{survey} package, using the \texttt{svyglm()} function. We can use the pseudo-MLEs estimated from the logistic regression model to construct estimates of ${}_{60}q_0$ using equation (\ref{eq:discretehaz}). Although the binomial likelihood does not reflect the exact data generating mechanism, many sampling schemes in LMICs (including that used by the DHS) allow data to be aggregated to binomial counts by cluster. 

The discrete hazards approach assumes a constant hazard within the specified age groups. Therefore, while we can estimate a full survival curve for children under the age of 5, we know its shape will not be realistic, as the probability of survival should change smoothly with age rather than make discrete jumps. To obtain a more ``continuous" survival curve, we could have 60 age groups for each 1-month breakdown in the discrete hazards approach, if the data were available at a monthly level for all 60 age groups. There is a balance here between flexibility and parsimony: the model fitted with more age groups better reflects the underlying smooth changes in hazard, but each hazard estimate is less precise than we might get fitting a more parsimonious model (if that model is appropriate). 

In contrast with the log-quad model, age-heaping can be handled in the discrete hazards model by construction of the age intervals. For example, one could consider age intervals $(J = 7)$: $[0,1)$, $[1,9)$, $[9,21)$, $[21,24)$, $[24,36)$, $[36, 48)$, $[48, 60)$, where we group deaths recorded between the ages of 9 and 21 months into a single age group. Additional notes on the discrete hazards model in conjunction with DHS surveys are in Section C.4 of the supplementary material.

\section{Application}
\label{sec:application}

We apply our proposed, parametric pseudo-likelihood approach to child mortality data from Burkina Faso, Malawi, Senegal, and Namibia. We chose single DHS surveys from each of these countries, and used the proposed approach to obtain continuous survival curves for the time periods $[2000,2005)$ and $[2005,2010)$ to demonstrate the ability of our approach to produce period estimates throughout time. The data used in the application is described in detail in Section \ref{subsec:data}, and all parametric models considered are catalogued in Section \ref{subsec:parmodels}. We additionally fit a survey-weighted version of the Turnbull estimator, with boostrapped confidence bands, to validate the parametric approaches, as described in Section C.1 of the supplementary material.

For further comparison, we compare our approach to estimates from the log-quad model using all 22 age inputs (calculated from the Turnbull estimate), and the discrete hazards model from \cite{mercer2015space}. 

For all parametric approaches, we estimate the survival curves in each time period, uncertainty bands surrounding each survival curve (95\% confidence bands based on finite population variances for all approaches other than log-quad, and the derived uncertainty band from \cite{guillot2022modeling} for the log-quad approach), and estimates of NMR, IMR, and U5MR from these survival curves. We note that the uncertainty band for the log-quad model does not have a clear interpretation, and point readers to the derivation in the Supplement of \cite{guillot2022modeling} for details.

Software for implementing the proposed methodology is available in the \texttt{R} package \texttt{pssst}, available at \url{https://github.com/taylorokonek/pssst}.

\subsection{Data}
\label{subsec:data}

All data used in our application comes from the Demographic and Health Surveys (DHS) programme. Child death data is collected via interviewing mothers, and asking them the birth and death dates of all children they have had. We treat deaths prior to one month as exact, and interval-censored afterwards with the interval given as a single month or a single year depending on when the child died (see Section \ref{sec:survival_framework}). 

It has previously been noted that DHS surveys are subject to potential biases that may negatively impact the resulting estimates of child mortality \citep{hill1995age, lawn2008four, guillot2022modeling}. The main concern for estimates of mortality under the age of 5 years is age-heaping at age 12 months, where more children are recorded as having died at 12 months than would otherwise be expected. \cite{lawn2008four} additionally note that age-heaping in DHS surveys may occur at 7 days, 14 days, and 1 month. As noted in Section \ref{subsec:par_approach}, in our proposed approach we address age-heaping at 12 months by interval-censoring all observations recorded as having died between 6 and 18 months for that entire 12 month period $[6,18)$. Additionally, we compare estimates from our proposed approach that account for age heaping in this way to estimates that do not make this adjustment. Further details relating to DHS survey design can be found in Section A of the supplementary material.

\subsection{Parametric Models}
\label{subsec:parmodels}

The parametric distributions considered for our proposed approach are listed in Table 1. The exponentially-truncated shifted-power (ETSP) family of hazards we consider is slightly different than that considered in \cite{scholey2019age}, as we set $c = 0$ as opposed to estimating it via profile likelihood. In \citet{scholey2019age}'s applications, $c$ was estimated to be very close to zero, typically around $6\times 10^{-4}$. 

The generalized Gamma distribution is parametrized as in the \texttt{flexsurv} package in \texttt{R}, as it is more numerically stable than the original parameterization \citep{prentice1974log}.

\begin{table}
	\caption{Parametric distributions considered and their characterizations in terms of a probability density function $f(x)$ or hazard $h(x)$. *The ETSP hazard as described in \cite{scholey2019age} contains four parameters, but in our applications we set $c = 0$.}
	\centering
	\renewcommand*{\arraystretch}{1.5}
	\resizebox{\columnwidth}{!}{%
		\begin{tabular}{l c l}
			Distribution                                                                           & Characterization & Parameters\\ \hline
			Exponential                                                                            & $f(x) = \beta e^{-\beta x}$                  & $\beta$                \\
			Piecewise Exponential                                                                  & $h(x) = \beta_0  I[x < 1] + \beta_1  I[1 \leq x < 12] + \beta_2  I[x \geq 12]$               & $\beta_0, \beta_1, \beta_2$            
			
			\\
			Weibull                                                                                & $f(x) = \beta k (\beta x)^{k-1} e^{-(\beta x)^k}$    & $\beta, k$                              \\
			Generalized Gamma                                                                      & $f(x) = \frac{|Q| (Q^{-2})^{Q^{-2}}}{\sigma x \Gamma(Q^{-2})} e^{Q^{-2}(Q \omega - e^{Q\omega})}$        & $Q, \sigma, \omega$                         \\
			Lognormal                                                                                                      & $f(x) = \frac{1}{x\sigma \sqrt{2\pi}} e^{-\frac{1}{2\sigma^2}(\log(x) - \mu)^2}$    & $\sigma, \mu$     \\
			Gompertz                                                                                                      & $f(x) = \beta k e^{k + \beta x - k e^{\beta x}}$    & $\beta, k$      \\
			\begin{tabular}[c]{@{}l}Exponentially-truncated\\ shifted power (ETSP)*\end{tabular}                          & $h(x) = a (x + c)^{-p} e^{-b x}$    & $a, b, c, p$    
		\end{tabular}
	}
	\label{tab:paramaetric_dists}
\end{table}

\subsection{Model Validation}
\label{sec:modelvalidation}

To assist with model validation, we fit a survey-weighted version of the Turnbull estimator, with bootstrapped confidence bands, to provide a guideline for how well each of the parametric distributions is able to capture the underlying survival curve in each time period. This is treated as a reasonable reference point for the underlying survival curve as it is free of parametric assumptions. However, despite our use of bootstrapped CIs there are no well-justified variance estimates for the Turnbull estimator (Section \ref{sec:turnbull}), making our comparisons to the Turnbull estimator only crudely calibrated.

Let a sample $k$ at age $x$ from the boostrapped distribution of the Turnbull estimate at age $x$ be denoted $\tilde{\theta}_{x}^{(k)}$, and a sample $k$ from the asymptotic distribution of the parametric survival curve at age $x$ be denoted $\hat{\theta}_x^{(k)}$. We obtain $k = 1, \dots, 500$ samples, and compute $\hat{\theta}_x^{(k)} - \tilde{\theta}_{x}^{(k)}$ to obtain samples from the empirical distribution of the difference between the Turnbull and parametric distribution at a given age $x$. 

We calculate the proportion of uncertainty intervals derived from $\hat{\theta}_x^{(k)} - \tilde{\theta}_{x}^{(k)}$ at ages $x$ that contain $0$ as a rough estimate of how closely each parametric model aligns with the Turnbull estimate. \textit{This is not a formal hypothesis test}, but rather a means of assessing how close the parametric estimate is to the Turnbull estimate while accounting for uncertainty in \textit{both} estimates. 

\subsection{Results}
\label{sec:results}

In this section, we display a subset of results from the application of the seven parametric models, log-quad model, and discrete hazards model to DHS data from Burkina Faso, Malawi, Senegal, and Namibia. The results shown for Malawi are representative of the results for other countries. Additional results can be found in Section D of the supplementary material, with comparisons to models where the data is not adjusted for age-heaping in Section E of the supplementary material.

In Figure \ref{fig:weibull_curves} we display the fitted survival curves for the Weibull and lognormal models using our proposed methodology in both time periods for Malawi, and compare them to the Turnbull estimator, log-quad model, and discrete hazards model. While other parametric models were estimated in addition to the Weibull and lognormal, we chose to include only these two in Figure \ref{fig:weibull_curves} for visual clarity (two parametric estimates as opposed to six), where Weibull was chosen because it is one of the most commonly used parametric survival models, and lognormal was chosen because the fit was particularly close to the Turnbull estimator. Additional visualizations for \textit{all} parametric models fit can be found in Section D of the supplementary material.

\begin{figure}
	\centering
	\resizebox{\columnwidth}{!}{%
		\includegraphics{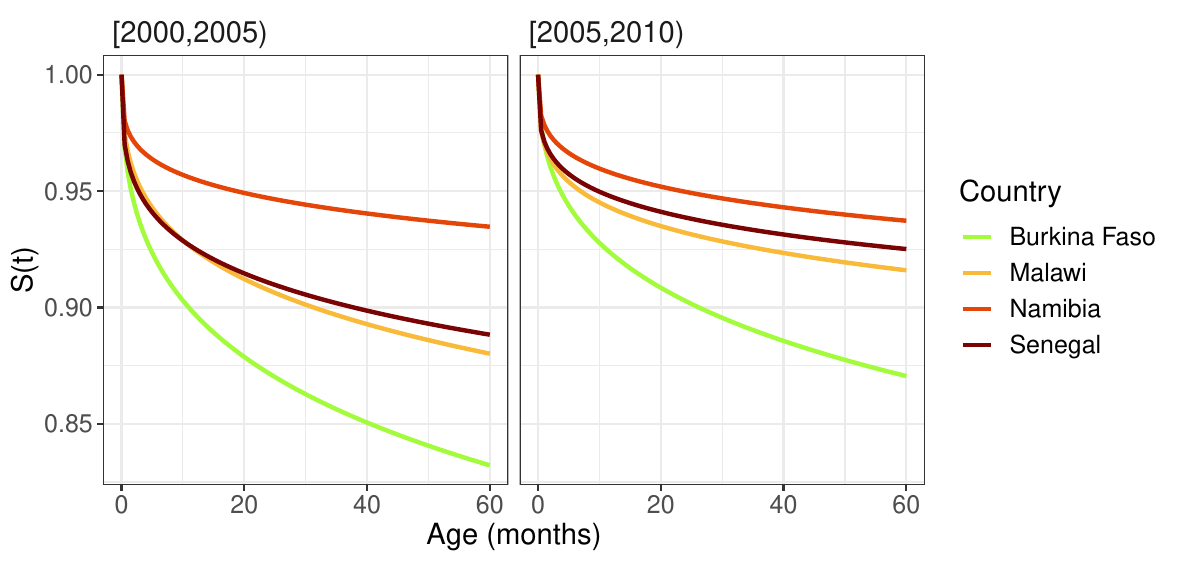}
	}
	\caption{Estimated lognormal survival curves for time periods $[2000,2005)$ (left) and $[2005,2010)$ (right) for Burkina Faso, Malawi, Namibia, and Senegal.}
	\label{fig:surv_compare_lognormal}
\end{figure}

Compared to the Turnbull estimator, the Weibull model tends to estimate higher survivorship at younger ages, and lower survivorship at older ages. The lognormal model captures the sharp increase in mortality within the first 12 months of life more accurately than the Weibull model. In Figure \ref{fig:surv_compare_lognormal} we compare estimated lognormal survival curves across all countries in our application and both time periods. 

\begin{figure}
	\centering
	\resizebox{\columnwidth}{!}{%
		\includegraphics[scale = 0.7]{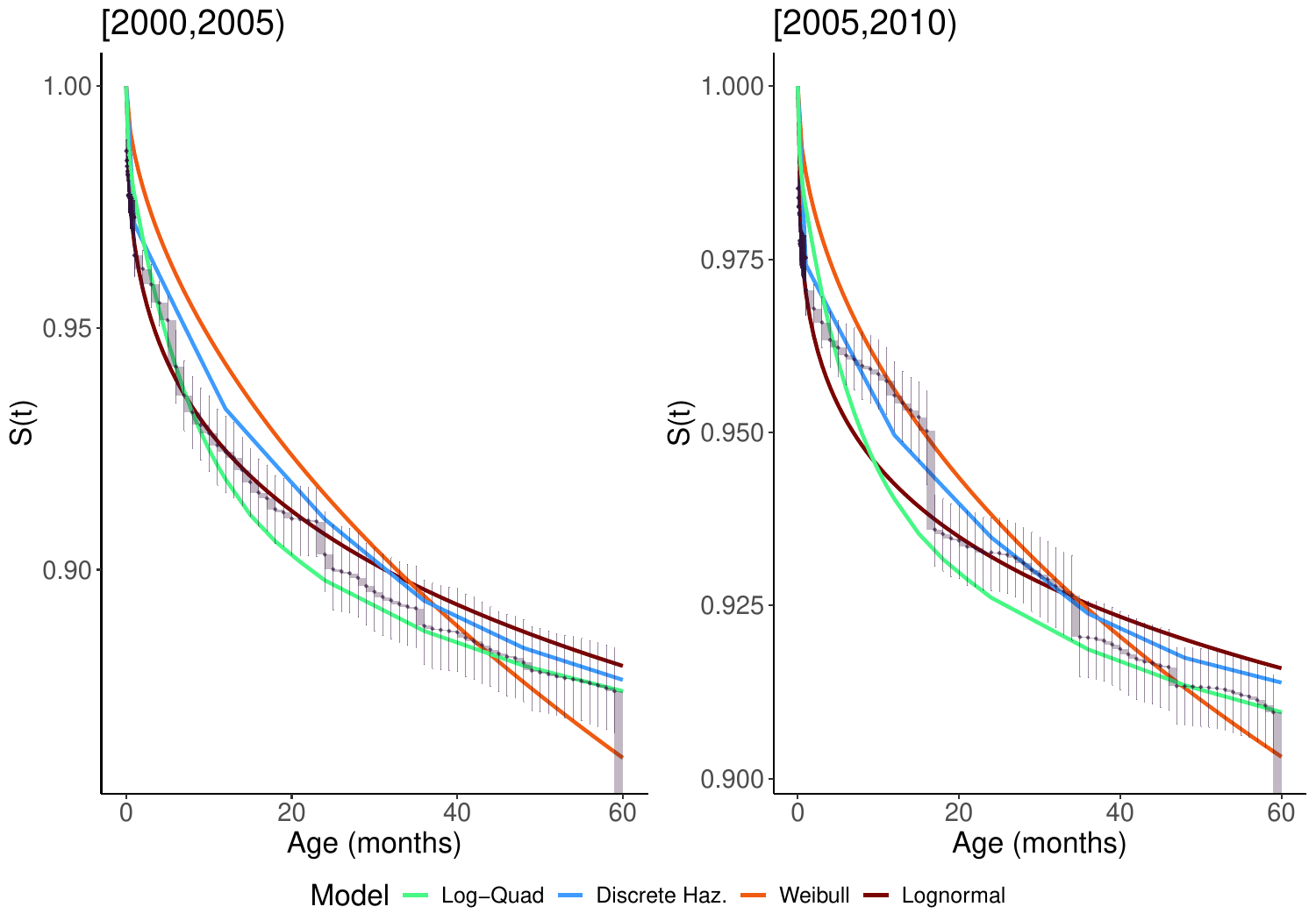}
	}
	\caption{Estimated Weibull and lognormal survival curves for time periods $[2000,2005)$ (left) and $[2005,2010)$ (right) for Malawi, compared to estimated survival curves from the discrete hazards and log quad approach. The Turnbull estimator is displayed by the gray step function, uncertainty boxes, and uncertainty intervals.}
	\label{fig:weibull_curves}
\end{figure}

\begin{figure}
	\centering
	\resizebox{\columnwidth}{!}{%
		\includegraphics[scale = 0.65, page = 1]{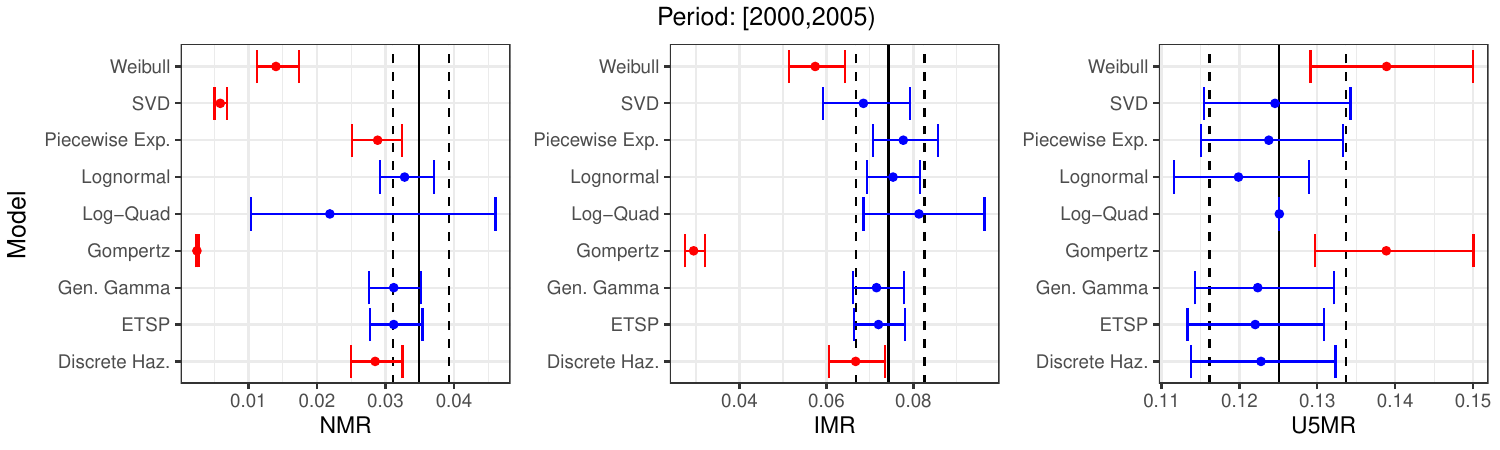}
	} \resizebox{\columnwidth}{!}{%
		\includegraphics[scale = 0.65, page = 2]{Plots/Malawi/nmr_imr_u5mr_compare_2000_2010.pdf}
	}
	\caption{Estimates of NMR, IMR, and U5MR for Malawi in periods $[2000,2005)$ (top) and $[2005,2010)$ (bottom). Turnbull point estimates are denoted by vertical black lines, with dashed vertical black lines denoting 95\% confidence intervals. Horizontal error bars are blue if the interval captures the monthly discrete hazards point estimate, or red if the interval does not capture the monthly discrete hazards point estimate. All 95\% confidence intervals are based on finite population variances, with the exception of the log-quad model where uncertainty is calculated as in \cite{guillot2022modeling}.}
	\label{fig:malawi_mr_mainbody}
\end{figure}

Showing just summary measures of mortality (NMR, IMR, U5MR), we see the same patterns in Figure \ref{fig:malawi_mr_mainbody}. The Weibull model in each time period underestimates NMR, and overestimates U5MR, relative to the Turnbull estimator, particularly for the period $[2000,2005)$. In contrast, the lognormal model confidence intervals cover NMR, IMR, and U5MR in both time periods, with the exception of IMR in $[2005,2010)$ where only the Weibull model captures the Turnbull estimate and U5MR in $[2005,2010)$ where the lognormal confidence interval is slightly too low to capture the Turnbull estimate of U5MR. As seen in Table 2, the differences between the Weibull estimates and Turnbull estimates capture zero for 44\% and 60\% of ages where the Turnbull estimate is defined, prior to age 60 months, for $[2000, 2005)$ and $[2005, 2010)$, respectively. In contrast, the differences between the lognormal estimates and Turnbull estimates capture zero for 93\% and 80\% of ages. This aligns with the visualizations to suggest that the lognormal model is a better parametric fit for the mortality curve for children under the age of 5 than the Weibull model. The log-quad model is not included in Table 2 because (1) it is not a model you can sample from based on the way the uncertainty bands are defined and therefore our model validation approach cannot apply, and (2) the model validation approach we use is intended specifically to compare the Turnbull estimator to our proposed parametric models, and not existing methods.

\begin{table}
	\label{tab:modelval}
	\caption{Model validation results. Percentage of samples (out of 500) from $\hat{\theta} - \tilde{\theta}$ that contain 0 for all parametric models, countries, and periods. Results that contain more than 70\% of samples noted in bold.}
	\centering
	\renewcommand*{\arraystretch}{1.5}
	\resizebox{\columnwidth}{!}{%
		\resizebox{.9\hsize}{!} {
			\begin{tabular}{l | l r r r r r r r}
				Country                       & \multicolumn{1}{c}{Period} & \multicolumn{1}{c}{Weibull} & \multicolumn{1}{c}{\begin{tabular}[c]{@{}c@{}}Piecewise\\ Exponential\end{tabular}} & \multicolumn{1}{c}{\begin{tabular}[c]{@{}c@{}}Generalized\\ Gamma\end{tabular}} & \multicolumn{1}{c}{Lognormal} & \multicolumn{1}{c}{Gompertz} & \multicolumn{1}{c}{ETSP} & \multicolumn{1}{c}{\begin{tabular}[c]{@{}c@{}}Discrete\\ Hazards\end{tabular}} \\ \hline
				\multirow{2}{*}{Burkina Faso} & {[}2000, 2005)              & 17                         & 40                                                                                & \textbf{93}                                                                            & 52                           & 12                          & 33          & 48           \\  
				& {[}2005, 2010)              & 14                         & 48                                                                                 & 70                                                                             & 65                           & 9                          & 57     & 60             \\ \hline
				\multirow{2}{*}{Malawi}       & {[}2000, 2005)              & 44                         & \textbf{76}                                                                               & \textbf{94}                                                                           & \textbf{93}                           & 18                          & \textbf{94}   & \textbf{73}                  \\  
				& {[}2005, 2010)              & 60                         & 61                                                                                 & \textbf{82}                                                                           & \textbf{80}                        & 12                          & \textbf{81}   & \textbf{73}                 \\ \hline
				\multirow{2}{*}{Senegal}      & {[}2000, 2005)              & 28                         & \textbf{73}                                                                                 & \textbf{84}                                                                             & \textbf{86}                        & 19                         & \textbf{85}    & \textbf{72}                 \\  
				& {[}2005, 2010)              & 38                         & \textbf{73}                                                                                 & 63                                                                             & \textbf{85}                           & 13                          & \textbf{87} & \textbf{72}                    \\ \hline
				\multirow{2}{*}{Namibia}      & {[}2000, 2005)              & 56                         & \textbf{86}                                                                                & \textbf{100}                                                                              & \textbf{100}                            & 25                         & \textbf{99}  & \textbf{86}                   \\ 
				& {[}2005, 2010)              & 53                         & \textbf{86}                                                                                 & \textbf{100}                                                                              & \textbf{100}                            & 29                         & \textbf{99}           & \textbf{85}          
		\end{tabular}}
	}
\end{table}

The log-quad approach performs adequately, though it is important to note that while the point estimates may be reasonable, the uncertainty quantification is less so. In particular, U5MR is assumed to be estimated with no uncertainty. This is not a desirable property of this approach since our estimates of U5MR that are input to the log-quad model are themselves estimated with uncertainty. Second, we note that the uncertainty bands around the log-quad point estimates are in general much wider than the confidence bands for the parametric models. The confidence bands surrounding the parametric models may be interpreted at each age $x$ with a 95\% confidence interval interpretation based on resampling observations from the finite population, whereas the uncertainty surrounding the log-quad model does not have as straightforward of an interpretation. Furthermore, out of all of the analyses conducted, only the log-quad models for Namibia (both time periods) provided estimates and confidence bands that would be considered reasonable by \cite{guillot2022modeling}. All other countries either had estimated values for certain parameters outside the range suggested by \citet{guillot2022modeling}, or an increasing hazard by age in the uncertainty interval computed, which is unrealistic.

In general, the discrete hazards approach performed well, though perhaps not sufficiently better than some of the proposed parametric models (such as lognormal or piecewise exponential) to justify the need for six parameters in estimating the survival curve. Furthermore, assuming a constant hazard over certain age intervals is not necessarily an assumption we wish to make, as it is unrealistic even at a fine scale of age groups. Additional comments on the results of the application can be found in Section D of the supplementary material.

\section{Discussion}
\label{sec:discussion}


In this paper, we propose two novel approaches to estimating full survival curves for child mortality that are well-suited to applications in LMICs. We detail existing methods that can be used to estimate full survival curves and explain how they fall short in this specific context. 

Our application suggests that there are potentially very large differences in model fit between parametric distributions, with the Weibull distributions and Gompertz distributions generally providing the worst fit compared to the Turnbull estimator, in terms of capturing the survival curve under the age of 5. In general, the lognormal model seems to fit the countries in our application reasonably well. We note that two of the three-parameter models we compared---the piecewise exponential and ETSP model---also adequately captured the survival curve provided by the Turnbull estimator, though the piecewise exponential model has the undesirable property of assuming constant hazards within prespecified age groups and the ETSP model is computationally challenging to fit. We conclue that for our application, the lognormal model outperforms other parametric models in terms of the ability to capture the point estimate provided by the Turnbull estimator while only requiring two parameters to define the survival curve.

The benefits of a parametric approach to under-5 mortality estimation, and in particular to estimating the full survival curve for children under the age of 5, are many. As laid out in \cite{scholey2019age}, correctly specified parametric assumptions about the shape of mortality may greatly assist estimation of the survival curve under the age of 5 in situations with little data. This becomes particularly relevant in a small area setting, where often little data are available at small administrative regions \citep{wakefield2020small}. As such, the methods proposed in this paper may serve as a guideline, or as prior information in a Bayesian setting, for small area estimation problems of child mortality when a full survival curve is desired. Further benefits of a continuous, parametric approach involve interpretability and parsimony. The Heligman-Pollard model \citep{heligman1980age}, a well-known parametric, demographic model for mortality estimation, provides informative interpretations of the parameters involved in the model, and the same is true of the models we propose. Of course, we rely on the assumption that the parametric distribution used is \textit{correctly} specified, which likely is not the case. In fact, it is likely that there is \textit{no} parametric distribution that can perfectly capture the age trend in mortality for every country. However, especially in scenarios with little data, \textit{reasonable} parametric assumptions may still be useful. Hence it is important to observe and test these parametric models in settings with more data, such as the national setting we use in our application. A meaningful question is: \textit{Is the fit of a continuous parametric model at least as good as the 6-parameter discrete hazards model currently used by the UN IGME and DHS?} When comparing to the Turnbull estimator, the lognormal model does outperform the 6-parameter discrete hazards model in terms of our model performance metric (see Table 2).

Limitations of our nonparametric proposed approach include that the Turnbull estimator lacks a well-justified variance estimate. As previously noted, a variance estimate is not readily available due to the non-Gaussian, cubed-root asymptotics, and a bootstrap estimate of the variance is not applicable for similar reasons. More work needs to be done before comparisons between the nonparametric and parametric approaches (and model validation procedures) can be made with some degree of calibration. 

In conclusion, we have provided a method for obtaining a complete, continuous survival curve for children under the age of 5 using assumed parametric models. Our method enables estimations using interval-censored, left-truncated observations, as is required for period estimates of mortality from DHS data. Furthermore, aspects of survey design, which are particularly relevant in LMICs, may be directly incorporated into our modeling framework to provide design-consistent estimates of mortality with finite population variances.

\bibliography{references}
	
	\begin{appendices}
		
		\section{Data}
		
		\subsection{Survey Design}
		\label{appendix:data}
		All DHS surveys used in our application follow a two-stage, stratified cluster design, and were designed to provide accurate estimates at the Administrative 1 (admin1) subnational level. Strata are defined by admin1 region and urban/rural status. Each sampling frame is established from a previous census. Primary sampling units (PSUs), or clusters, are selected across strata, and the second stage of sampling consists of households within PSUs. GPS coordinates are displaced by up to 2km for urban clusters and 5km for rural clusters, and are not displaced outside of their strata. Information related to the sampling design for the surveys used in our application is given in Table \ref{tab:surveystuff}.
		
		\begin{table}
			\caption{\label{tab:surveystuff}Sampling information from DHS surveys. Census year is the year of the census upon which the sampling frame for the survey is based upon. PSUs and Households listed are the number of PSUs and Households in the sample, not the sampling frame, and counts are additionally disaggregated by Urban/Rural (U/R). *At the time of survey, Malawi's 28 districts were considered Admin2 regions, with Northern, Central, and Southern regions being Admin1. Some shapefiles now consider the 28 districts to be Admin1, with a finer grid as 243 Admin2 subregions.}
			\centering
			\resizebox{\columnwidth}{!}{%
				\begin{tabular}{l r r r r r}
					Country      & Survey Year & Census & Admin1 Regions & PSUs (U/R)    & Households (U/R)   \\ \hline
					Burkina Faso & 2010        & 2006   & 13             & 574 (176/398) & 14924 (4576/10348) \\
					Malawi       & 2016        & 2008   & *28            & 850 (173/677) & 27531 (5190/22341) \\
					Senegal      & 2010        & 2002   & 14             & 392 (147/245) & 8232 (3087/5145)   \\
					Namibia      & 2013        & 2011   & 13             & 554 (269/285) & 11080 (5380/5700) 
				\end{tabular}
			}
		\end{table}
		
		\subsection{Age-heaping in DHS surveys}
		\label{sec:ah_evidence}
		
\begin{figure}
			\centering
			\includegraphics[scale = 0.45]{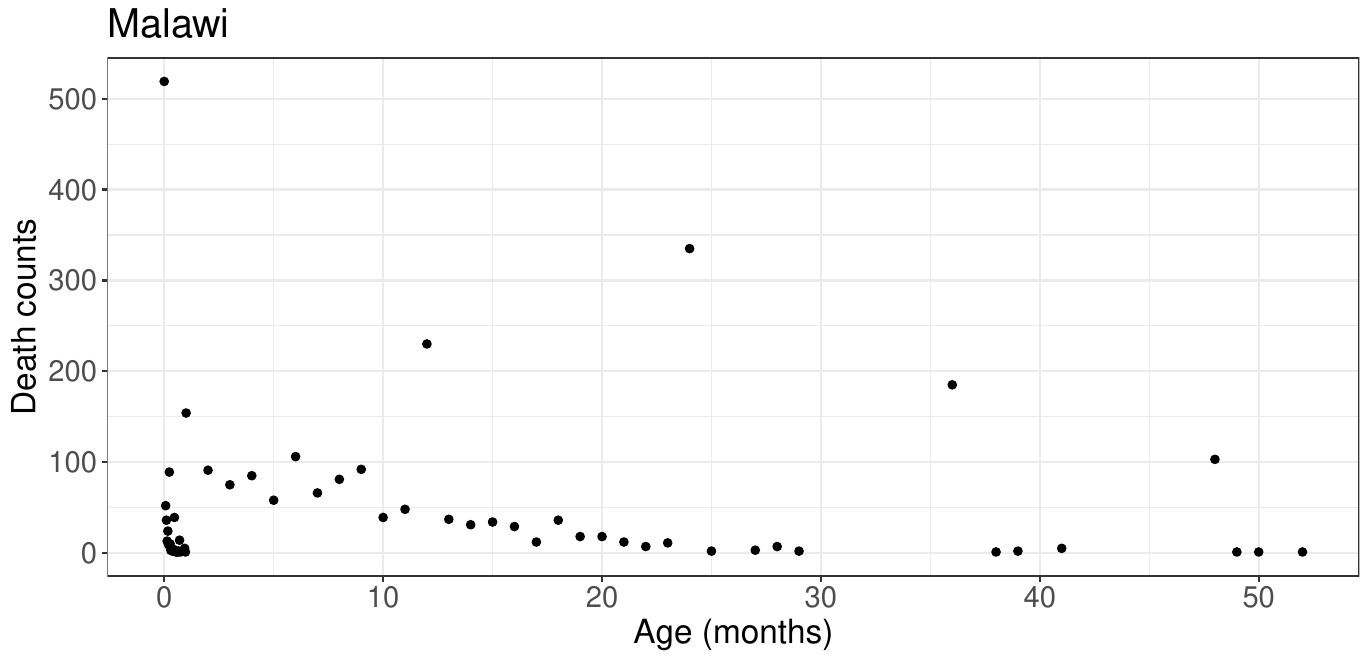}
			\includegraphics[scale = 0.45]{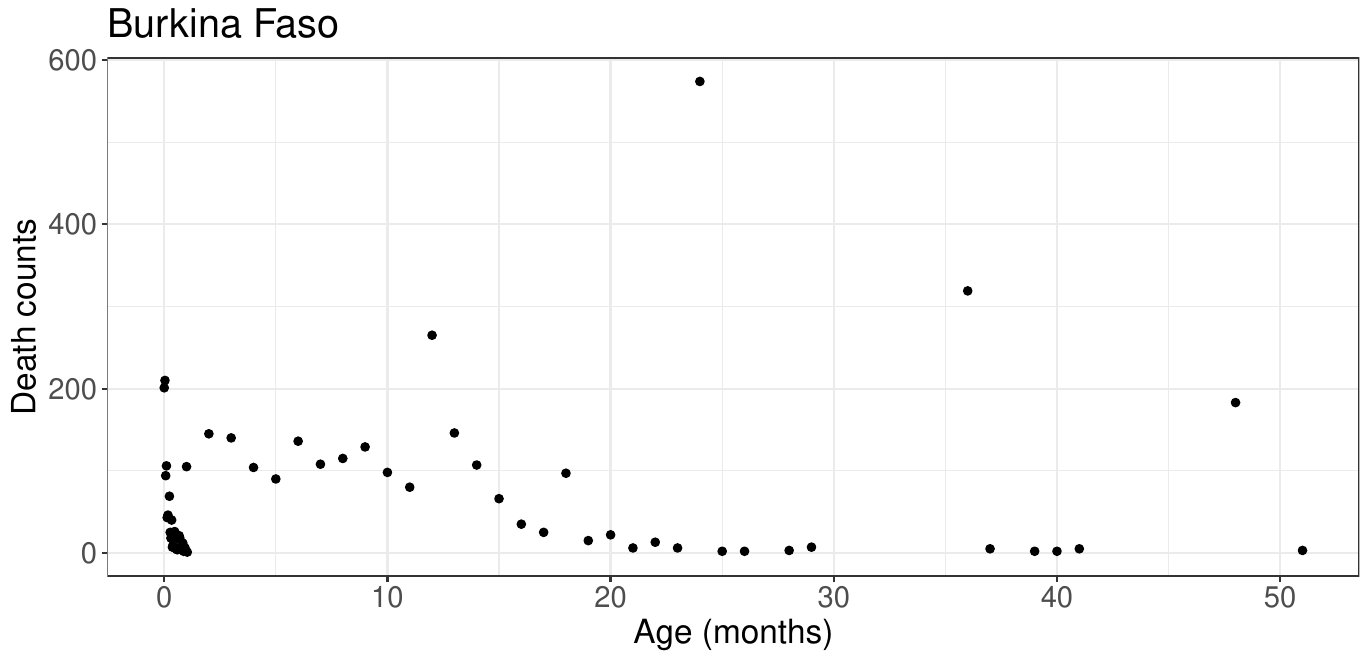}
			\includegraphics[scale = 0.45]{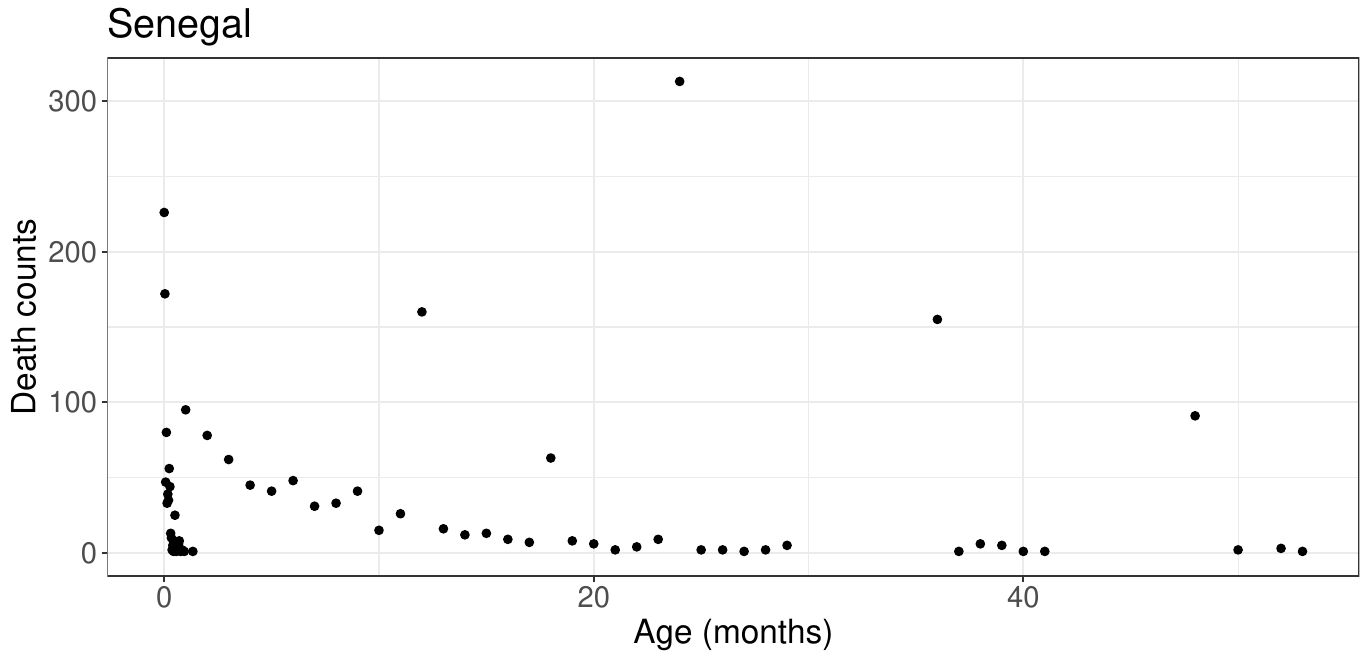}
			\includegraphics[scale = 0.45]{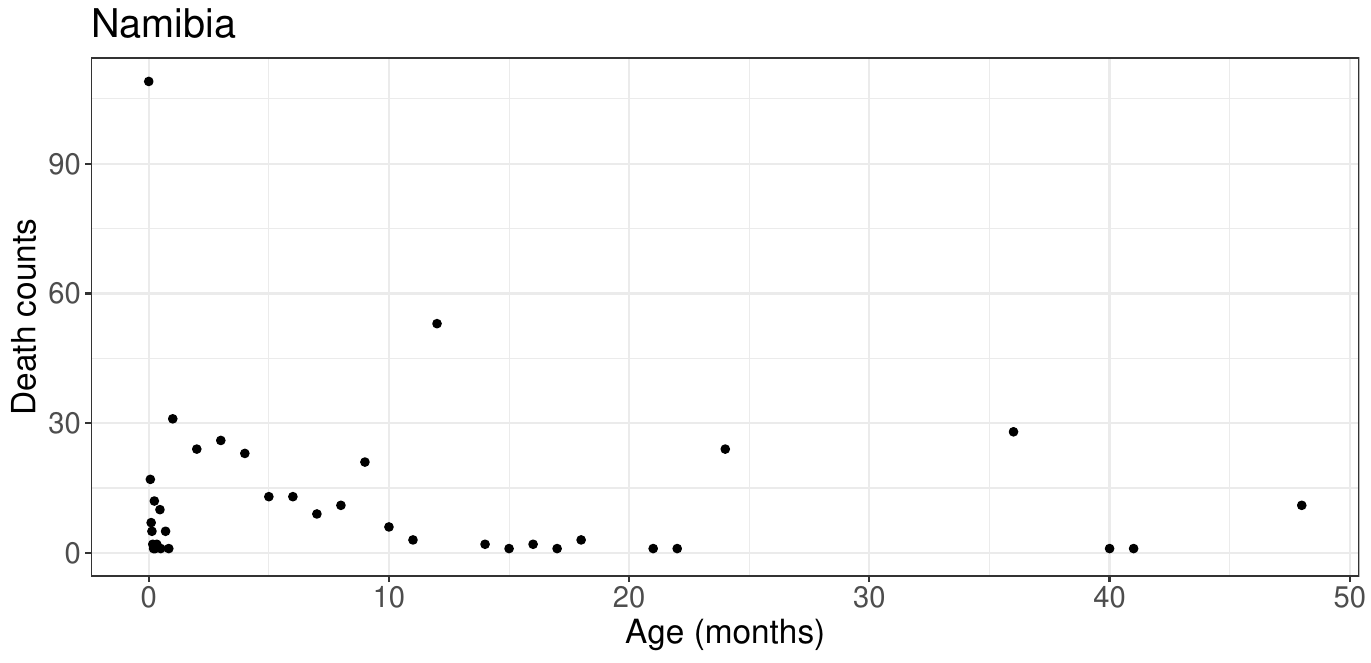}
			\caption{Total death counts at each age group within DHS surveys.}
			\label{fig:ah_evidence}
		\end{figure}

		For the four DHS surveys we consider in our application (Malawi 2015-2016, Burkina Faso 2010, Senegal 2010, Namibia 2013), we display the total death counts at each age recorded in the entire survey in Figure \ref{fig:ah_evidence}. Note that we expect peaks at 24, 36, and 48 months because they capture a full year of deaths as opposed to only single months, but the peaks observed at 12 months reflect age heaping as they cover the same age span as the age groups surrounding it. The small number of counts observed at unexpected age months (25 months, for example), are the few exceptions to the typical interval-censoring scheme used in DHS surveys.
		
		\section{Pseudo-likelihood}
		\label{appendix:binder}
		
		Consider a vector of superpopulation parameters $\boldsymbol{\theta}$ and observations $\textbf{y}$ that can be written as the solution to the score equations $\mathcal{S}(\boldsymbol{\theta}, \textbf{y}) = \textbf{0}$. In a finite population setting, we are interested in the finite population parameters $\boldsymbol{\theta}'$, obtained by solving $\sum_{i = 1}^N \mathcal{S}(\boldsymbol{\theta}', \textbf{y}) = \textbf{0}$, where $i = 1, \dots, N$ denote all individuals in the finite population. Rather than observe all $N$ individuals in the population, we instead take a probability sample of $j = 1, \dots, U \leq N$ individuals, with weights $w_j$ equal to the inverse of their inclusion probabilities in the sample. We obtain survey-weighted estimates of the finite population parameters $\boldsymbol{\theta}'$ via the finite population score equations
		\begin{align*}
			\sum_{j = 1}^U w_j \times \mathcal{S}(\boldsymbol{\theta}', \textbf{y}_j) = \textbf{0}.
		\end{align*}
		This framework, developed in \cite{binder1983variances}, works for sample designs and populations that admit asymptotically normal estimators, with certain regularity conditions (see \cite{binder1983variances} for details, as well as a more general case of score equations). As an example, for linear regression we would set $\mathcal{S}(\boldsymbol{\theta}, \textbf{y}) = -(y_i - \textbf{x}_i^\top \boldsymbol{\theta})\textbf{x}_i$. The variance of $\boldsymbol{\theta}'$ then has a sandwich form obtained via a Taylor expansion of the score equations at $\boldsymbol{\theta}' = \boldsymbol{\theta}$.

		\section{Methodological details}
		
		\subsection{Turnbull estimator}
		\label{appendix:turnbull_details}
		
		Suppose we are interested in estimating a full mortality schedule (survival curve) from ages 0 to 5 at a national level in time periods $[2000, 2005)$ and $[2005, 2010)$. For simplicity, we consider age in full years rather than months in this example. Let $[t_0, t_1]$ denote the interval in which an observation is censored, with the convention $[t_0, \infty)$ for right-censored observations. Our data is recorded as follows:
		
		\begin{table}
			\caption{(Left) Example data for children alive between the ages of 0 and 5 years in the time periods $[2000,2005)$ and $[2005,2010)$. (Right) Example data, separated into (truncated) observations for each time period.}
			\centering
			\resizebox{\columnwidth}{!}{%
				\begin{tabular}{l r r r}
					Individual & Year born & $t_0$ & $t_1$ \\ \hline
					1          & 2002      & 2 & 4 \\
					2          & 1998      & 5 & $\infty$ \\
					3 & 2007 & 1 & 1 \\ 
					\vdots & \vdots & \vdots & \vdots \\ 
					&  & &  \\
				\end{tabular} \hspace{0.5cm}
				\begin{tabular}{l r r r r}
					Individual & Year born & $B_i$          & $A_i$            & Period         \\ \hline
					1          & 2002      & $(-\infty, \infty)$     & {[}2, 3{]}   & {[}2000, 2005) \\
					1          & 2002      & {[}3, $\infty$) & {[}3, 4{]}   & {[}2005, 2010) \\
					2          & 1998      & {[}2, $\infty$) & {[}5, $\infty${)} & {[}2000, 2005) \\
					3          & 2007      & $(-\infty, \infty)$    & {[}1, 1{]}   & {[}2005, 2010) \\
					\vdots & \vdots & \vdots & \vdots& \vdots 
				\end{tabular}
			}
			\label{tab:exampledata1}
		\end{table}
		Individual 1 is interval censored to the age range $[2,4]$, individual 2 is right censored at age $5$, individual 3 is observed to die at exactly 1 year, and so forth. To account for time period as a time-varying covariate in our model, we rearrange our data by splitting each observation into multiple observations, one for each time period in which they contribute to the risk set. The observations are left-truncated at the beginning of each time period by $\max(0,$ age at which they enter the time period), where we denote this truncation in set notation as $B_i$. If $B_i = (-\infty, \infty)$, this indicates that no truncation has occurred, which in this context means the individual was born in the given time period and not before it. Individuals are censored according to the sets $A_i = [L_i, R_i]$, where if $L_i = R_i$ the observation is exactly observed (uncensored), and if $A_i = [L_i, \infty)$ the observation is right censored at $L_i$. The number of observations in our expanded dataset is $N \equiv \sum_{i = 1}^n p_i $, where $n$ is the number of individuals in our dataset and $p_i$ is the number of time periods in which individual $i$ contributes to the risk set.
		
		In the above example, for $i = 1, \dots, N$ observations, let $A_i$ denote an individual's censoring set, and let $B_i$ denote an individuals truncation set such that the likelihood contribution for an individual can be denoted $P(X_i \in A_i \mid X_i \in B_i)$, where $X_i$ is the random variable corresponding to the death of child $i$. The data is thus in the form of pairs $(A_1, B_1), \dots, (A_n, B_n)$. 
		
		The Turnbull estimator is the NPLME of the cumulative distribution function $\hat{F}$ of $F$, and therefore, also produces the NPMLE of the survival curve $\hat{S} = 1 - \hat{F}$. We write the likelihood in a convenient way that allows us to optimize the proportions of the CDF that lie within given intervals subject to the constraint that the proportions sum to $1$ and are greater than $0$. This allows us to view the likelihood maximization as a constrained optimization problem that can be solved using an Expectation-Maximization (EM) algorithm \citep{dempster1977maximum}, which we now describe.
		
		Assume that each $A_i$ can be written as a finite union of disjoint, closed intervals, with a single point $A_i = X_i$ being written as the closed interval $[X_i, X_i]$. Then for each censoring set $A_i$ we can write,
		\begin{align*}
			A_i = \bigcup_{j = 1}^{k_i} [L_{ij}, R_{ij}],
		\end{align*}
		for $i = 1, \dots, n$. Then the likelihood for all observations can be written as 
		\begin{align}
			\label{eq:turnbull_lik}
			\text{Likelihood} = \prod_{i = 1}^n \frac{\left[ \sum_{j = 1}^{k_i} F(R_{ij}) - F(L_{ij})\right]}{P(B_i)}.
		\end{align}
		Let $[q_1, r_1], \dots, [q_m, r_m]$ denote all unique intervals defined by $[L_{ij}, R_{ij}]$, and $s_j = F(r_j) - F(q_j)$. These $s_j$ define the proportions of the CDF that lie within an interval $[q_j, r_j]$.
		
		\begin{figure}
			\centering
			\includegraphics[scale = 0.5]{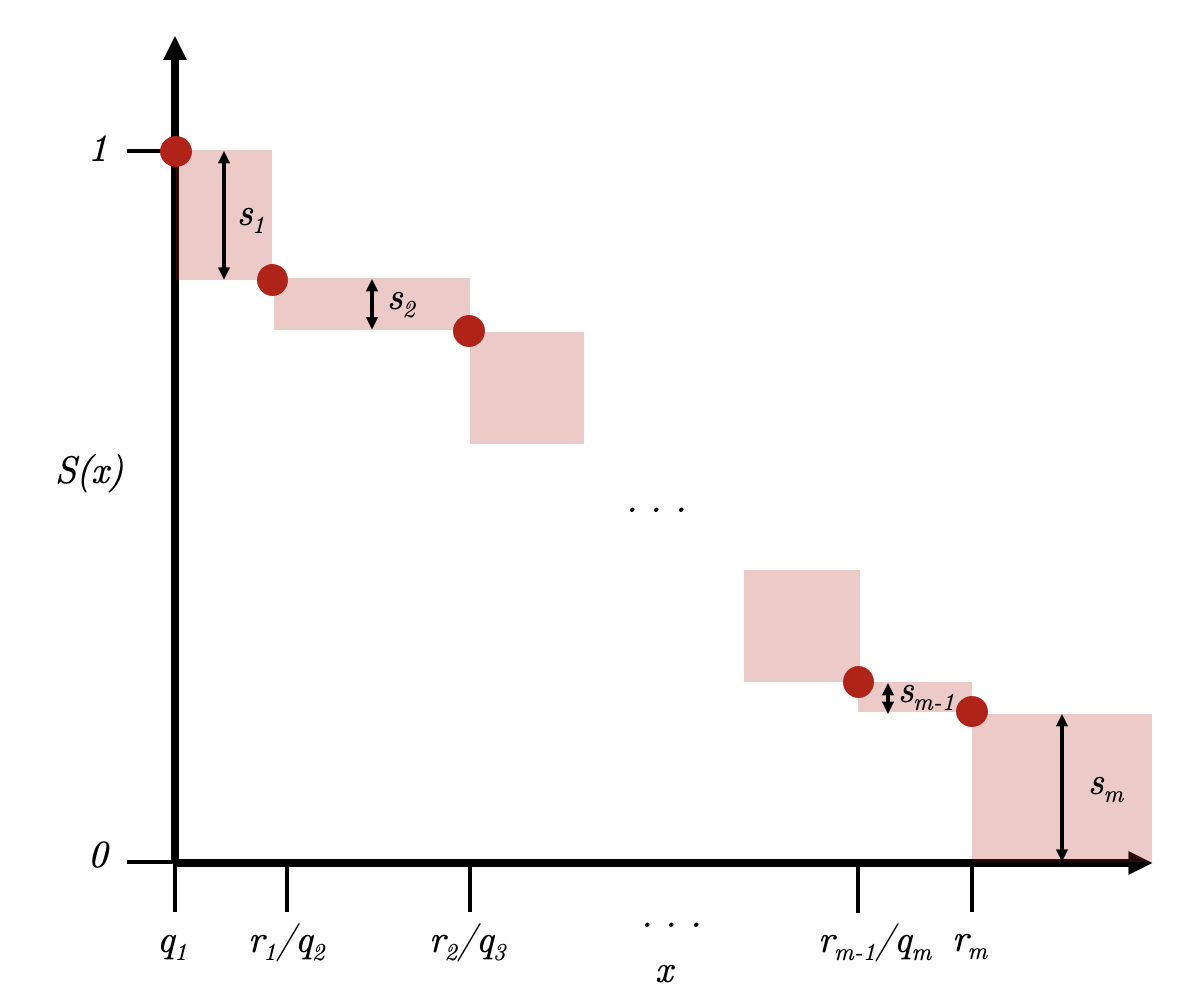}
			\caption{A visual representation of the values $s_j$ that make up the Turnbull estimator.}
			\label{fig:turnbull_viz}
		\end{figure}
		We can rewrite the likelihood defined in Equation (\ref{eq:turnbull_lik}) as
		\begin{align*}
			\text{Likelihood} = \prod_{i = 1}^n \left( \frac{\sum_{j = 1}^m I\{ [q_j, r_j] \in A_i \} s_j}{\sum_{j = 1}^m I\{ [q_j, r_j] \in B_i \} s_j} \right).
		\end{align*}
		Maximizing the above subject to the constraints $s_j \geq 0$, $\sum_{j = 1}^m s_j = 1$ then corresponds to maximizing the likelihood for arbitrarily censored and truncated observations, and provides us with a nonparametric estimate of the MLE. A visual description of the Turnbull estimator is provided in Figure \ref{fig:turnbull_viz}.
		
		\cite{turnbull1976empirical} suggests the following procedure for obtaining the MLE:
		
		\begin{enumerate}
			\item Obtain initial values for $\textbf{s}^0 = s_1^0, \dots, s_m^0$ subject to $\sum_{j = 1}^m s_j^0 = 1$, $s_j^0 \geq 0$
			\item Compute
			\begin{align*}
				\mu_{ij}(\textbf{s}) & = \frac{I\{ [q_j, r_j] \in A_i \} s_j}{\sum_{k = 1}^m, I\{ [q_k, r_k] \in A_i \} s_k} \\
				\nu_{ij}(\textbf{s}) & = \frac{(1 - I\{ [q_j, r_j] \in B_i \})s_j}{\sum_{k = 1}^m I\{ [q_k, r_k] \in B_i \} s_k} \\
				\pi_j(\textbf{s}) & = \frac{\sum_{i = 1}^n \left( \mu_{ij}(\textbf{s}) + \nu_{ij}(\textbf{s}) \right)}{\sum_{i = 1}^n \sum_{j = 1}^m \left( \mu_{ij}(\textbf{s}) + \nu_{ij}(\textbf{s}) \right)}
			\end{align*}
			\item Set $s_j^1 = \pi_j(\textbf{s}^0)$.
			\item Return to Step 1.
		\end{enumerate}
		The procedure exits once some predetermined, required tolerance is achieved.
		
		Incorporating survey weights into the Turnbull estimator is straightforward, as the likelihood contribution for each individual is simply multiplied by their survey weight. This corresponds to altering Step 2 in the algorithm described, replacing $\pi_j(\textbf{s})$ with $\tilde{\pi}_j(\textbf{s})$, where $\tilde{\pi}_j(\textbf{s})$ is
		\begin{align*}
			\tilde{\pi}_j(\textbf{s}) & = \frac{\sum_{i = 1}^n w_i \left( \mu_{ij}(\textbf{s}) + \nu_{ij}(\textbf{s}) \right)}{\sum_{i = 1}^n w_i \sum_{j = 1}^m \left( \mu_{ij}(\textbf{s}) + \nu_{ij}(\textbf{s}) \right)},
		\end{align*}
		and $w_i$ is the survey weight for a given individual. 
		
		\subsubsection{A note on the Turnbull estimator}
		\label{subsec:note_on_turnbull}
		
		
		When an individual is interval-censored across the boundary of a time period, that individual cannot be split into separate observations that are left truncated at the beginning of a given time period. Intuitively, this would imply that a single individual could contribute more than one death to the risk set, which biases estimates of mortality upwards, and also overstates the amount of information present. When working on a 5-year period scale, few observations are interval censored across the boundary of a time period. For yearly periods (or shorter), we expect this to be a more prevalent occurrence. For 2000-2009 data from Malawi, roughly 0.3\% of all individuals are interval-censored across a time period boundary, which constitutes approximately 4\% of all observed deaths throughout 2000-2009. Table 3 gives the exact breakdown of these percentages for the additional countries we consider in our application.
		
		\begin{table}
			\caption{Percentages of individuals who are interval-censored across a time period boundary out of all individuals at risk, and percentages of individuals who are interval-censored across a time period boundary out of all observed deaths for 2000-2009.}
			\centering
			\begin{tabular}{l r r}
				\multicolumn{1}{c}{Country} & \multicolumn{1}{c}{\begin{tabular}[c]{@{}c@{}}Percent across boundary\\ out of all individuals\end{tabular}} & \multicolumn{1}{c}{\begin{tabular}[c]{@{}c@{}}Percent across boundary\\ out of all deaths\end{tabular}} \\ \hline
				Burkina Faso                        & 0.7                                                                                                             & 6                                                                                                      \\ 
				Malawi                      &      0.3                                                                                                        &           4                                                                                            \\ 
				Namibia                     &      0.1                                                                                                       &                   3                                                                                    \\ 
				Senegal                    &        0.3                                                                                                     &     4
			\end{tabular}
			\label{tab:percentA}
		\end{table}
		
		To account for an individual in our application who is interval-censored across time period boundaries in the Turnbull estimator, we split the individual into separate, left truncated observations at the beginning of the time period, where the last two observations for this individual will each contain an interval censored observation. We then down-weight these last two observations by the proportion of the length of the original individual's interval that is included in that time period. Note that this assumes that the age distribution of deaths in each of these two time periods is the same. Though we know this assumption will not hold (due to cohort effects of conflicts, for example), the resulting survey-weighted Turnbull estimator is still a useful comparator, as it is a nonparametric estimator that estimates rates more robustly than a parametric estimator.

		\subsection{Influence Functions}
		\label{appendix:infl_funcs}
		
		For a generic parametric model (in a non-survey context) with $j = 1, \dots, J$ parameters, let $\hat{\boldsymbol{\theta}} = (\hat{\theta}_1, \dots, \hat{\theta}_J)$ denote the MLE. We can write $\hat{\theta}_j$ as asymptotically linear, meaning 
		$$
		\hat{\theta}_j - \theta_j = \frac{1}{n} \sum_{i = 1}^n \Delta_i + o_p(n^{-1/2}),
		$$
		with influence functions $\Delta_i$ given by
		$$
		\Delta_i =  \left[ \frac{\partial}{\partial \boldsymbol{\theta}} \log(L_i (\hat{\boldsymbol{\theta}})) \right] \left[ \textbf{H}_{\log(L_i)} \right],
		$$
		where $\frac{\partial}{\partial \boldsymbol{\theta}} \log(L_i (\hat{\boldsymbol{\theta}}))$ is an $n \times J$ dimensional matrix of score functions for each individual $i = 1, \dots n$ with respect to parameters $j = 1, \dots J$, and $\textbf{H}_{\log(L_i)}$ denotes the Hessian of the log likelihood. Then for an influence function of a pseudo-MLE with weights $w_i$, we can write
		\begin{align*}
			\hat{\theta}_j - \theta_j =  \frac{1}{n} \sum_{i = 1}^n  w_i \Delta_i + o_p(n^{-1/2}).
		\end{align*}
		To estimate the finite population variance of $\hat{\theta}_j$, calculating the finite population variance of \\ $\sum_{i = 1}^n w_i \Delta_i$ corresponds to the Taylor-linearization method described in \cite{binder1983variances}. The convenience here lies in that $\sum_{i = 1}^n w_i \Delta_i$ is a survey total, and the influence functions are simple to obtain since we have a parametric model. In fact, only the gradient of the log likelihood evaluated at the pseudo-MLE and the Hessian obtained during optimization of the weighted log likelihood are needed to calculate the finite population variance.

		Since our log likelihood is given by
		$$
		\log(\text{L}) = \sum_{i = 1}^n \left[ (1 - I_i) (-H_i(t_i)) + I_i \log(\exp(-H_i(t_{0i})) - \exp(-H_i(t_{1i}))) \right]
		$$
		we can then obtain the score for an individual $i$ for each parameter $\theta_j$ as
		\begin{align*}
			\frac{\partial}{\partial \theta_j}  & \log(\text{L}) (\hat{\boldsymbol{\theta}}) = \\
			& (1 - I_i) (-\frac{\partial}{\partial \theta_j} H_i(t_i)) + I_i \left( \frac{-\exp(-H_i(t_{0i})) \frac{\partial}{\partial \theta_j}H_i(t_{0i}) + \exp(-H_i(t_{1i})) \frac{\partial}{\partial \theta_j}H_i(t_{1i})}{\exp(-H_i(t_{0i})) - \exp(-H_i(t_{1i}))} \right).
		\end{align*}

		In practice, it is more computationally efficient to calculate the gradient analytically, though we may calculate the gradient numerically as well.

		\subsection{Log-quad model}
		\label{appendix:logquad_details}
		
		The only parameter that is estimated in the modeling step when predicting the U5MR pattern for a new country is $k$, unless the average pattern of mortality observed across data in the HMD is desired, in which case $k$ is set to 0. (Of note, \cite{guillot2022modeling} consider ${}_{60}\hat{q}_0$ to be an additional parameter for the log-quad model. Here we consider ${}_{60}\hat{q}_0$ to be data rather than a parameter, as ${}_{60}\hat{q}_0$ is input to the model as a fixed value rather than estimated during the modeling step.) The parameter $k$ is estimated in one of two ways: 
		
		\begin{enumerate}
			\item \textbf{Option 1}: If only a single value $x$ is supplied for ${}_xq_0$ to the model as a fixed constant, $k$ is estimated as
			\begin{align*}
				\hat{k} = \frac{e(x)}{v_x},
			\end{align*}
			where $e(x)$ is the difference between the predicted and observed values of ${}_xq_0$ when the the model is fit with $k = 0$.
			\item \textbf{Option 2}: If more than one $x$ is supplied for ${}_xq_0$ to the model as data, $k$ is estimated as
			\begin{align*}
				\hat{k}^*= \frac{\sum_x w(x) e(x) v_x}{\sum_x w(x) v_x^2},
			\end{align*}
			where $\hat{k}^*$ is the value of $k$ that minimizes the root-mean-square error (RMSE) of predicted values of ${}_xq_0$ to observed values of ${}_xq_0$ across all values of $x$ supplied, and $w(x)$ is the weight corresponding to the length of the previous age interval ending at age $x$ (i.e. $w(1) = 7d$, $w(2) = 7d$, \dots, $w(22) = 1y$). 
		\end{enumerate}
		When all 22 possible values for $x$ are supplied to the model, \cite{guillot2022modeling} propose an uncertainty band around the estimated survival curve that can be computed as
		\begin{align*}
			\hat{k}^* & \pm 1.96 \times \sqrt{Var(\hat{k}^*)}, \\
			Var(\hat{k}^*) & = \frac{22}{21} \left( \frac{\sum_x w(x) e(x)^2}{\sum_x w(x) v_x^2} - (\hat{k}^*)^{2} \right).
		\end{align*}
		They propose a separate uncertainty band when only one value for $x$ is supplied to the model, but we exclude it in our summary as that scenario is of little importance in our applications. In the derivation of this variance estimator, \cite{guillot2022modeling} assume that the errors $e(x)$ are independent across values $x$, that the weighted errors $w(x)e(x)$ are homoskedastic, and that $\hat{k}^*$ is approximately normally distributed. 
		
		Additionally, they note that almost all data used to estimate the age-specific coefficients $\{ a_x$, $b_x$, $c_x$, $v_x\}$ in the U5MD are estimated with values of $k$ that fall in $(-1,1)$. Due to this observed range of values, they state that values of $k$ that are estimated outside the range $(-1.1270, 1.5047)$ (the exact range of all observed values) have no ``empirical basis" and may in fact produce estimates of ${}_xq_0$ that progress nonmonotically for children under the age of 5, whereas actual survival curves must be monotonically nonincreasing. Therefore, they suggest a rule of thumb that the estimates should only be used when $k$ is estimated in the range $(-1.1, 1.5)$.

		\subsection{Discrete hazards approach}
		\label{appendix:dischaz_details}
		
		We make two further notes on using the discrete hazards model in conjunction with DHS surveys. First, in DHS surveys deaths are recorded at exact days between ages $0$ and $1$ month, monthly until $24$ months, and yearly onwards. The discrete hazards model does not take advantage of the fine-scale, daily data available prior to 1 month, and instead groups those deaths together to form a neonatal age group. If NMR is the smallest demographic rate we wish to estimate, this grouping is not inherently an issue. However, daily recorded deaths may be informative of the overall \textit{pattern} of mortality before the age of 5, so if we instead want to estimate an accurate survival curve over the entire age range from 0 to 5, grouping all deaths within the first month of life together will not capture the expected sharp decline in survival in the first week of life, or even the first two weeks of life. 
		
		A second benefit of aggregating our data across age groups is that, especially at small levels of spatial aggregation, we may have very little data available on the hazard in some age groups. Consequently, if data is sparse we may prefer to use fewer age groups in the discrete hazards model, as if no deaths were present in a certain age-period group (which is common for small regions in small area estimation, or fine-scale time periods), their estimated hazard will be exactly zero. Hazards of exactly zero are undesirable, as they are both implausible and it is difficult to get inference around such an estimate.

		\section{Additional Results}
		\label{sec:additional_results}
		
		One thing to note from the results of our application is that both the ETSP and generalized Gamma models performed reasonably well, but again perhaps not significantly better than some of the two-parameter models. This can be seen for Namibia in Table 2 in the main paper in particular, where both differences between the models' respective parametric estimates and the Turnbull estimates capture zero for 99\% (ETSP) and 100\% (generalized Gamma) of ages where the Turnbull estimate is defined prior to 60 months. However, the lognormal model performs just as well in Namibia according to this metric. Furthermore, fitting both models (generalized Gamma and ETSP) results in computational complexities that may make them less desirable than other parametric options. The ETSP model, for example, does not have a closed form cumulative hazard, and therefore requires numerical integration at every step of the likelihood optimization. Therefore, this model takes more time to fit and is potentially less numerically stable than others with closed form cumulative hazards. The generalized Gamma distribution, on the other hand, occasionally produces unreasonably wide confidence bands. In some countries and time periods, Burkina Faso $[2000,2005)$, for example, the shape parameter $Q$ is estimated with a large variance relative to the other parameters in the model. This results in the confidence bands produced being highly asymmetric, and so wide as to be unusable in practice. As such, we believe that in certain cases there may not be enough data under the age of 5 to reliably estimate all three parameters that define the generalized Gamma survival curve.
		
\begin{figure}
			\centering
			\resizebox{\columnwidth}{!}{%
				\includegraphics[scale = 0.7]{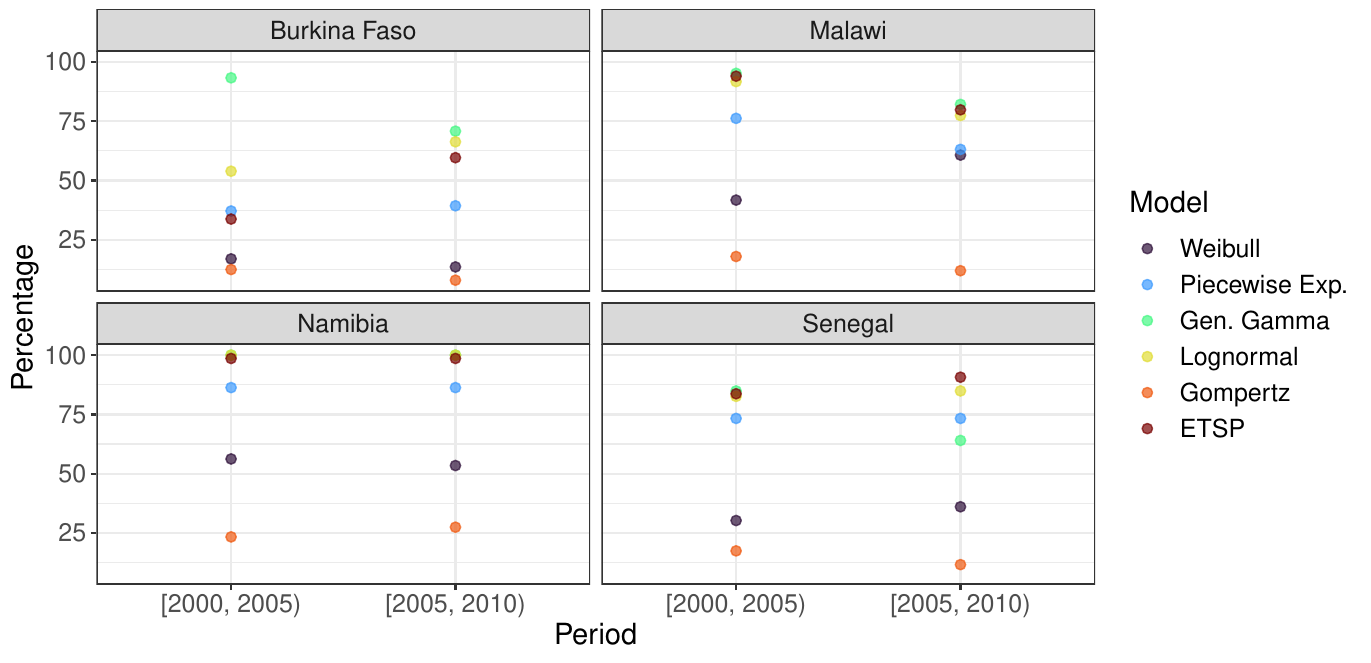}
			}
			\caption{Model validation results. Percentage of samples (out of 500) from $\hat{\theta} - \tilde{\theta}$ that contain 0 for all parametric models, countries, and periods.}
			\label{fig:modelva}
		\end{figure}

		A visual representation of model validation results is presented in \ref{fig:modelva}.
		
		Below we display additional results from Weibull, generalized Gamma, piecewise exponential, lognormal, Gompertz, and exponentially-truncated shifted power (ETSP) models for Burkina Faso, Malawi, Senegal, and Namibia.
		
		
		\begin{figure}
			\centering
			\resizebox{\columnwidth}{!}{\includegraphics[page=1,scale = 0.65]{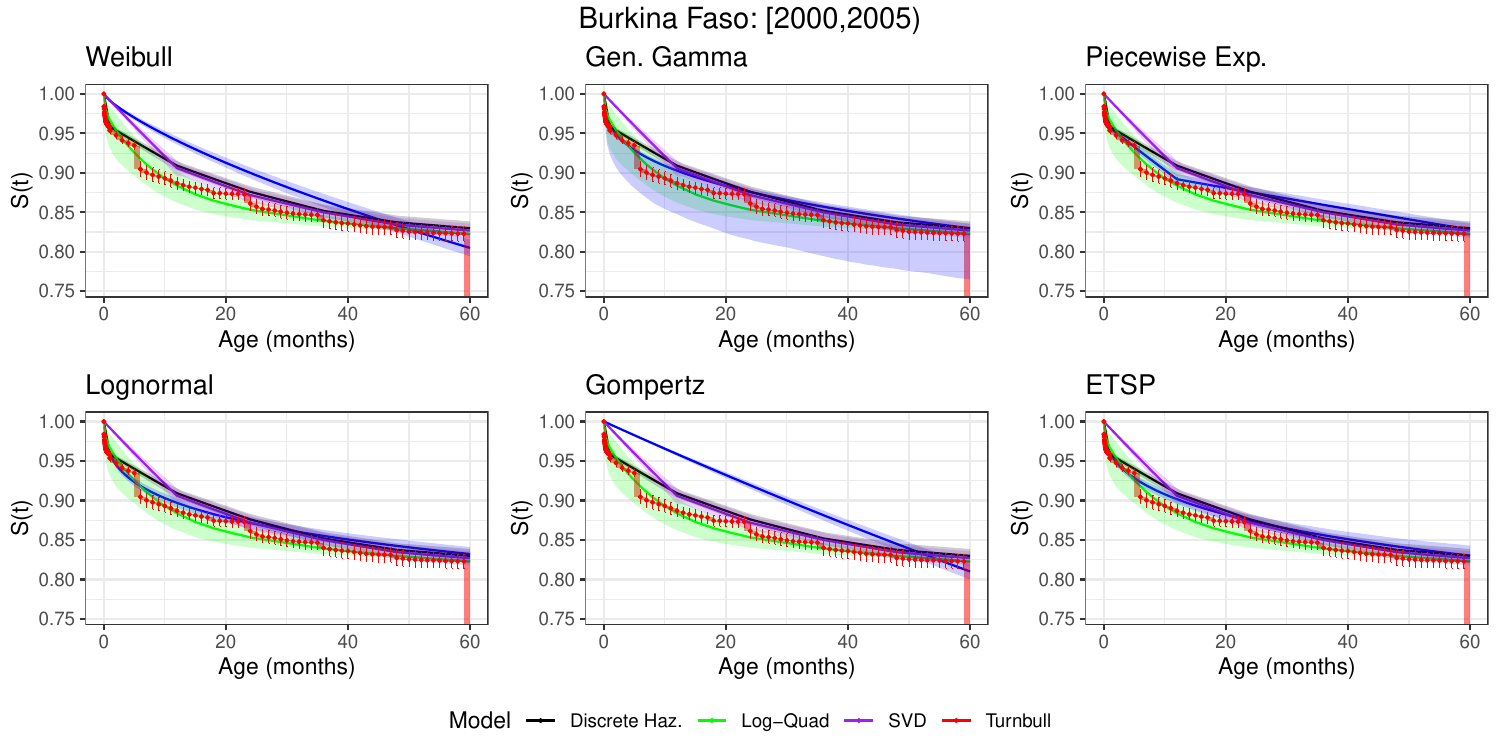}}
			\resizebox{\columnwidth}{!}{\includegraphics[page=2,scale = 0.65]{Plots/BurkinaFaso/curves_compare_separated_2000_2010.pdf}}
			\caption{Estimated survival curves for Burkina Faso in $[2000,2005)$ (top) and $[2005,2010)$ (bottom) from ages 0 to 60 months. Parametric, pseudo-likelihood estimates are in blue. All confidence bands are 95\% confidence intervals based on finite population variances, with the exception of the log-quad model where uncertainty is calculated as in \citet{guillot2022modeling}.}
			\label{fig:bf_curves}
		\end{figure}
		
		\begin{figure}
			\centering
			\resizebox{\columnwidth}{!}{\includegraphics[scale = 0.65, page = 1]{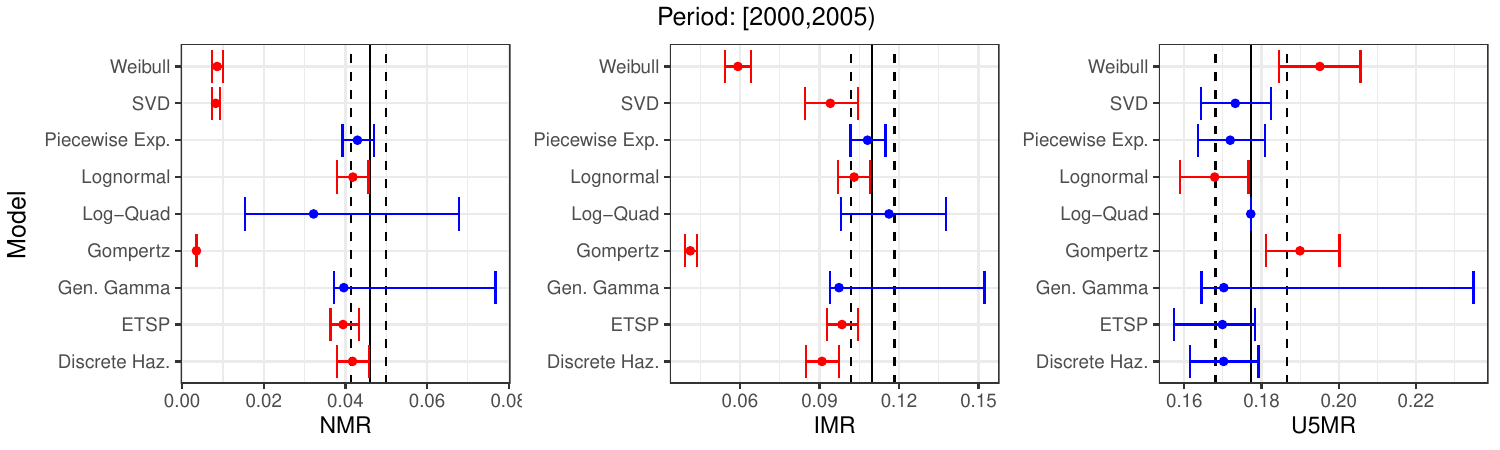}}
			\resizebox{\columnwidth}{!}{\includegraphics[scale = 0.65, page = 2]{Plots/BurkinaFaso/nmr_imr_u5mr_compare_2000_2010.pdf}}
			\caption{Estimates of NMR, IMR, and U5MR for Burkina Faso in periods $[2000,2005)$ (top) and $[2005,2010)$ (bottom). Turnbull point estimates are denoted by vertical black lines. All 95\% confidence intervals are based on finite population variances, with the exception of the log-quad model where uncertainty is calculated as in \citet{guillot2022modeling}.}
			\label{fig:bf_mr}
		\end{figure}
		
		\begin{figure}
			\centering
			\resizebox{\columnwidth}{!}{\includegraphics[scale = 0.65, page = 1]{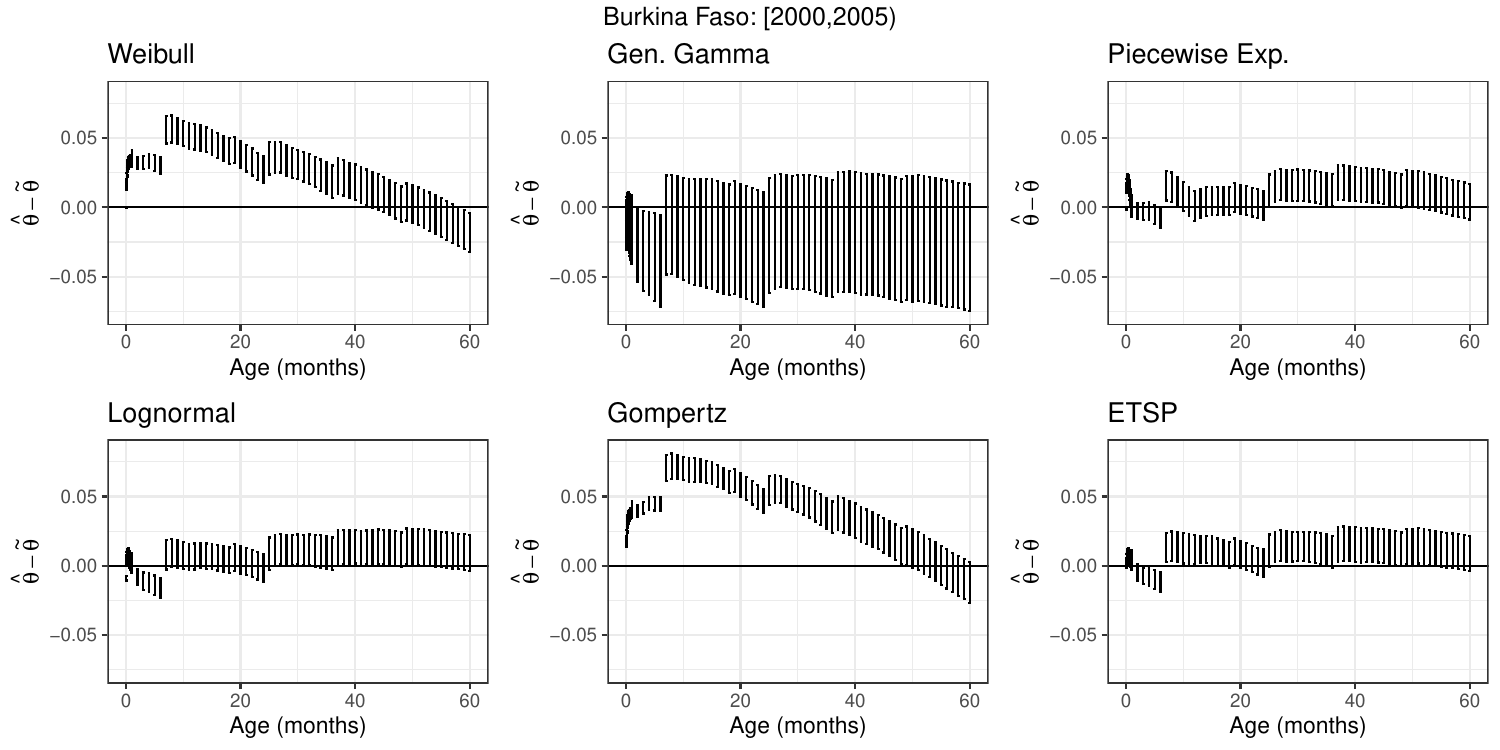}}
			\resizebox{\columnwidth}{!}{ \includegraphics[scale = 0.65, page = 2]{Plots/BurkinaFaso/turnbull_diffs.pdf}}
			\caption{Empirical distributions of differences in survival curves for Burkina Faso in $[2000,2005)$ (top) and $[2005,2010)$ (bottom) from ages 0 to 60 months between parametric estimates $\hat{\theta}$ and the Turnbull estimate $\tilde{\theta}$.}
			\label{fig:bf_turn_diffs}
		\end{figure}
		
		
		\begin{figure}
			\centering
			\resizebox{\columnwidth}{!}{\includegraphics[scale = 0.65,page=1]{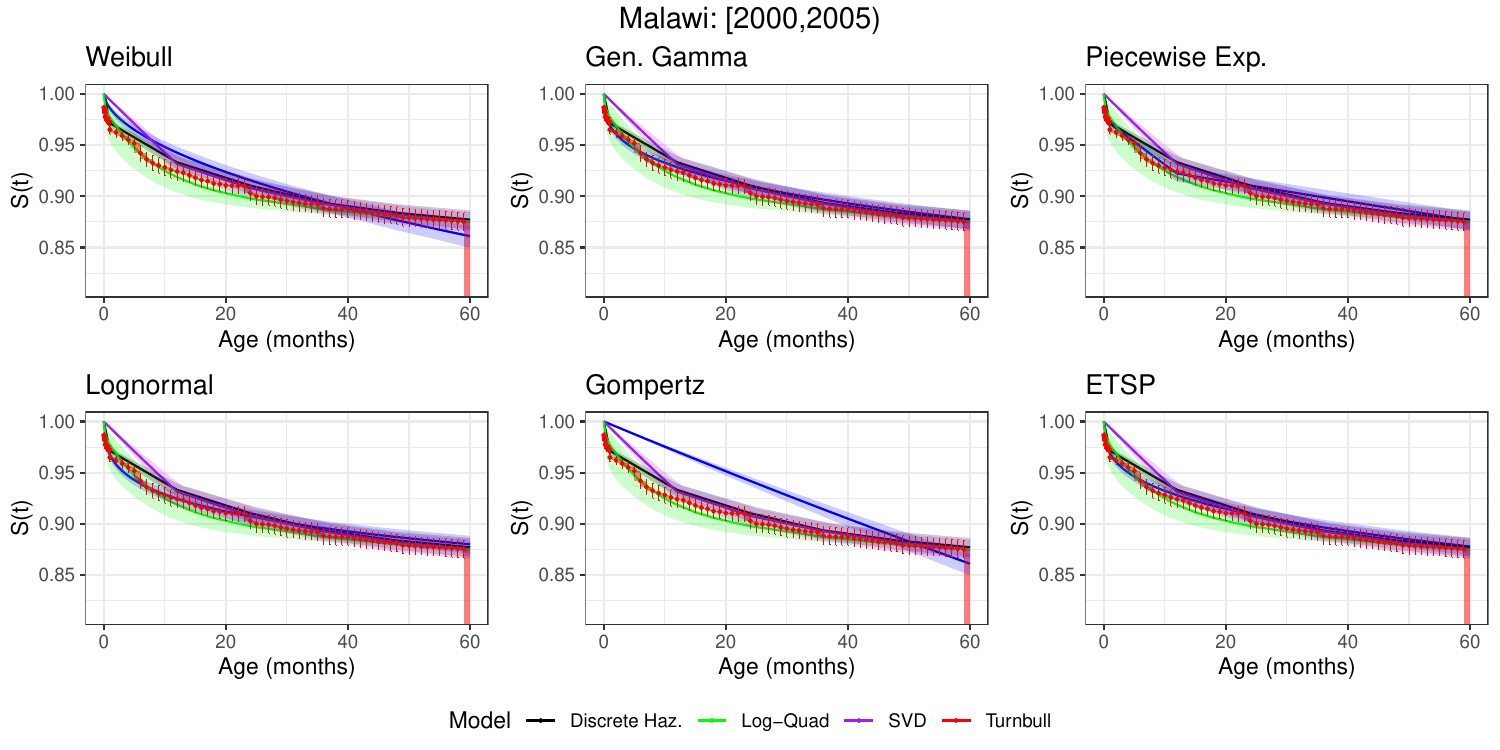}}
			\resizebox{\columnwidth}{!}{\includegraphics[scale = 0.65,page=2]{Plots/Malawi/curves_compare_separated_2000_2010.pdf}}
			\caption{Estimated survival curves for Malawi in $[2000,2005)$ (top) and $[2005,2010)$ (bottom) from ages 0 to 60 months. Parametric, pseudo-likelihood estimates are in blue. All confidence bands are 95\% confidence intervals based on finite population variances, with the exception of the log-quad model where uncertainty is calculated as in \citet{guillot2022modeling}.}
			\label{fig:malawi_curves}
		\end{figure}
		
		\begin{figure}
			\centering
			\resizebox{\columnwidth}{!}{\includegraphics[scale = 0.65, page = 1]{Plots/Malawi/nmr_imr_u5mr_compare_2000_2010.pdf}}
			\resizebox{\columnwidth}{!}{\includegraphics[scale = 0.65, page = 2]{Plots/Malawi/nmr_imr_u5mr_compare_2000_2010.pdf}}
			\caption{Estimates of NMR, IMR, and U5MR for Malawi in periods $[2000,2005)$ (top) and $[2005,2010)$ (bottom). Turnbull point estimates are denoted by vertical black lines. All 95\% confidence intervals are based on finite population variances, with the exception of the log-quad model where uncertainty is calculated as in \citet{guillot2022modeling}.}
			\label{fig:malawi_mr}
		\end{figure}
		
		\begin{figure}
			\centering
			\resizebox{\columnwidth}{!}{\includegraphics[scale = 0.65, page = 1]{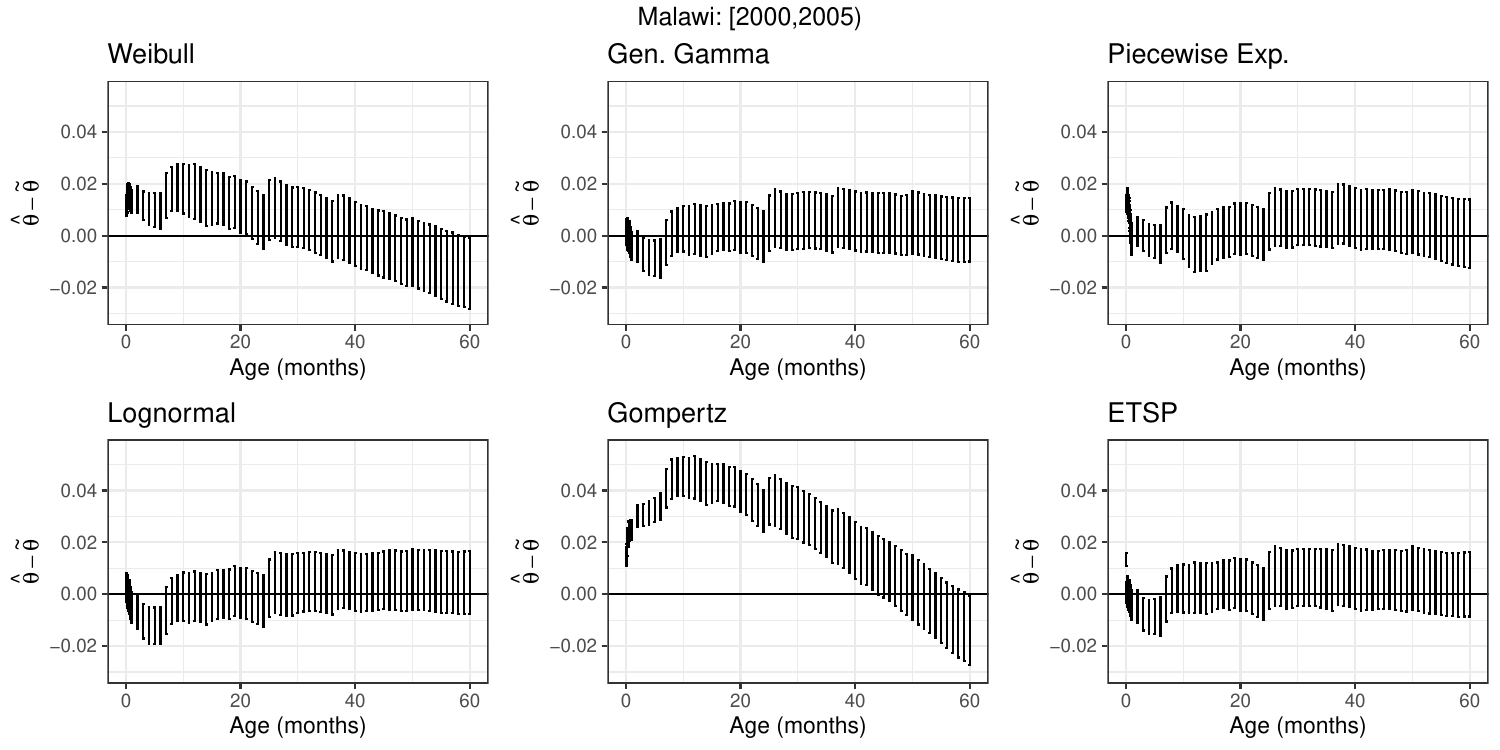}}
			\resizebox{\columnwidth}{!}{ \includegraphics[scale = 0.65, page = 2]{Plots/Malawi/turnbull_diffs.pdf}}
			\caption{Empirical distributions of differences in survival curves for Malawi in $[2000,2005)$ (top) and $[2005,2010)$ (bottom) from ages 0 to 60 months between parametric estimates $\hat{\theta}$ and the Turnbull estimate $\tilde{\theta}$.}
			\label{fig:malawi_turn_diffs}
		\end{figure}
		
		
		\begin{figure}
			\centering
			\resizebox{\columnwidth}{!}{\includegraphics[scale = 0.65,page=1]{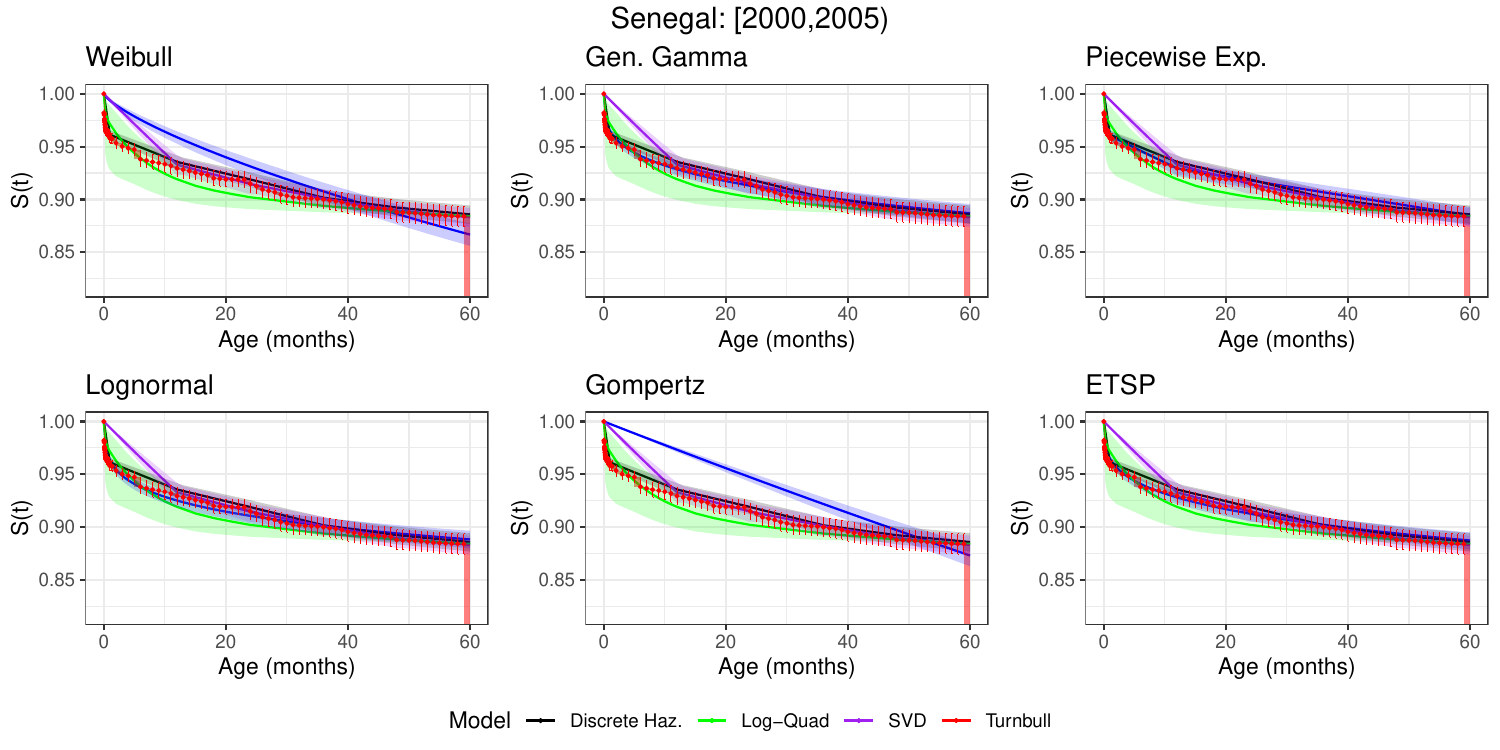}}
			\resizebox{\columnwidth}{!}{\includegraphics[scale = 0.65,page=2]{Plots/Senegal/curves_compare_separated_2000_2010.pdf}}
			\caption{Estimated survival curves for Senegal in $[2000,2005)$ (top) and $[2005,2010)$ (bottom) from ages 0 to 60 months. Parametric, pseudo-likelihood estimates are in blue. All confidence bands are 95\% confidence intervals based on finite population variances, with the exception of the log-quad model where uncertainty is calculated as in \citet{guillot2022modeling}.}
			\label{fig:senegal_curves}
		\end{figure}
		
		\begin{figure}
			\centering
			\resizebox{\columnwidth}{!}{\includegraphics[scale = 0.65, page = 1]{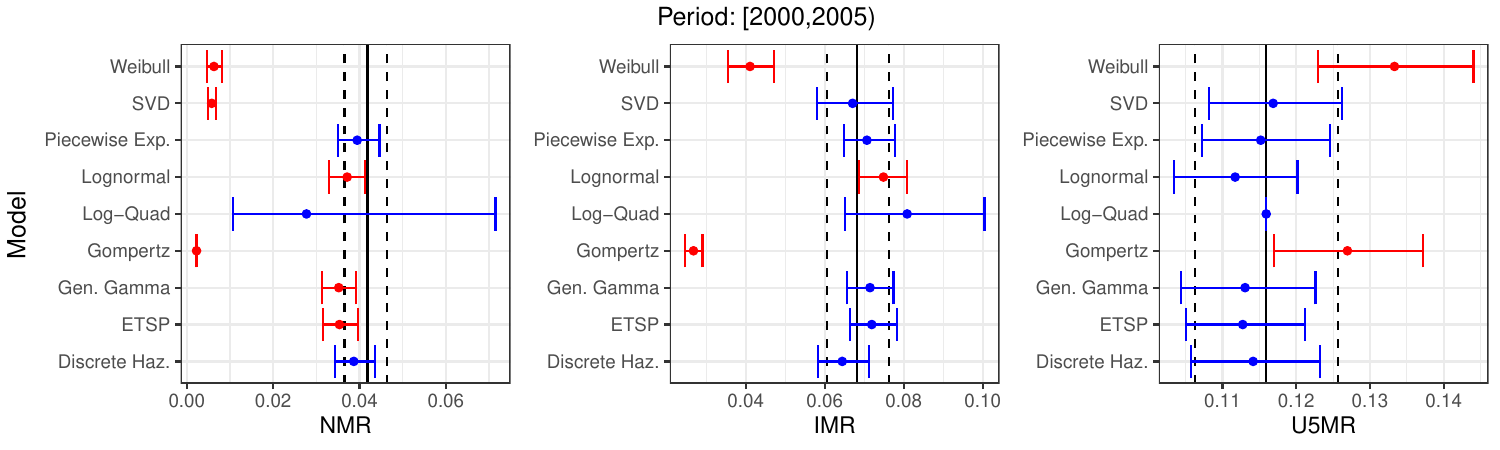}}
			\resizebox{\columnwidth}{!}{\includegraphics[scale = 0.65, page = 2]{Plots/Senegal/nmr_imr_u5mr_compare_2000_2010.pdf}}
			\caption{Estimates of NMR, IMR, and U5MR for Senegal in periods $[2000,2005)$ (top) and $[2005,2010)$ (bottom). Turnbull point estimates are denoted by vertical black lines. All 95\% confidence intervals are based on finite population variances, with the exception of the log-quad model where uncertainty is calculated as in \citet{guillot2022modeling}.}
			\label{fig:senegal_mr}
		\end{figure}
		
		\begin{figure}
			\centering
			\resizebox{\columnwidth}{!}{\includegraphics[scale = 0.65, page = 1]{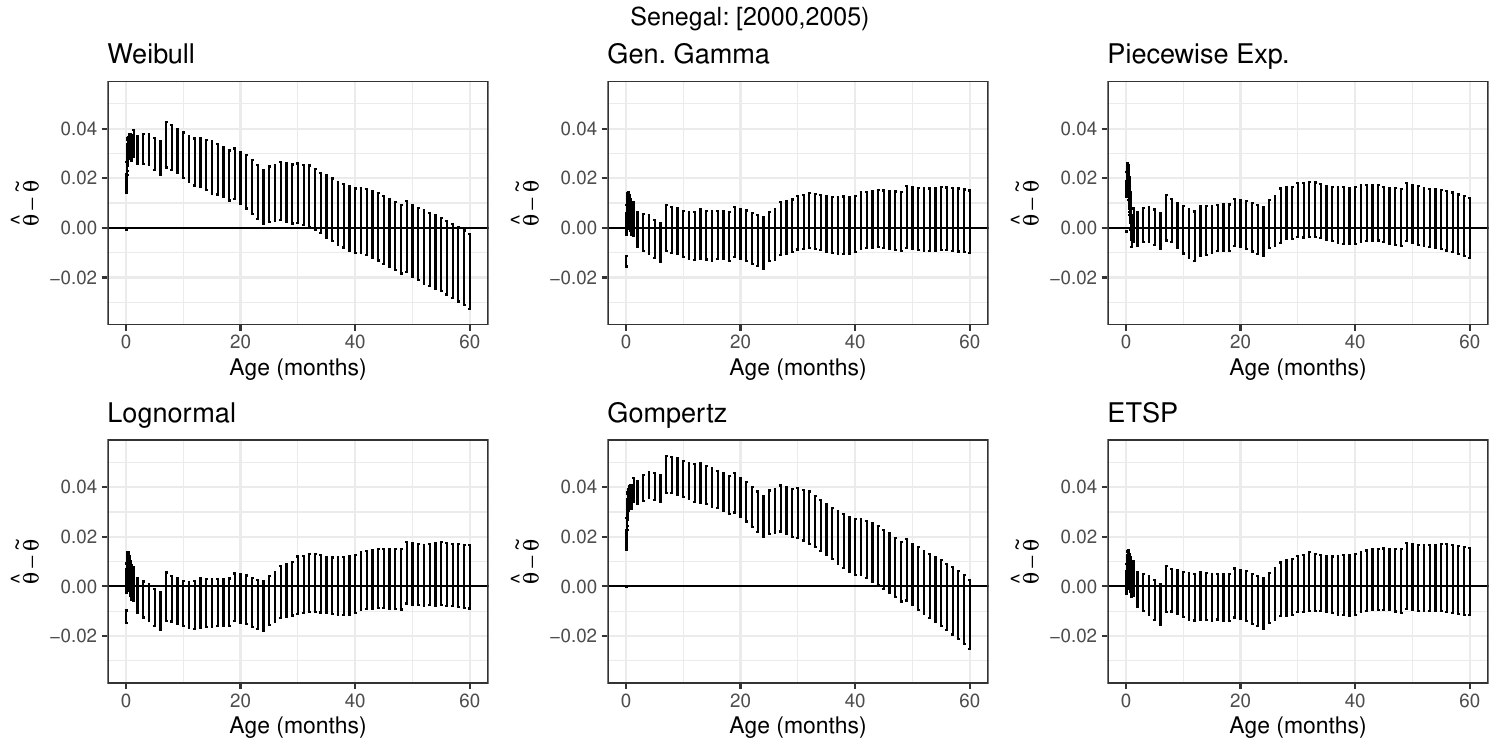}}
			\resizebox{\columnwidth}{!}{\includegraphics[scale = 0.65, page = 2]{Plots/Senegal/turnbull_diffs.pdf}}
			\caption{Empirical distributions of differences in survival curves for Senegal in $[2000,2005)$ (top) and $[2005,2010)$ (bottom) from ages 0 to 60 months between parametric estimates $\hat{\theta}$ and the Turnbull estimate $\tilde{\theta}$. Note that for $[2005, 2010)$ the differences have been cut off at $-0.03$ for clarity, though the differences extend much further negative for the generalized gamma model.}
			\label{fig:senegal_turn_diffs}
		\end{figure}
		
		
		\begin{figure}
			\centering
			\resizebox{\columnwidth}{!}{\includegraphics[scale = 0.65,page=1]{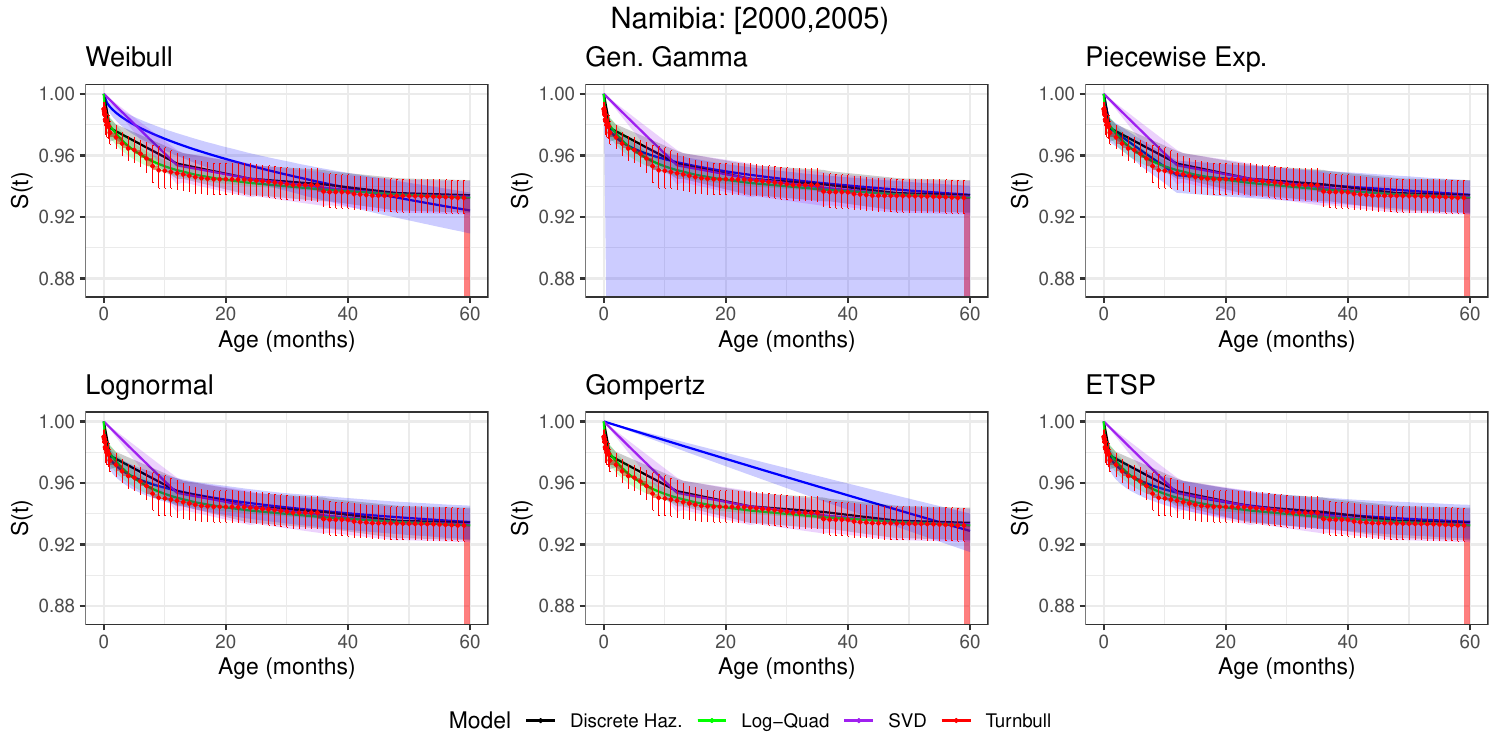}}
			\resizebox{\columnwidth}{!}{\includegraphics[scale = 0.65,page=2]{Plots/Namibia/curves_compare_separated_2000_2010.pdf}}
			\caption{Estimated survival curves for Namibia in $[2000,2005)$ (top) and $[2005,2010)$ (bottom) from ages 0 to 60 months. Parametric, pseudo-likelihood estimates are in blue. All confidence bands are 95\% confidence intervals based on finite population variances, with the exception of the log-quad model where uncertainty is calculated as in \citet{guillot2022modeling}.}
			\label{fig:namibia_curves}
		\end{figure}
		
		\begin{figure}
			\centering
			\resizebox{\columnwidth}{!}{\includegraphics[scale = 0.65, page = 1]{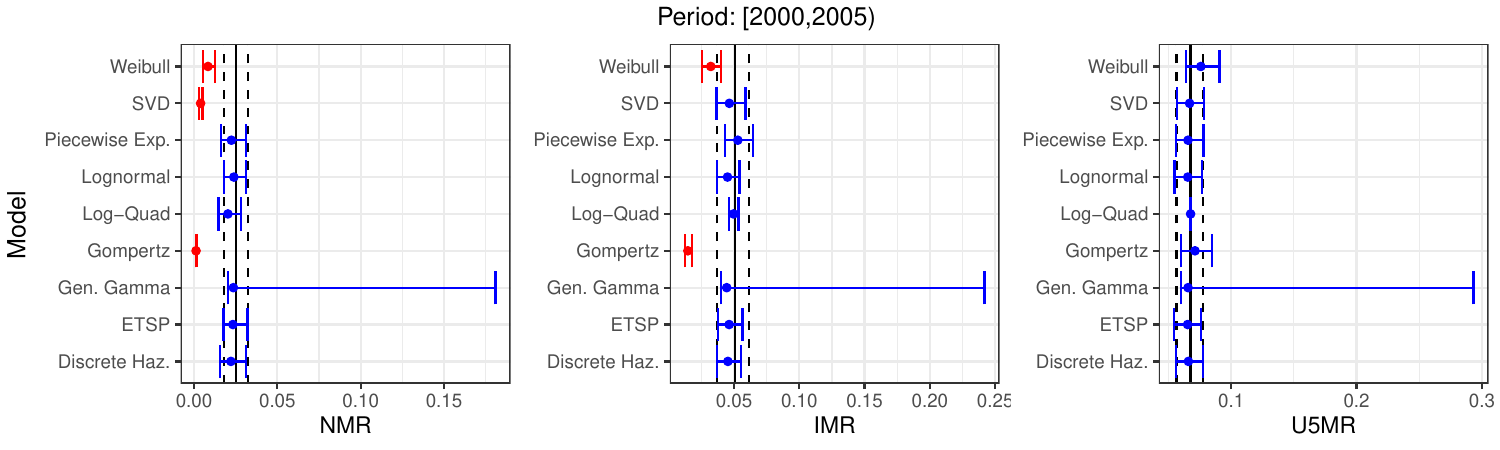}}
			\resizebox{\columnwidth}{!}{ \includegraphics[scale = 0.65, page = 2]{Plots/Namibia/nmr_imr_u5mr_compare_2000_2010.pdf}}
			\caption{Estimates of NMR, IMR, and U5MR for Namibia in periods $[2000,2005)$ (top) and $[2005,2010)$ (bottom). Turnbull point estimates are denoted by vertical black lines. All 95\% confidence intervals are based on finite population variances, with the exception of the log-quad model where uncertainty is calculated as in \citet{guillot2022modeling}.}
			\label{fig:namibia_mr}
		\end{figure}
		
		\begin{figure}
			\centering
			\resizebox{\columnwidth}{!}{\includegraphics[scale = 0.65, page = 1]{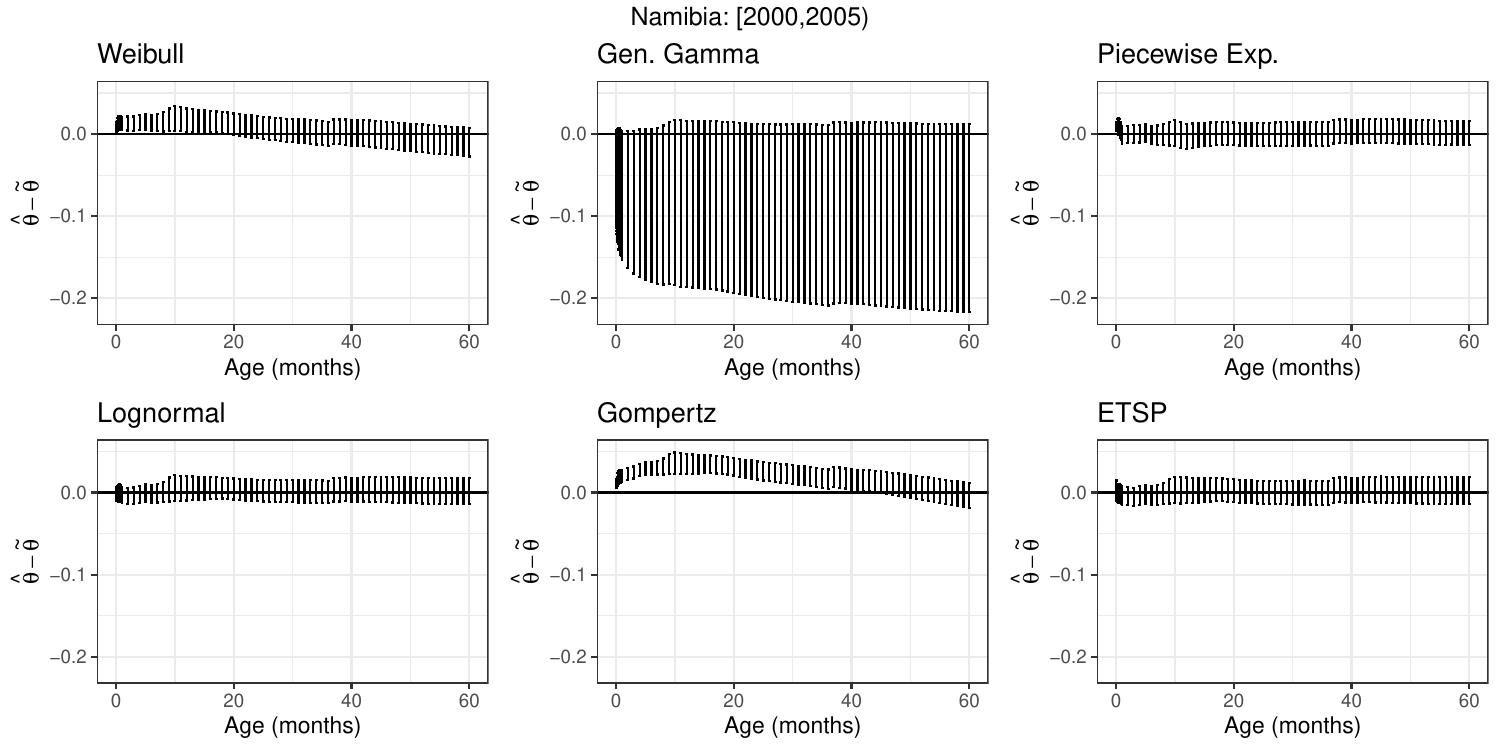}}
			\resizebox{\columnwidth}{!}{\includegraphics[scale = 0.65, page = 2]{Plots/Namibia/turnbull_diffs.pdf}}
			\caption{Empirical distributions of differences in survival curves for Namibia in $[2000,2005)$ (top) and $[2005,2010)$ (bottom) from ages 0 to 60 months between parametric estimates $\hat{\theta}$ and the Turnbull estimate $\tilde{\theta}$.}
			\label{fig:namibia_turn_diffs}
		\end{figure}
		
		\section{Comparison to models unadjusted for age heaping}
		\label{sec:nah}
		
		We repeat our application, fitting the same models without addressing age-heaping at 12 months (by interval-censoring observations recorded as having died between 6 and 18 months for that entire 12 month period). 
		
		
		\begin{figure}
			\centering
			\resizebox{\columnwidth}{!}{\includegraphics[page=1,scale = 0.65]{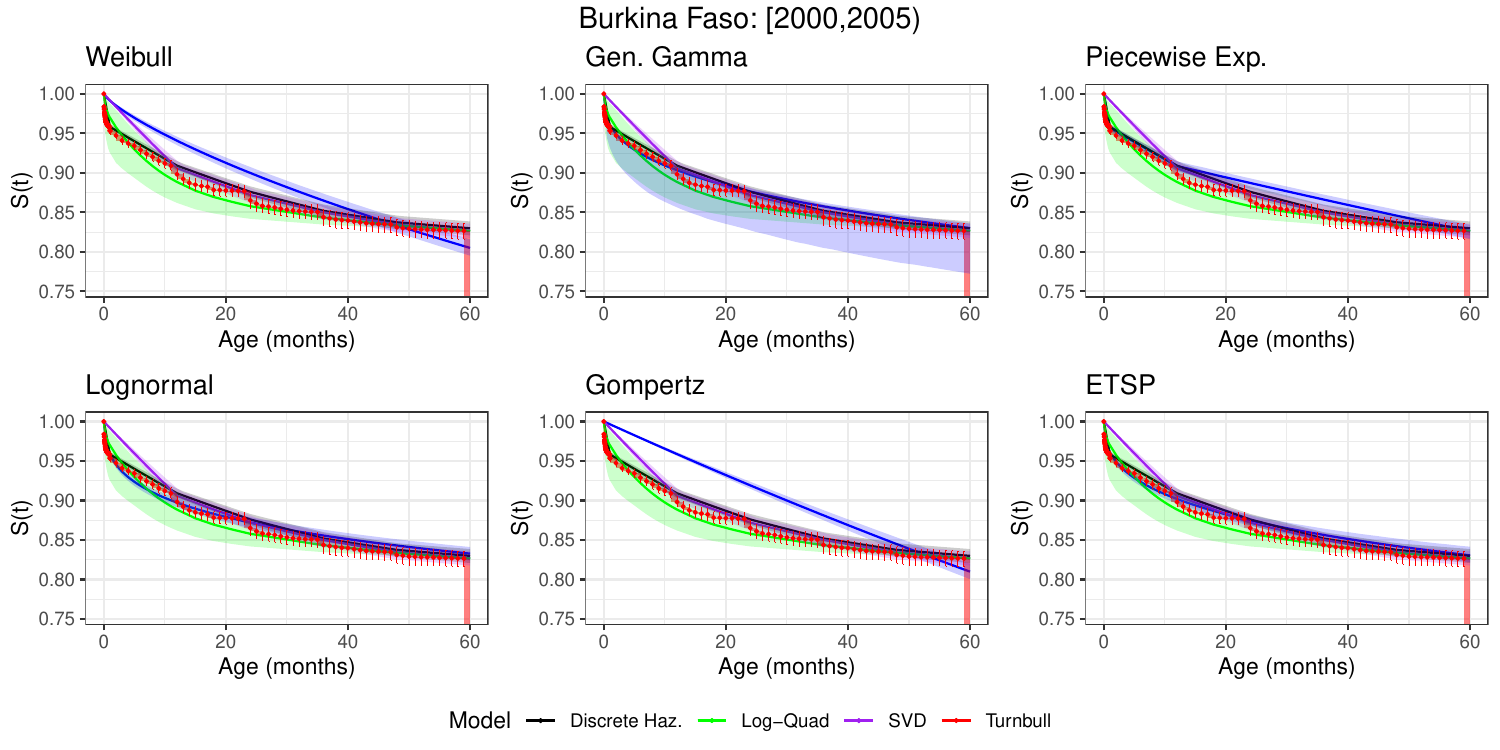}}
			\resizebox{\columnwidth}{!}{\includegraphics[page=2,scale = 0.65]{Plots/BurkinaFaso/curves_compare_separated_2000_2010_nah.pdf}}
			\caption{Estimated survival curves for Burkina Faso in $[2000,2005)$ (top) and $[2005,2010)$ (bottom) from ages 0 to 60 months, not adjusted for age heaping at 12 months. Parametric, pseudo-likelihood estimates are in blue. All confidence bands are 95\% confidence intervals based on finite population variances, with the exception of the log-quad model where uncertainty is calculated as in \citet{guillot2022modeling}.}
			\label{fig:bf_curves_nah}
		\end{figure}
		
		\begin{figure}
			\centering
			\resizebox{\columnwidth}{!}{\includegraphics[scale = 0.65, page = 1]{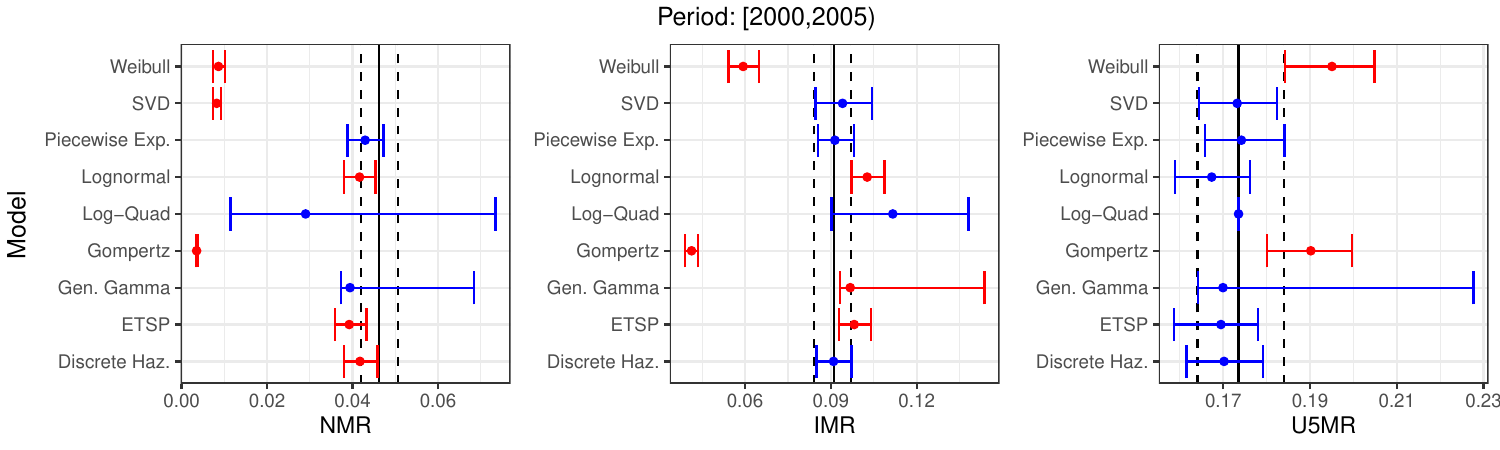}}
			\resizebox{\columnwidth}{!}{\includegraphics[scale = 0.65, page = 2]{Plots/BurkinaFaso/nmr_imr_u5mr_compare_2000_2010_nah.pdf}}
			\caption{Estimates of NMR, IMR, and U5MR for Burkina Faso in periods $[2000,2005)$ (top) and $[2005,2010)$ (bottom), not adjusted for age heaping at 12 months. Turnbull point estimates are denoted by vertical black lines. All 95\% confidence intervals are based on finite population variances, with the exception of the log-quad model where uncertainty is calculated as in \citet{guillot2022modeling}.}
			\label{fig:bf_mr_nah}
		\end{figure}
		
		\begin{figure}
			\centering
			\resizebox{\columnwidth}{!}{\includegraphics[scale = 0.65, page = 1]{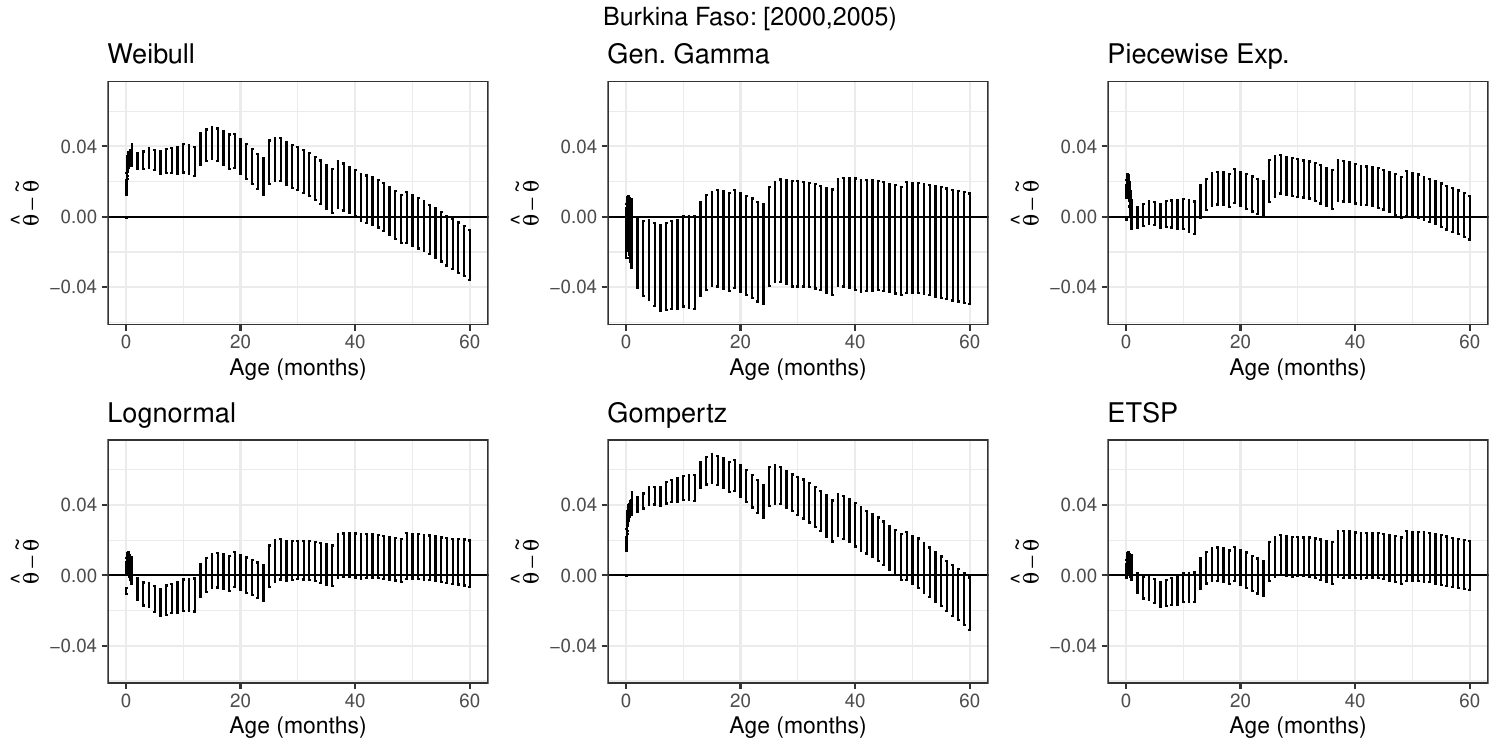}}
			\resizebox{\columnwidth}{!}{\includegraphics[scale = 0.65, page = 2]{Plots/BurkinaFaso/turnbull_diffs_nah.pdf}}
			\caption{Empirical distributions of differences in survival curves for Burkina Faso in $[2000,2005)$ (top) and $[2005,2010)$ (bottom) from ages 0 to 60 months between parametric estimates (not adjusted for age heaping) $\hat{\theta}$ and the Turnbull estimate $\tilde{\theta}$.}
			\label{fig:bf_turn_diffs_nah}
		\end{figure}
		
		
		\begin{figure}
			\centering
			\resizebox{\columnwidth}{!}{\includegraphics[page=1,scale = 0.65]{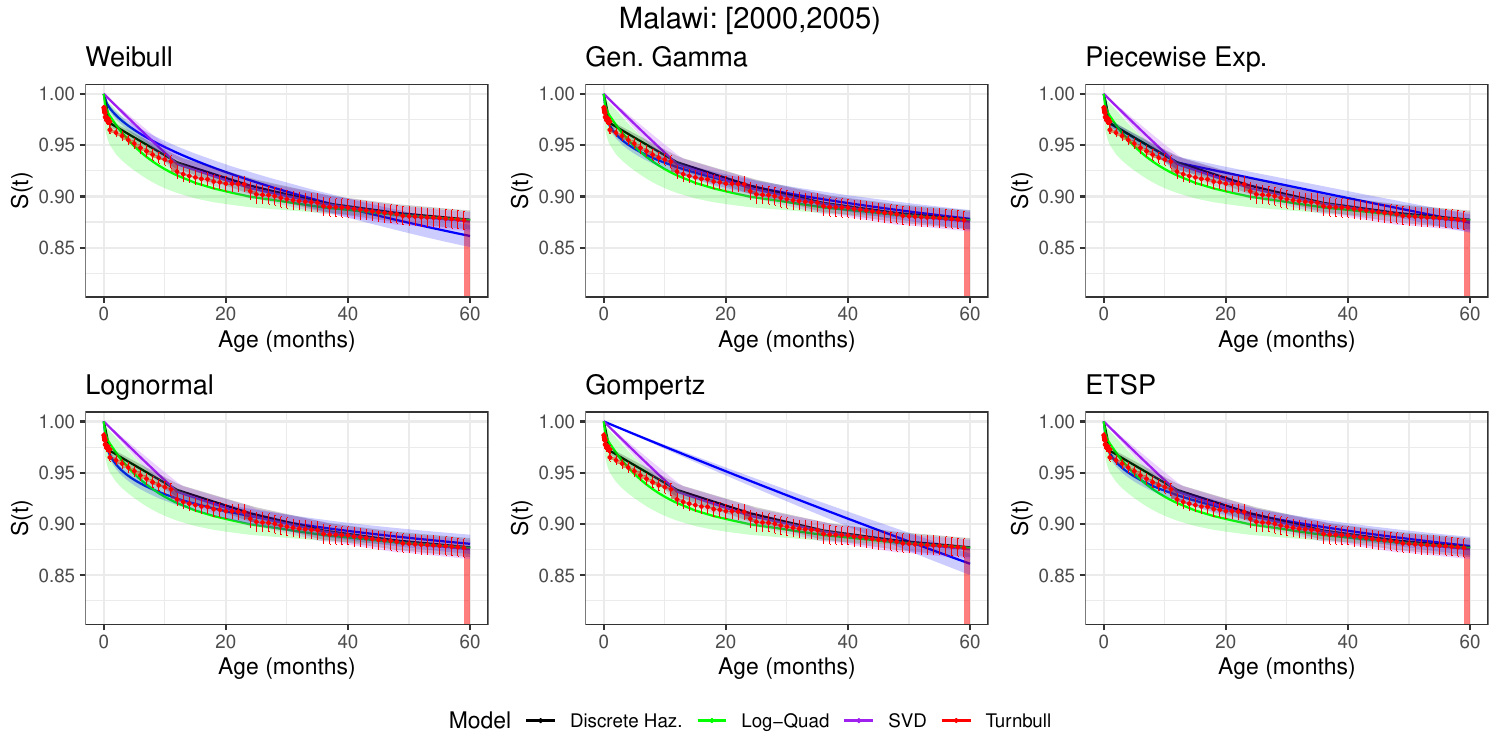}}
			\resizebox{\columnwidth}{!}{\includegraphics[page=2,scale = 0.65]{Plots/Malawi/curves_compare_separated_2000_2010_nah.pdf}}
			\caption{Estimated survival curves for Malawi in $[2000,2005)$ (top) and $[2005,2010)$ (bottom) from ages 0 to 60 months, not adjusted for age heaping at 12 months. Parametric, pseudo-likelihood estimates are in blue. All confidence bands are 95\% confidence intervals based on finite population variances, with the exception of the log-quad model where uncertainty is calculated as in \citet{guillot2022modeling}.}
			\label{fig:mwi_curves_nah}
		\end{figure}
		
		\begin{figure}
			\centering
			\resizebox{\columnwidth}{!}{\includegraphics[scale = 0.65, page = 1]{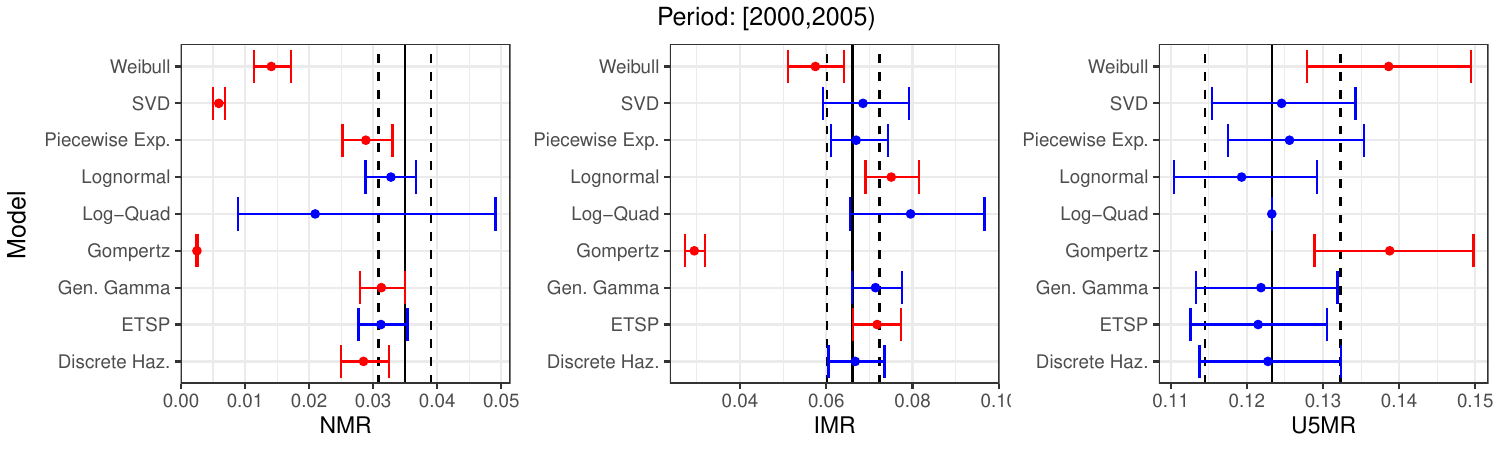}}
			\resizebox{\columnwidth}{!}{\includegraphics[scale = 0.65, page = 2]{Plots/Malawi/nmr_imr_u5mr_compare_2000_2010_nah.pdf}}
			\caption{Estimates of NMR, IMR, and U5MR for Malawi in periods $[2000,2005)$ (top) and $[2005,2010)$ (bottom), not adjusted for age heaping at 12 months. Turnbull point estimates are denoted by vertical black lines. All 95\% confidence intervals are based on finite population variances, with the exception of the log-quad model where uncertainty is calculated as in \citet{guillot2022modeling}.}
			\label{fig:mwi_mr_nah}
		\end{figure}
		
		\begin{figure}
			\centering
			\resizebox{\columnwidth}{!}{\includegraphics[scale = 0.65, page = 1]{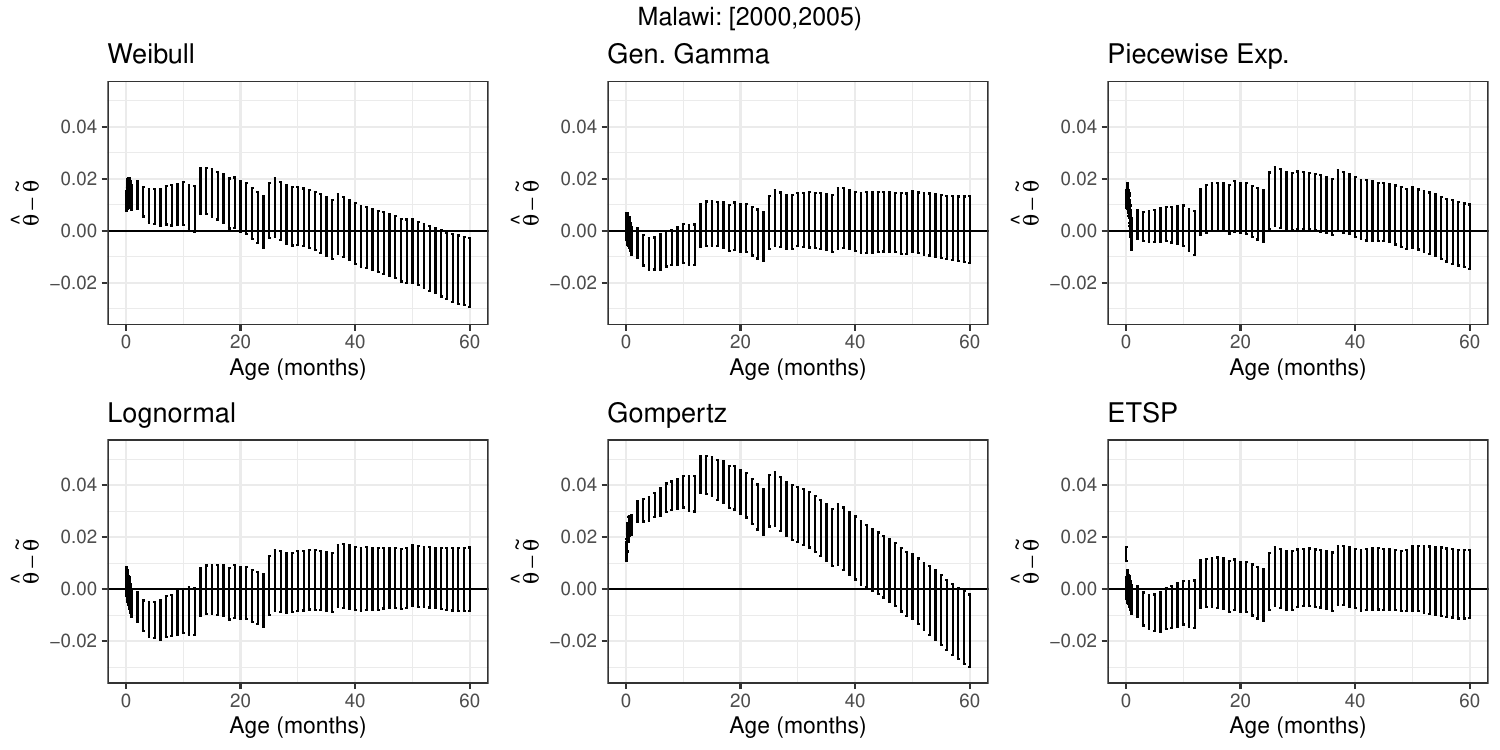}}
			\resizebox{\columnwidth}{!}{\includegraphics[scale = 0.65, page = 2]{Plots/Malawi/turnbull_diffs_nah.pdf}}
			\caption{Empirical distributions of differences in survival curves for Malawi in $[2000,2005)$ (top) and $[2005,2010)$ (bottom) from ages 0 to 60 months between parametric estimates (not adjusted for age heaping) $\hat{\theta}$ and the Turnbull estimate $\tilde{\theta}$.}
			\label{fig:mwi_turn_diffs_nah}
		\end{figure}
		
		
		\begin{figure}
			\centering
			\resizebox{\columnwidth}{!}{\includegraphics[page=1,scale = 0.65]{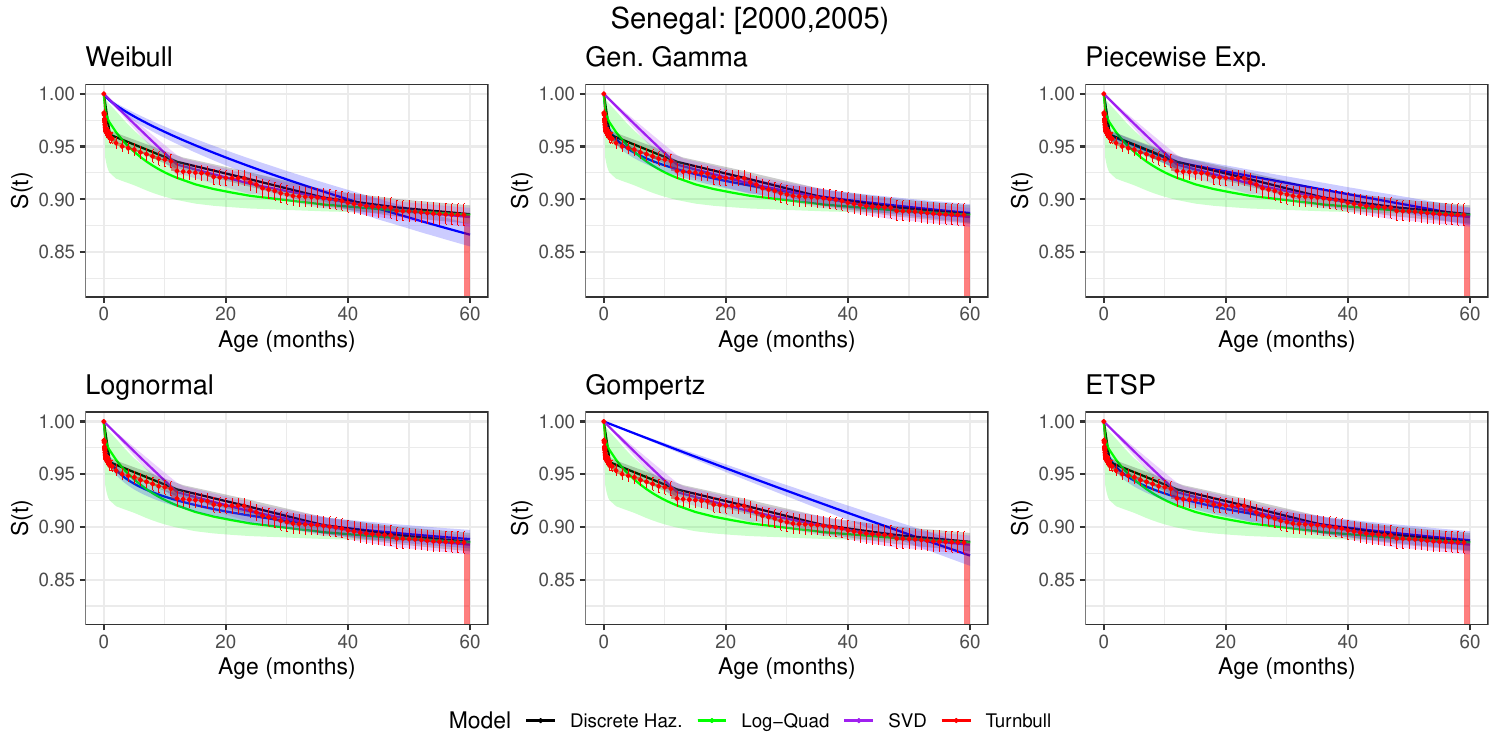}}
			\resizebox{\columnwidth}{!}{\includegraphics[page=2,scale = 0.65]{Plots/Senegal/curves_compare_separated_2000_2010_nah.pdf}}
			\caption{Estimated survival curves for Senegal in $[2000,2005)$ (top) and $[2005,2010)$ (bottom) from ages 0 to 60 months, not adjusted for age heaping at 12 months. Parametric, pseudo-likelihood estimates are in blue. All confidence bands are 95\% confidence intervals based on finite population variances, with the exception of the log-quad model where uncertainty is calculated as in \citet{guillot2022modeling}.}
			\label{fig:sen_curves_nah}
		\end{figure}
		
		\begin{figure}
			\centering
			\resizebox{\columnwidth}{!}{\includegraphics[scale = 0.65, page = 1]{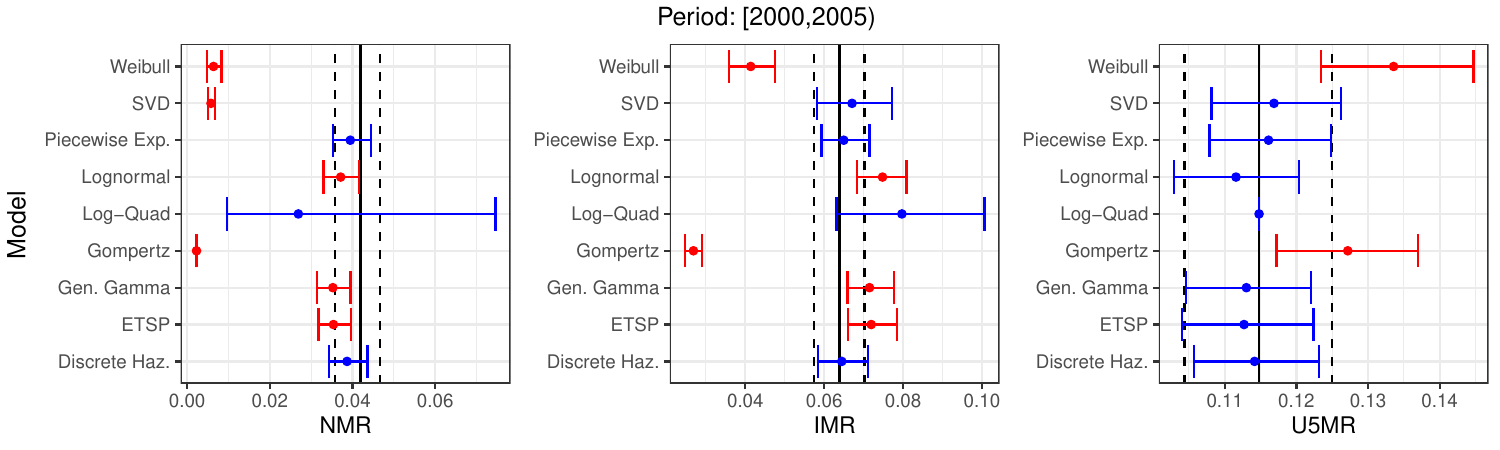}}
			\resizebox{\columnwidth}{!}{\includegraphics[scale = 0.65, page = 2]{Plots/Senegal/nmr_imr_u5mr_compare_2000_2010_nah.pdf}}
			\caption{Estimates of NMR, IMR, and U5MR for Senegal in periods $[2000,2005)$ (top) and $[2005,2010)$ (bottom), not adjusted for age heaping at 12 months. Turnbull point estimates are denoted by vertical black lines. All 95\% confidence intervals are based on finite population variances, with the exception of the log-quad model where uncertainty is calculated as in \citet{guillot2022modeling}.}
			\label{fig:sen_mr_nah}
		\end{figure}
		
		\begin{figure}
			\centering
			\resizebox{\columnwidth}{!}{\includegraphics[scale = 0.65, page = 1]{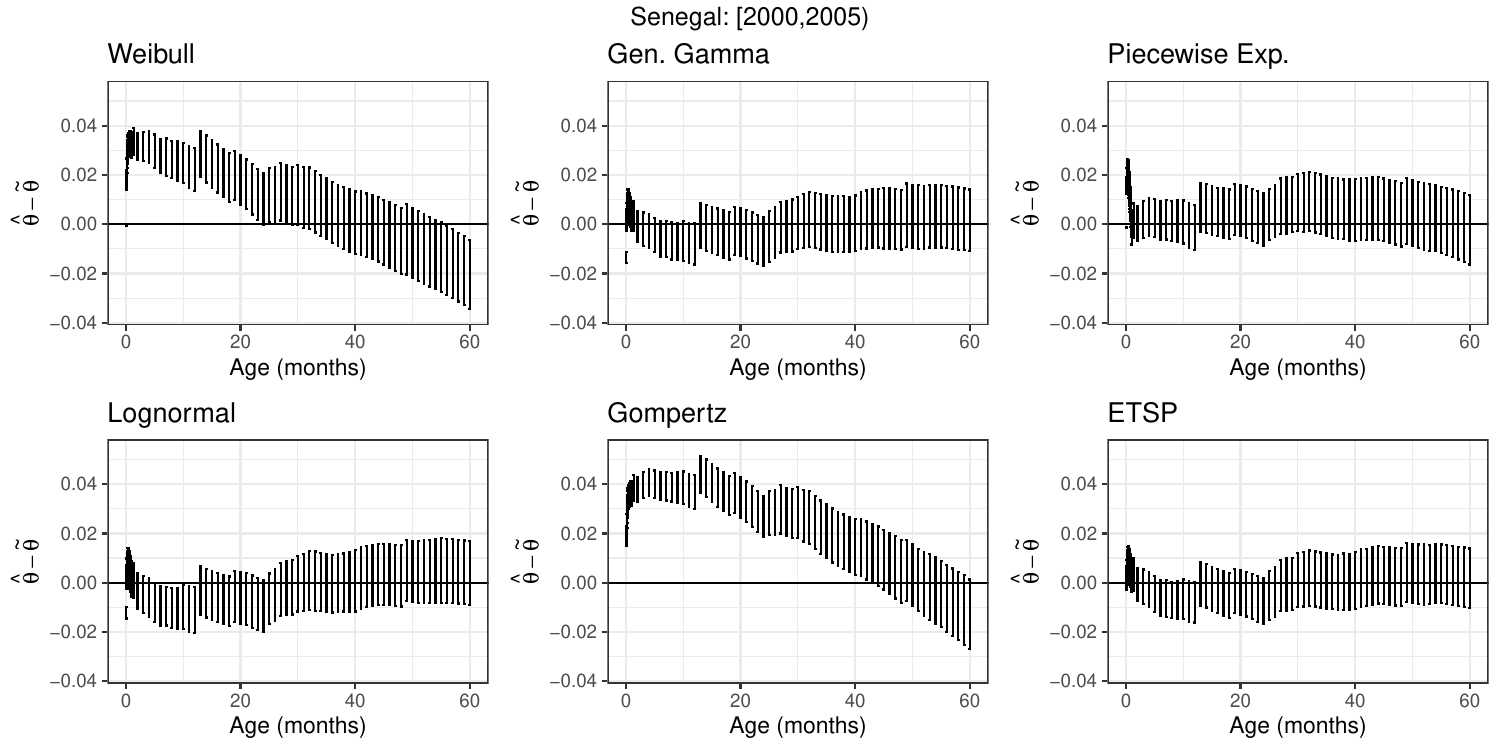}}
			\resizebox{\columnwidth}{!}{\includegraphics[scale = 0.65, page = 2]{Plots/Senegal/turnbull_diffs_nah.pdf}}
			\caption{Empirical distributions of differences in survival curves for Senegal in $[2000,2005)$ (top) and $[2005,2010)$ (bottom) from ages 0 to 60 months between parametric estimates (not adjusted for age heaping) $\hat{\theta}$ and the Turnbull estimate $\tilde{\theta}$. Note that for $[2005, 2010)$ the differences have been cut off at $-0.03$ for clarity, though the differences extend much further negative for the generalized gamma model.}
			\label{fig:sen_turn_diffs_nah}
		\end{figure}
		
		
		\begin{figure}
			\centering
			\resizebox{\columnwidth}{!}{\includegraphics[page=1,scale = 0.65]{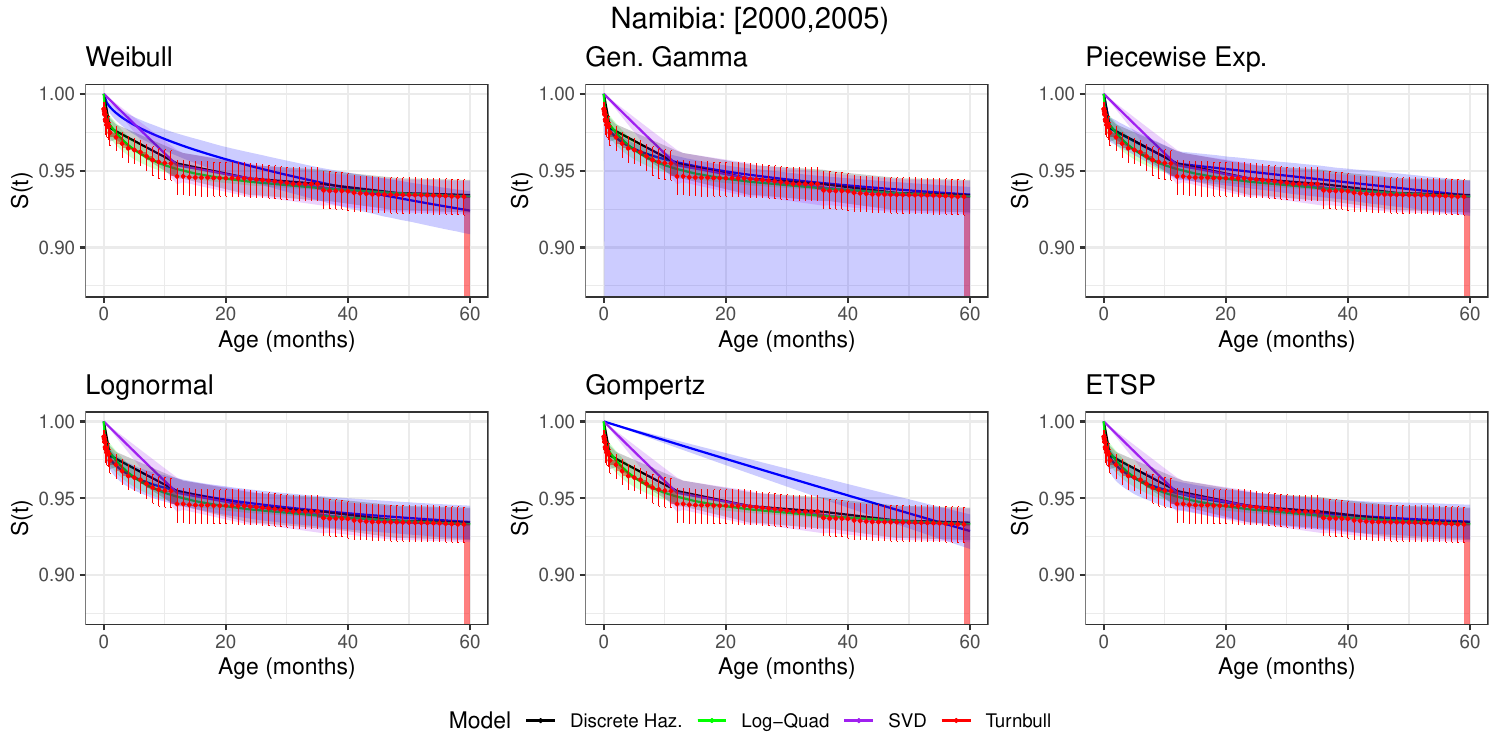}
			}    \resizebox{\columnwidth}{!}{\includegraphics[page=2,scale = 0.65]{Plots/Namibia/curves_compare_separated_2000_2010_nah.pdf}}
			\caption{Estimated survival curves for Namibia in $[2000,2005)$ (top) and $[2005,2010)$ (bottom) from ages 0 to 60 months, not adjusted for age heaping at 12 months. Parametric, pseudo-likelihood estimates are in blue. All confidence bands are 95\% confidence intervals based on finite population variances, with the exception of the log-quad model where uncertainty is calculated as in \citet{guillot2022modeling}.}
			\label{fig:nam_curves_nah}
		\end{figure}
		
		\begin{figure}
			\centering
			\resizebox{\columnwidth}{!}{\includegraphics[scale = 0.65, page = 1]{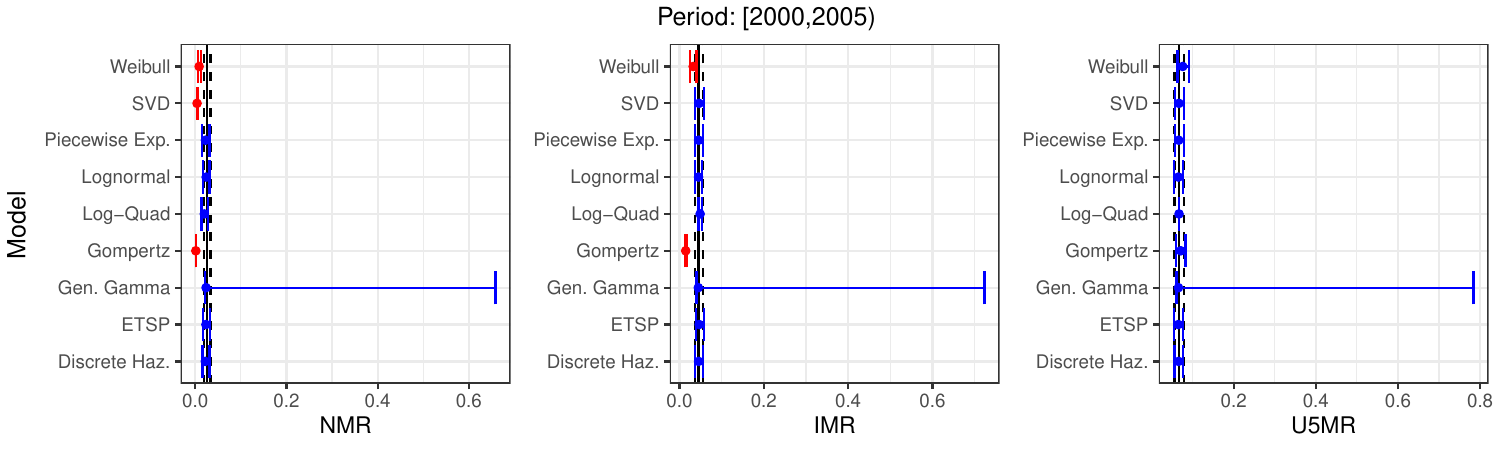}}
			\resizebox{\columnwidth}{!}{\includegraphics[scale = 0.65, page = 2]{Plots/Namibia/nmr_imr_u5mr_compare_2000_2010_nah.pdf}}
			\caption{Estimates of NMR, IMR, and U5MR for Namibia in periods $[2000,2005)$ (top) and $[2005,2010)$ (bottom), not adjusted for age heaping at 12 months. Turnbull point estimates are denoted by vertical black lines. All 95\% confidence intervals are based on finite population variances, with the exception of the log-quad model where uncertainty is calculated as in \citet{guillot2022modeling}.}
			\label{fig:nam_mr_nah}
		\end{figure}
		
		\begin{figure}
			\centering
			\resizebox{\columnwidth}{!}{\includegraphics[scale = 0.65, page = 1]{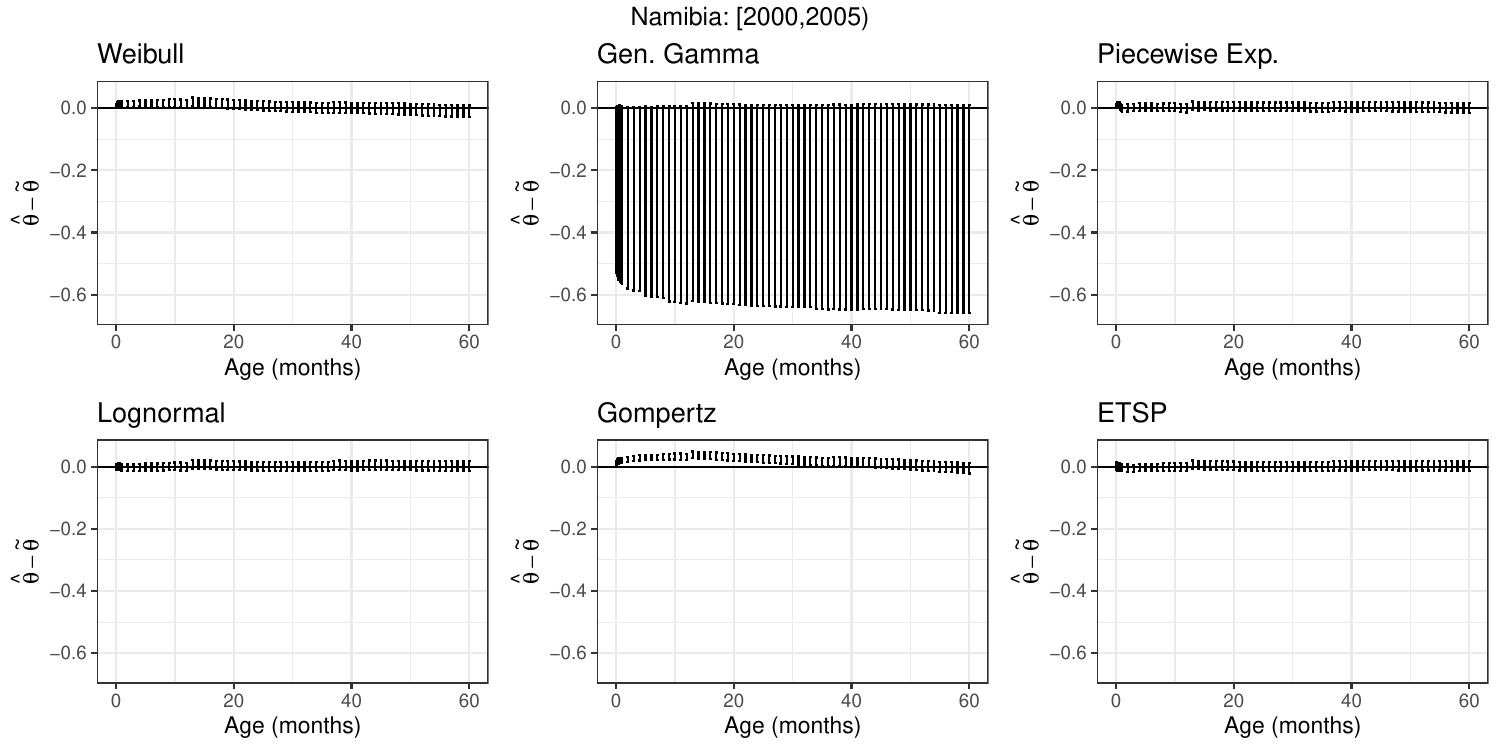}}
			\resizebox{\columnwidth}{!}{\includegraphics[scale = 0.65, page = 2]{Plots/Namibia/turnbull_diffs_nah.pdf}}
			\caption{Empirical distributions of differences in survival curves for Namibia in $[2000,2005)$ (top) and $[2005,2010)$ (bottom) from ages 0 to 60 months between parametric estimates (not adjusted for age heaping) $\hat{\theta}$ and the Turnbull estimate $\tilde{\theta}$.}
			\label{fig:nam_turn_diffs_nah}
		\end{figure}
		
		\begin{table}
			\caption{Model validation results. Percentage of samples (out of 500) from $\hat{\theta} - \tilde{\theta}$ that contain 0 for all parametric models, countries, and periods, for models that do not adjust for age-heaping. Results that contain more than 70\% of samples noted in bold. }
			\centering
			\renewcommand*{\arraystretch}{1.5}
			\resizebox{\columnwidth}{!}{
				\begin{tabular}{l|r r r r r r r r}
					Country                       & \multicolumn{1}{c}{Period} & \multicolumn{1}{c}{Weibull} & \multicolumn{1}{c}{\begin{tabular}[c]{@{}c@{}}Piecewise\\ Exponential\end{tabular}} & \multicolumn{1}{c}{\begin{tabular}[c]{@{}c@{}}Generalized\\ Gamma\end{tabular}} & \multicolumn{1}{c}{Lognormal} & \multicolumn{1}{c}{Gompertz} & \multicolumn{1}{c}{ETSP} & \multicolumn{1}{c}{\begin{tabular}[c]{@{}c@{}}Discrete\\ Hazards\end{tabular}}\\ \hline
					\multirow{2}{*}{Burkina Faso} & {[}2000, 2005)              & 18                        & 37                                                                                 & \textbf{91}                                                                             & 67                           & 15                          & 67    & 64                \\ 
					& {[}2005, 2010)              & 27                         & \textbf{73}                                                                                & \textbf{75}                                                                             & 65                           & 12                          & 70   &   \textbf{74}               \\ \hline
					\multirow{2}{*}{Malawi}       & {[}2000, 2005)              & 42                         & 66                                                                                 & \textbf{94}                                                                             & \textbf{85}                           & 19                          & \textbf{92}           &      \textbf{76}   \\ 
					& {[}2005, 2010)              & 52                         & \textbf{76}                                                                                & \textbf{92}                                                                             & \textbf{86}                           & 17                          & \textbf{88}           &     \textbf{76}       \\ \hline
					\multirow{2}{*}{Senegal}      & {[}2000, 2005)              & 29                        & \textbf{73}                                                                                 & \textbf{85}                                                                             & \textbf{78}                           & 20                         & \textbf{85}             &   \textbf{72}      \\ 
					& {[}2005, 2010)              & 40                         & \textbf{73}                                                                                 & 62                                                                             & \textbf{90}                           & 14                          & \textbf{94}                 &   \textbf{72}   \\ \hline
					\multirow{2}{*}{Namibia}      & {[}2000, 2005)              & 58                         & \textbf{86}                                                                                & \textbf{100}                                                                              & \textbf{100}                           & 27                          & \textbf{99}           &    \textbf{86}        \\ 
					& {[}2005, 2010)              & 56                         & \textbf{86}                                                                                 & \textbf{100}                                                                              & \textbf{100}                            & 29                         & \textbf{99}   & \textbf{85}                 
				\end{tabular}
			}
			\label{tab:modelval_nah}
		\end{table}
		
		In the following plots, we compare the parametric survival curves and Turnbull estimators when age heaping is adjusted for vs. unadjusted. Note that the point estimates for the survival curves are extremely similar for the Weibull, lognormal, Gompertz, generalized Gamma, and ETSP models, only differing in the third or fourth decimal place. This suggests that age-heaping occurring between 6 and 18 months does not greatly impact the overall shape of the survival curve. The age-heaping-adjusted piecewise exponential model differs quite significantly from the unadjusted piecewise exponential model at age 12 months, which is to be expected based on how we interval censored the data in the adjusted model. Also note that in all cases, the uncertainty surrounding the age-heaping-adjusted model is slightly larger than the uncertainty surrounding the unadjusted models, though perhaps not meaningfully larger.
		
		\begin{figure}
			\centering
			\resizebox{\columnwidth}{!}{\includegraphics[page = 1,scale = 0.5]{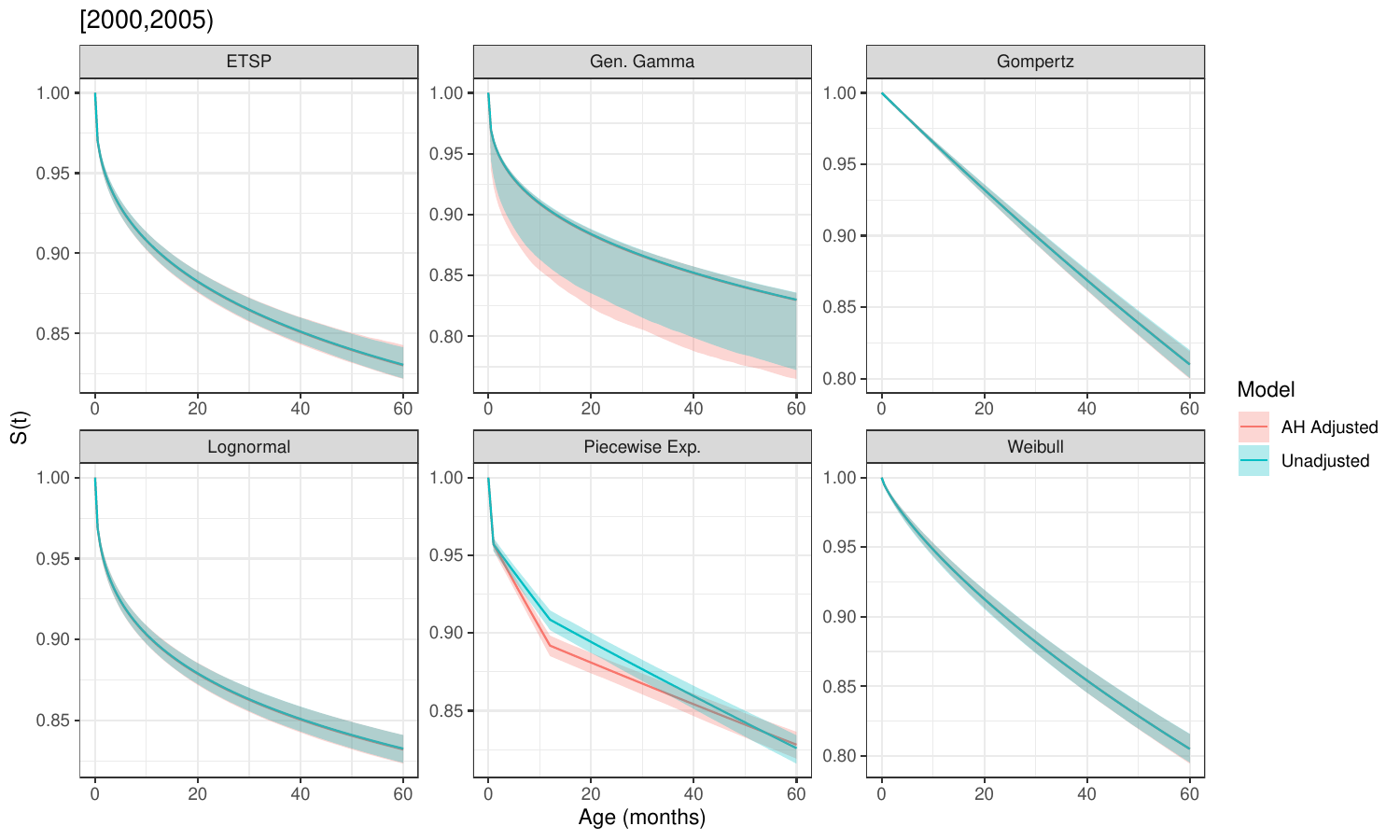}}
			\resizebox{\columnwidth}{!}{ \includegraphics[page = 2,scale = 0.5]{Plots/BurkinaFaso/nah_compare_parametric.pdf}}
			\caption{Comparison of parametric models where data is adjusted for age-heaping at 12 months versus not for Burkina Faso in periods $[2000,2005)$ (top) and $[2005,2010)$ (bottom).}
			\label{fig:bf_nah_compare_par}
		\end{figure}
		
		\begin{figure}
			\centering
			\resizebox{\columnwidth}{!}{\includegraphics[scale = 0.7]{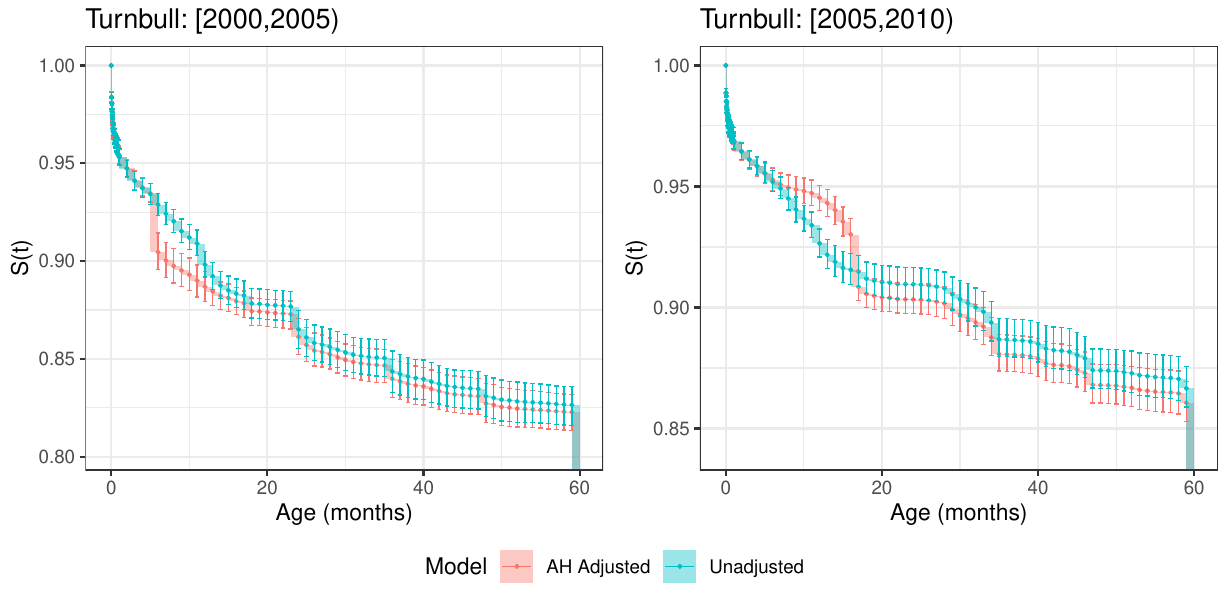}}
			\caption{Comparison of Turnbull estimator where data is adjusted for age-heaping at 12 months versus not for Burkina Faso in periods $[2000,2005)$ (left) and $[2005,2010)$ (right).}
			\label{fig:bf_nah_compare_turn}
		\end{figure}
		
		\begin{figure}
			\centering
			\resizebox{\columnwidth}{!}{\includegraphics[page = 1,scale = 0.5]{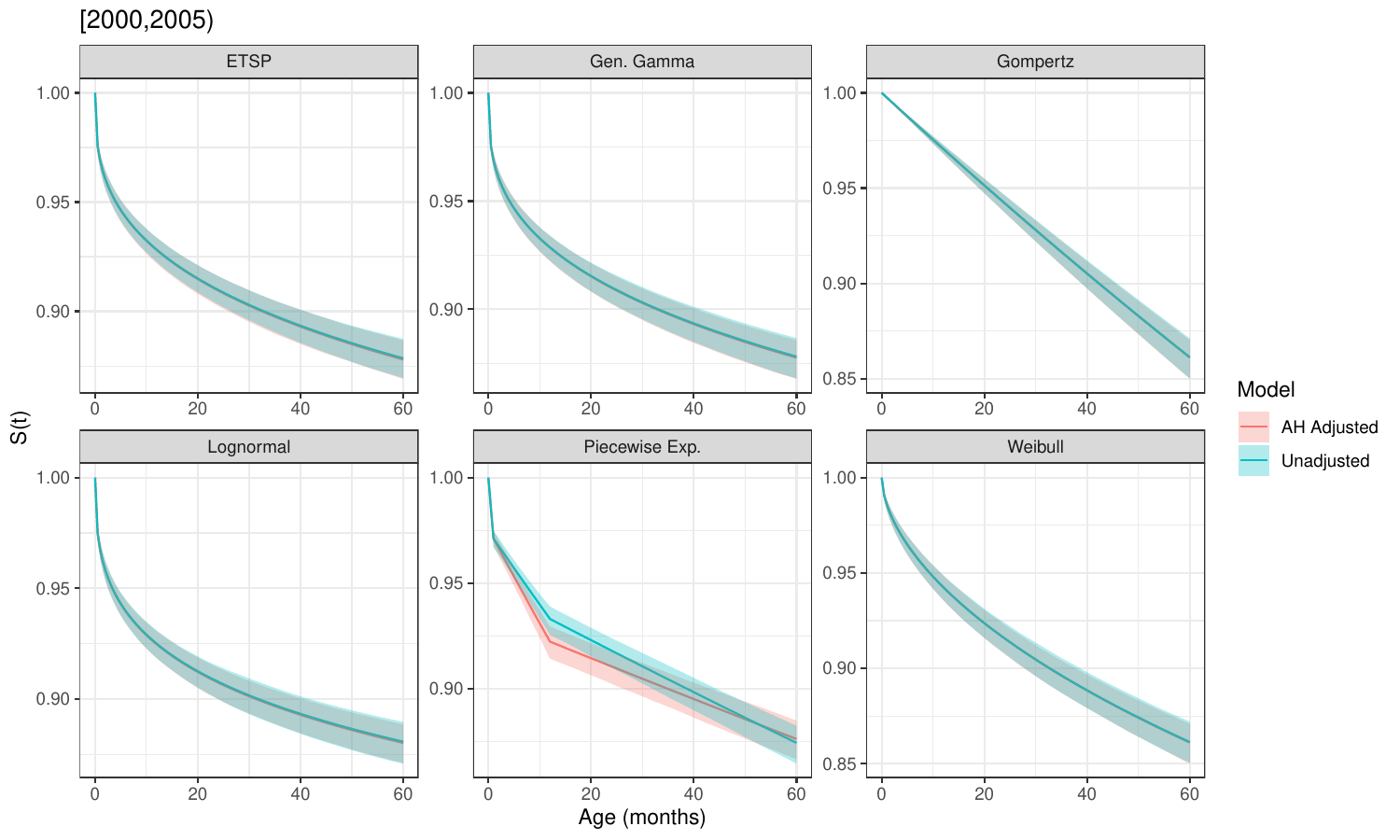}}
			\resizebox{\columnwidth}{!}{\includegraphics[page = 2,scale = 0.5]{Plots/Malawi/nah_compare_parametric.pdf}}
			\caption{Comparison of parametric models where data is adjusted for age-heaping at 12 months versus not for Malawi in periods $[2000,2005)$ (top) and $[2005,2010)$ (bottom).}
			\label{fig:mwi_nah_compare_par}
		\end{figure}
		
		\begin{figure}
			\centering
			\resizebox{\columnwidth}{!}{\includegraphics[scale = 0.7]{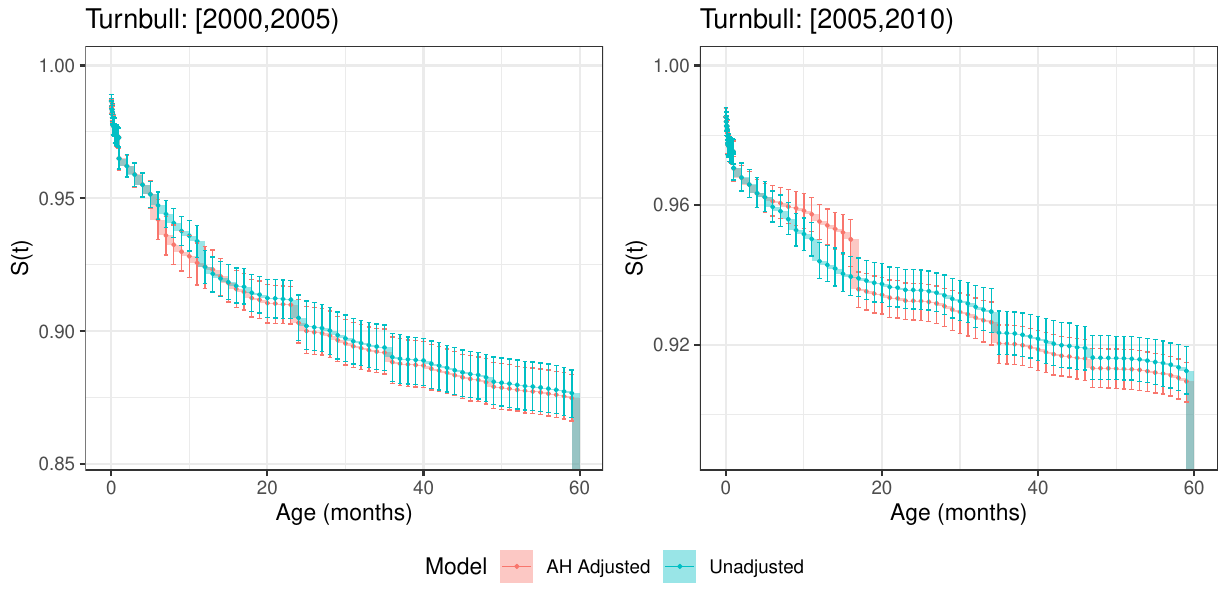}}
			\caption{Comparison of Turnbull estimator where data is adjusted for age-heaping at 12 months versus not for Malawi in periods $[2000,2005)$ (left) and $[2005,2010)$ (right).}
			\label{fig:mwi_nah_compare_turn}
		\end{figure}
		
		\begin{figure}
			\centering
			\resizebox{\columnwidth}{!}{
				\includegraphics[page = 1,scale = 0.5]{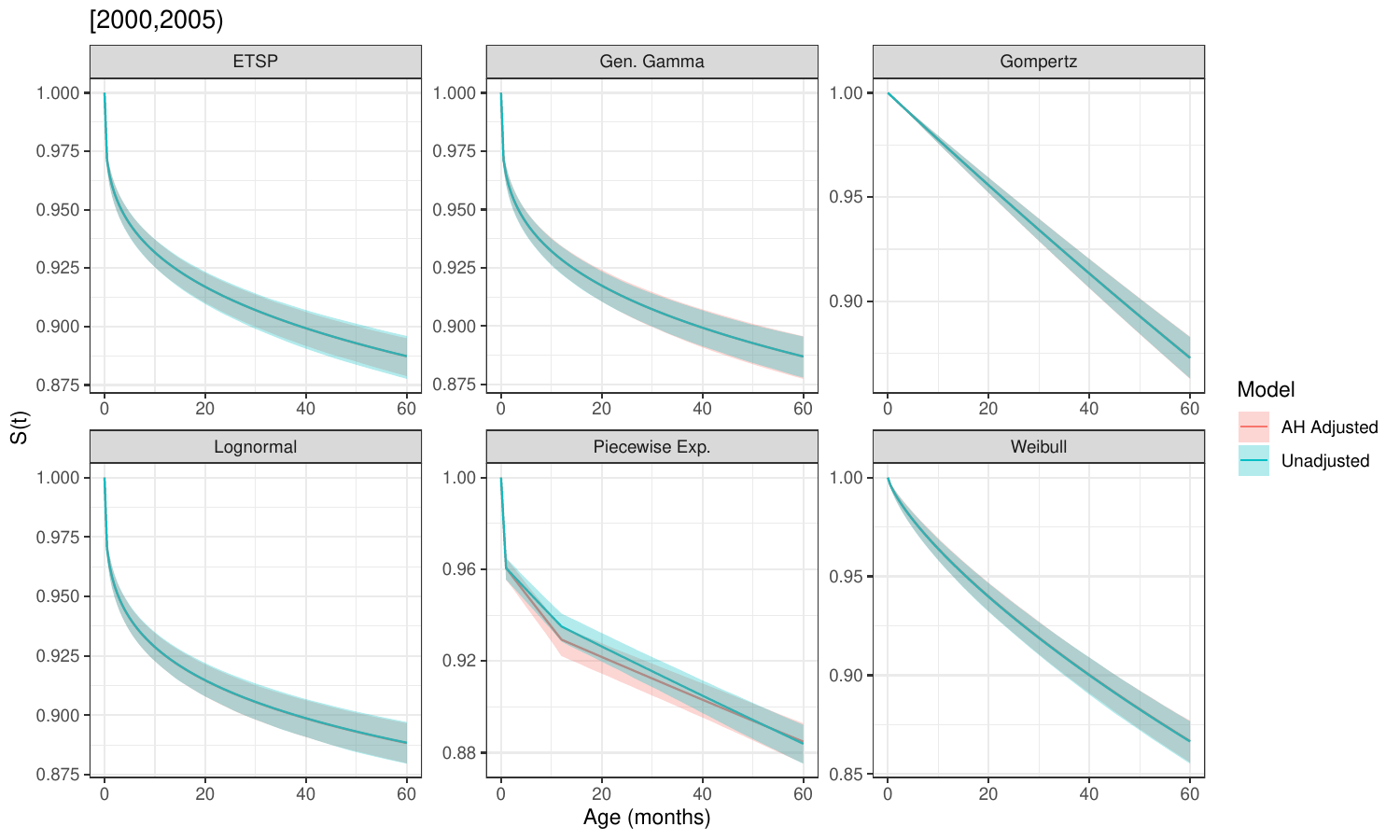}}
			\resizebox{\columnwidth}{!}{
				\includegraphics[page = 2,scale = 0.5]{Plots/Senegal/nah_compare_parametric.pdf}}
			\caption{Comparison of parametric models where data is adjusted for age-heaping at 12 months versus not for Senegal in periods $[2000,2005)$ (top) and $[2005,2010)$ (bottom).}
			\label{fig:sen_nah_compare_par}
		\end{figure}
		
		\begin{figure}
			\centering
			\resizebox{\columnwidth}{!}{\includegraphics[scale = 0.7]{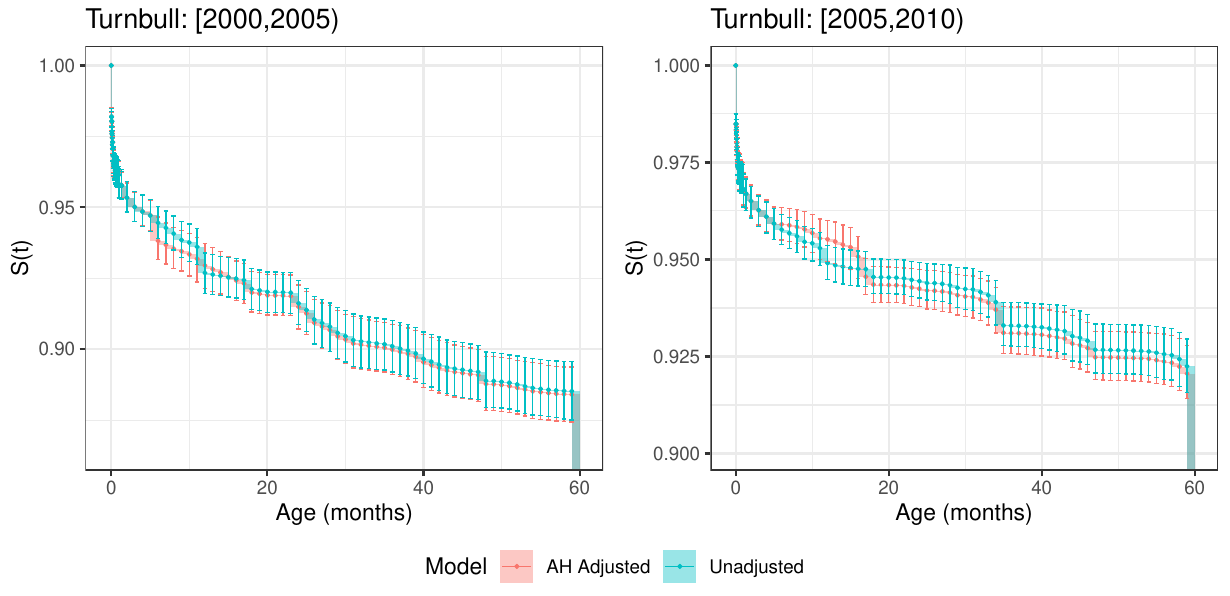}}
			\caption{Comparison of Turnbull estimator where data is adjusted for age-heaping at 12 months versus not for Senegal in periods $[2000,2005)$ (left) and $[2005,2010)$ (right).}
			\label{fig:sen_nah_compare_turn}
		\end{figure}
		
		\begin{figure}
			\centering
			\resizebox{\columnwidth}{!}{
				\includegraphics[page = 1,scale = 0.5]{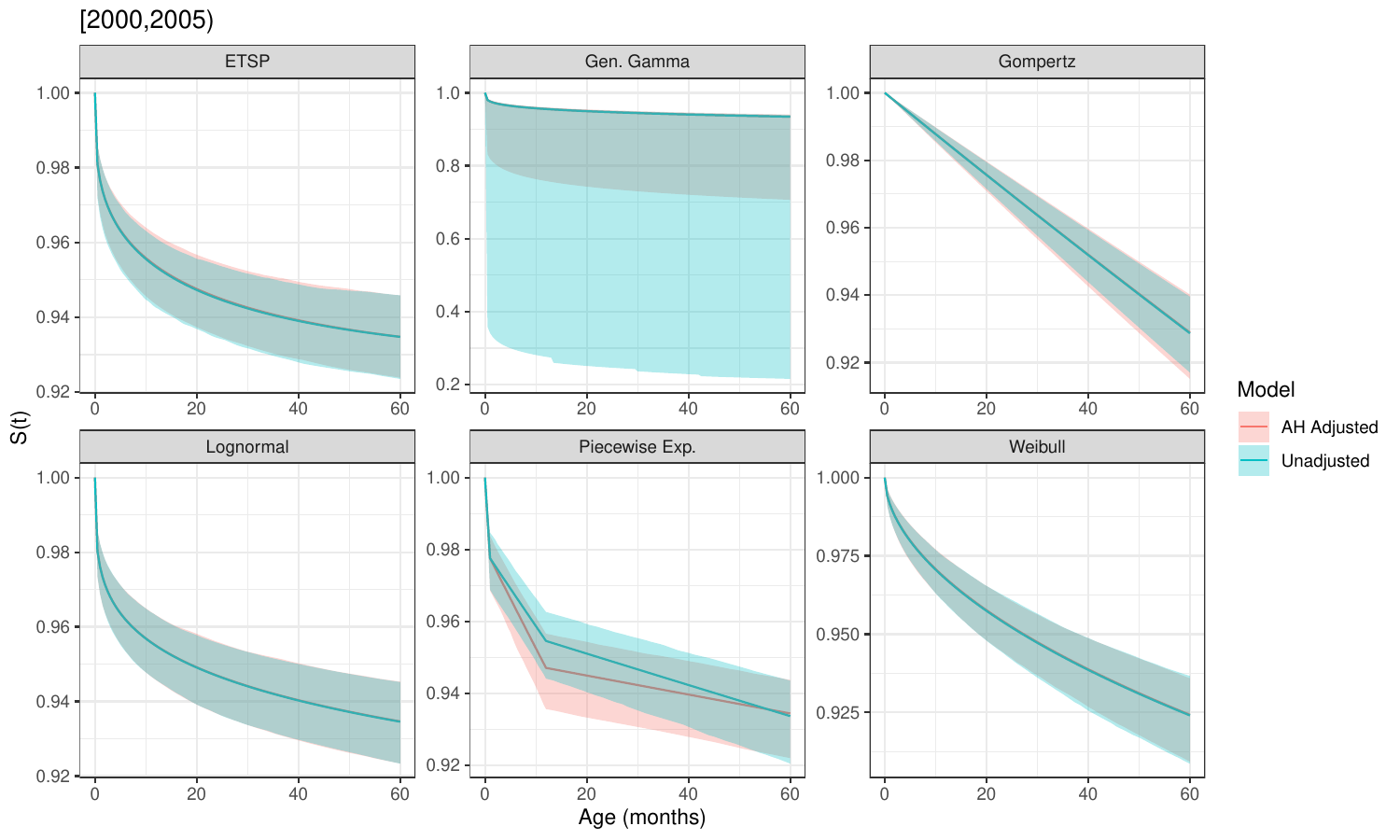}}
			\resizebox{\columnwidth}{!}{
				\includegraphics[page = 2,scale = 0.5]{Plots/Namibia/nah_compare_parametric.pdf}}
			\caption{Comparison of parametric models where data is adjusted for age-heaping at 12 months versus not for Namibia in periods $[2000,2005)$ (top) and $[2005,2010)$ (bottom).}
			\label{fig:nam_nah_compare_par}
		\end{figure}
		
		\begin{figure}
			\centering
			\resizebox{\columnwidth}{!}{%
				\includegraphics[scale = 0.7]{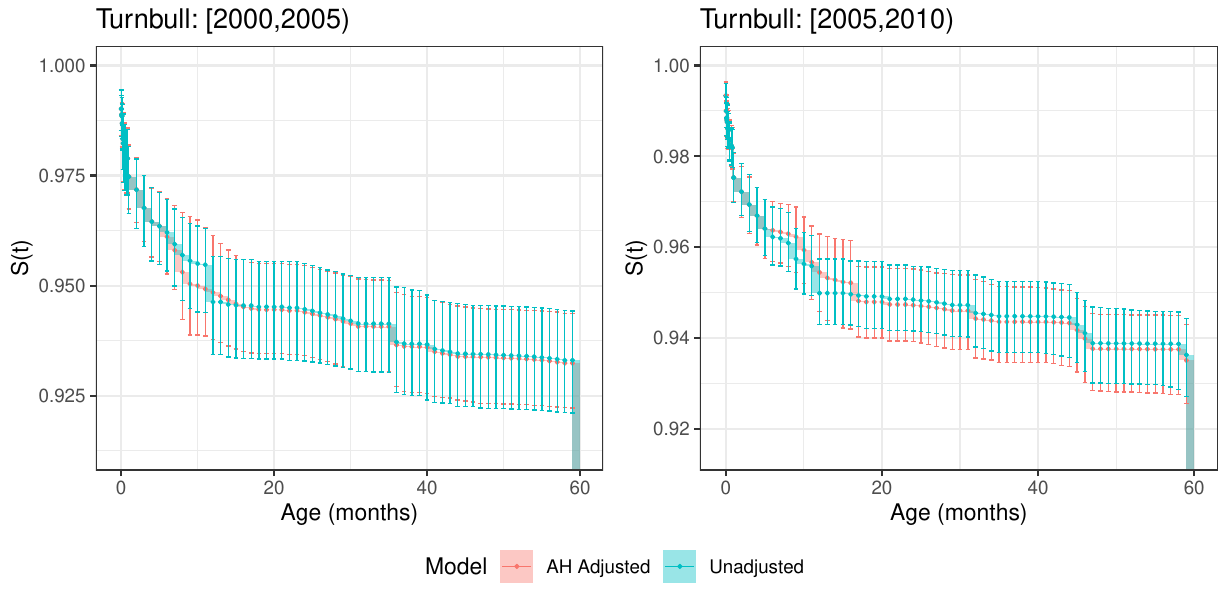}}
			\caption{Comparison of Turnbull estimator where data is adjusted for age-heaping at 12 months versus not for Namibia in periods $[2000,2005)$ (left) and $[2005,2010)$ (right).}
			
			\label{fig:nam_nah_compare_turn}
		\end{figure}
		
		\clearpage
		
	\end{appendices}

\end{document}